\documentclass[a4paper,11pt]{article}
\pdfoutput=1 

\usepackage{jheppub} 
\usepackage{jheppub}
    \usepackage{etoolbox}
    \makeatletter
    \patchcmd{\maketitle}{\@fpheader}{}{}{}
    \makeatother
    
\usepackage[T1]{fontenc} 

\usepackage{graphics} 

\usepackage{epstopdf}
\usepackage{float}
\usepackage{caption}
\usepackage{subcaption}
\usepackage{color}
\makeatletter
\ifcsname phantomsection\endcsname
    \newcommand*{\qrr@gobblenexttocentry}[5]{}
\else
    \newcommand*{\qrr@gobblenexttocentry}[4]{}
\fi
\newcommand*{\addsubsubsection}{%
    \addtocontents{toc}{\protect\qrr@gobblenexttocentry}%
    \subsubsection}
\makeatother

\title{\boldmath Inhomogeneous Thermal Quenches}

\author[1,2]{Kiyoumars A. Sohrabi}
\affiliation[1]{Albert Einstein Center for Fundamental Physics,\\Universit\"{a}t Bern,\\ Sidlerstrasse 5, CH-3012 Bern\\}
\affiliation[2]{Department of Physics, Brandon University,\\ Brandon, Manitoba, R7A 6A9 Canada}

\emailAdd{SohrabiK@BrandonU.CA}

\abstract{
We describe holographic thermal quenches that are inhomogeneous in space.  The main characteristic of the quench is to take the system far from
its equilibrium configuration. Except special extreme cases, the problem has no analytic solution.
Using the numerical holography methods, we study different observables that measure thermalization such as the time evolution of the apparent horizon,
two-point Wightman function and entanglement entropy (EE). Having an extra nontrivial spacial direction, allows us to study this peculiar generalization
since we categorize the problem based on whether we do the measurements along this special direction or perpendicular to it.
Exciting new features appear that are absent in the common computations in the literature; the appearance of negative EE valleys surrounding the
 positive EE hills and abrupt quenches that occupy the whole space at their universal limit are some of the results of this paper. 
 Physical explanation is given and connections to the Cardy's idea of thermalization are discussed.  }

\usepackage{amsmath}
\usepackage{hyperref}

\def\Eq(#1){Eq.~(\ref{#1})}
\def\be{\begin{equation}}
\def\ee{\end{equation}}
\def\bea{\begin{eqnarray}}
\def\eea{\end{eqnarray}}
\def\half{\frac{1}{2}}

\begin{document}

\maketitle
\flushbottom
\section{Introduction and motivation}

Experiments of the heavy-ion collisions have provided a magnificent opportunity to study strongly coupled systems 
\cite{experiments}.
An important part of this study is to understand the physics of the thermalization in which the fascinating
state of matter ``quark-gluon plasma'' has formed \cite{ideal_hydro}.

In the last decade,  extensive studies of the hot plasmas close to equilibrium using the weakly coupled field theories have been performed.
While the regime of the validity of those results is limited, they have contributed a great deal to our physical interpretation \cite{Arnold:2000dr} and have been the motivation for more complex computational toolboxes.

Gauge$/$gravity duality \cite{Maldacena:1997re} together with spectral methods  have become a successful phenomenological framework  \cite{Chesler:2013lia} \cite{OthersNumerics}  to study the above mentioned systems in the regime where they can be arbitrarily far from equilibrium while the theory is experiencing strongly coupled behaviors. This is indeed the regime that we are mostly interested to study the physics of thermalization which allows us to gather information
about subtle and more realistic setups that were  seemingly out of reach. Example of such scenarios often includes breaking of symmetries to
incorporate the realistic features. This can be conformality, supersymmetry or a simple time and spatial translational invariance.

An easy way to construct such a setup that can have the above attributions is deduced by  simply making an abrupt change in one or some of the couplings of a microscopic theory, in our context a  quantum field theory, that governs the dynamics of the system. Then the theory is said to undergo a quantum quench \cite{Calabrese:2005in}\cite{0808.0116}\cite{2dcft}. The most common type of quench which in part is also very simple to interpret is to change the mass of the QFT i.e to produce a mass gap artificially. As the goal of studying quenches is to observe thermalization, one can see that a rapid change in the mass of the action or the corresponding Hamiltonian will correspond to excess of energy that has to be shared among  new degrees of freedom in the new system. The physics of how the quantum system will manage to reach this new state which can or cannot be accompanied by a thermal process, will be of great importance to us \cite{Calabrese:2005in}.

Of course, our primary interest is the non-Abelian QCD plasma which has a strongly coupled dynamics. QCD's  long distance
behavior at high temperature is hoped to be more or less described by the pure $\mathcal{N}=4$ super Yang-Mills. In light of this connection, attempts have been made to mimic some aspects of the
QCD which maybe enable us to use the AdS/CFT duality. The maximally supersymmetric content of the theory contains degrees of freedom such as adjoint fields that are absent in QCD but still has a
good resemblance to the quark-gluon plasma that we are interested in. It turns out that we can modify the $\mathcal{N}=4$ SYM further to overcome some of the physically unwanted features
of the theory. One example, in this regard, is  breaking the conformality in $\mathcal{N}=4$ SYM by adding a bare mass term \cite{Donagi:1995cf}. The resulting theory is $\mathcal{N}=2$ \footnote{
This should not be confused by a closely related model of  $\mathcal{N}=1^{\ast}$ $SU(N)$ gauge theory which is another possibility of softly breaking $\mathcal{N}=4$ by a chiral multiplet mass term.
}
with massive hypermultiplets in the adjoint representation i.e $\mathcal{N}=2^{\ast}$ with a nontrivial RG flow \cite{Pilch:2000ue}. Note again that at high temperatures this mass deformation  will become irrelevant. The superpotential for the hypermultiplet mass term then will consist structures such as  $\text{Tr} Q^{2}+\text{Tr} \tilde{Q}^2$ and $\text{Tr}\left(\left[Q,\tilde{Q}\right]\Phi\right)$ with $Q$, $\tilde{Q}$ the $\mathcal{N}=2$ hypermultiplets and $\Phi$ is an adjoint chiral superfield which is related to a gauge field under $\mathcal{N}=2$. These superpotential terms have been expanded in terms of their matter content simply in the form \cite{Buchel:2007vy},
\be\label{massdeform}
\delta S=-2\int d^{4}x\left(m^{2}_{b}\mathcal{O}_{2}+m_{f}\mathcal{O}_{3}\right)\,,
\ee
with operators $\mathcal{O}_{2}$ and $\mathcal{O}_{3}$ defined according to
\bea\label{operators1}
\mathcal{O}_{2}&=&\frac{1}{3}\text{Tr}\left(|\phi_{1}|^{2}+|\phi_{2}|^{2}-2|\phi_{3}|^{2}\right)\,,\\
\label{operators2}
\mathcal{O}_{3}&=&-\text{Tr}\left(i\psi_{1}\psi_{2}-\sqrt{2}g_{\text{YM}}\phi_{3}\left[\phi_{1},\phi^{\dagger}_{1}\right]
+\sqrt{2}g_{\text{YM}}\phi_{3}\left[\phi^{\dagger}_{2},\phi_{2}\right]+\text{h.c.}\right)
\nonumber\\&&+\frac{2}{3}m_{f}\text{Tr}\left(|\phi_{1}|^{2}+|\phi_{2}|^{2}+|\phi_{3}|^{2}\right)\,,
\eea 
and $m_{b}$ and $m_{f}$ are bosonic and fermionic masses that will be determined below.

The holographic dual (supergravity)  of the above theory was studied elegantly by Pilch and Warner in \cite{Pilch:2000ue}. In their work,
the supergravity scalar fields dual to the operators defined in \Eq(operators1)-\Eq(operators2) named as $\alpha$ and $\chi$
satisfy a potential and kinetic term given by:
\bea\label{superpotential}
\mathcal{V}&=&-\frac{g^{2}}{4}e^{-4\alpha}-\frac{g^{2}}{2}e^{2\alpha}\cosh\left(2\chi\right)+\frac{g^{2}}{16}e^{8\alpha}\sinh^{2}\left(2\chi\right)\,,\\
T&=&-3\left(\partial \alpha\right)^{2}-\left(\partial\chi\right)^{2}\,.
\eea
For more details of the construction and the RG flow refer to \cite{Khavaev:1998fb,Evans:2000ct} . Having this dictionary
for the AdS/CFT duality, made exploration of different aspects of the theory that has great resemblance   to QCD   possible \cite{Donagi:1995cf}. Particularly, at finite temperatures,  thermodynamics of $\mathcal{N}=2^{\ast}$ $SU(N)$ gauge theory at large 't Hooft coupling has been at the center of various works. Buchel,  Deakin, Kerner and Liu showed that at temperatures that are near the mass scale of the theory, thermodynamics attributed   to the mass deformation is irrelevant  and derived the finite temperature version of the Pilch-Warner flows at the boundaries \cite{Buchel:2007vy}. This latter study was then extended to find the behavior of the thermal screening masses of the QGP and beyond to lower temperatures \cite{Hoyos}. Various aspects of the free energy of the $\mathcal{N}=2^{\ast}$ were reported in \cite{Buchel:2003ah} and further on,  corrections to the transport coefficients were derived \cite{Buchel:2004hw}. For a work on finite baryon density in this context refer to 
\cite{Kobayashi:2006sb}.

An enlightening simplicity  appears in the regime where $m_{b,f}/T\ll1$ since in this limit  a black hole has formed inside and the boundary of the bulk space will be asymptotically an AdS space. This motivates us \cite{Buchel:2012gw} to expand the scalar fields in \Eq(superpotential)
to obtain
\be\label{action}
S_{5}=\frac{1}{16\pi G_{5}}\int d^{5}x\sqrt{-g}\left(R+12-\half(\partial\phi)^{2}-\half m^{2}\phi^{2}+\mathcal{O}(\phi^{3})\right)\,,
\ee
where in the above $\phi\in\left\{2\sqrt{6}\alpha,2\sqrt{2}\chi\right\}$ with the corresponding masses $m^{2}\in\left\{-4,-3\right\}$ and $G_{5}\equiv\frac{\pi}{2N_{c}^{2}}$. Note that we have put the radius of AdS in \Eq(action) equal to one.
It must be clear that in the above range of temperatures, we're looking at large scale  black holes and it is reasonable to treat the amplitudes of the scalar fields perturbatively with respect to the former length scales and the length $l\sim m_{f}/T$ will be used to truncate the backreaction.

Now, we are at the position to make the connection to the quench picture more concrete. As mentioned above, the result of the mass deformation is to map  our starting point i.e $S_{\text{SYM}}$ of $\mathcal{N}=4$  into  $S_{\text{SYM}}+\delta S$ with $\delta S$ defined already in \Eq(massdeform). The operators $\mathcal{O}_{2}$ and $\mathcal{O}_{3}$ that  are dual to the scalar field $\phi$, with different masses,
have different dimensions based on their structures in the superpotential. If $\Delta$ is the dimension of each operator, then the corresponding mass of the dual scalar field will satisfy \cite{Hoyos} $\Delta(\Delta-4)=m^{2}$. In other words, in the boundary theory, one of the operators namely $\mathcal{O}_{3}$ couples to a fermionic mass $m_{f}$ and $\mathcal{O}_{2}$ couples to a bosonic mass. Similar to \cite{Alex2014}, we will concentrate only on the fermionic operator in this paper and fix the dual mass of the scalar field to $m^{2}=-3$.

By fixing the parameters of the bulk theory, it was  remarkably suggested \cite{Buchel:2012gw} to use a toy profile for $m_{f}$.
Among various choices, the profile that produces a mass gap is particularly interesting. This evolution can be simply written in terms of the step function, $m_{f}=m_{0}\theta(\tau)$, as a function of real time or a more smooth and  articulated variation of it
\be\label{quenchcoupling}
m_{f}=\half m_{0}\left[1\pm\tanh(\tau)\right]\,.
\ee
Either way, the system can start  from  a massless (massive) ground state and end up in a massive (massless) eventual state after thermalization \cite{Alex2014}. We refer to this setup as the \textit{homogeneous} scenario. 
Calabrese and Cardy came up with an attractive idea to describe the effect of such an evolution of a mass gap \cite{Calabrese:2005in}. In their ``horizon effect'' picture, semi-classical propagations (quasiparticles) \footnote{
The concept of quasiparticles has an old history in thermal QFT and it has been used successfully in the perturbative and close to equilibrium physics, but not at far from equilibrium and strongly coupled systems.
} 
at the initial state or in fact,  every imaginary Cauchy surface that was satisfying causality, was responsible for the later thermalization of the system. A key point that came up in their discussion, was to associate with each coherent set of particles an effective temperature $T_{eff}$. Then at later times, interference of incoherent quasiparticles that sets off their journey in an uncorrelated fashion, derives the system to thermalization. It was further speculated by the authors that this can be a thermal process such as a thermal diffusion. To clarify this idea further, in \cite{0808.0116} they studied the evolution of the mass deformation with an inhomogeneous initial state in models such as conformal and free field theory.

These ideas are worth a second look. We're curious to know if the final stationary state of matter depends in any way on the initial state to begin with. Having an extra toy dimension that affects the dynamics  will help us in this direction. If the theory is very symmetric, motion of trajectories will be confined to a specific section of the phase space, this should be compared with a less symmetric case that trajectories will occupy the whole space of  solutions and therefore a more realistic situation to study in the case of the  thermalization. Reference \cite{0708.1324} has looked into this point with different settings.

We will not consider an inhomogeneous initial state but rather  extend \Eq(quenchcoupling) to include the following form 
\be\label{p0-intro}
m_{f}=\frac{1}{2}\left[1+\tanh\left(\frac{\tau}{\alpha}\right)\right]\,e^{-\frac{x^{2}}{\sigma^{2}}}\,.
\ee
This is the \textit{inhomogeneous} scenario that we will consider. The response of the strongly coupled $\mathcal{N}=4$ supersymmetric Yang-Mills thermal plasma will be studied while it is quenched by 
 tuning parameters $\alpha$ and $\sigma$ that play the role of different scales for perturbations in time and space respectively.
Note that the natural scale of the problem is set by the initial scale of the horizon, $\pi T$\footnote{For numerical purposes, we factor out scales of the coordinates such as $\rho_{new}=\frac{\pi T}{r_{old}}$ , $x_{new}= \frac{x_{old}}{\pi T}$ and $\tau_{new}=\pi T \tau_{old}$. And we will be working with the ``new'' variables. This factorization also affects components of the metric for instance $A_{new}=(\pi T)^{2}A_{old}$ $\,,\cdots$ }.

 We will consider a cherry picked range of $\alpha$ and $\sigma$.  In this way, we can have more control and a better insight into the physics of thermalization.  The chosen values for the parameters in \Eq(p0-intro) in the text, correspond to interesting physics such as the limit of slow/fast quenches with various sizes of spacial inhomogeneity.           
 
 To solve the problem, we will be using an ansatz with 4 arbitrary\footnote{Please refer to \cite{YaffeLong}.}
  functions of space and time with $x$ being the coordinate that profiles are inhomogeneous with respect to it,
\be\label{fullmetric1}
ds^{2}_{5}=-A(\tau,\rho,x)d\tau^{2}+\Sigma_{d} (\tau,\rho,x)^{2}dx^{2}+\Sigma_{b} (\tau,\rho,x)^{2}d\vec{y}^{2}
+2\Xi(\tau,\rho,x)d\tau dx-2\frac{d\rho d\tau}{\rho^{2}}\,,
\ee
and if for the brevity of argument, we neglect the logarithmic corrections and higher order terms here,  the boundary could be written as\footnote{The complete list is outlined in the appendix.}
\bea
\phi&=&l\left(\rho\, p_{0}
+\rho^{2}\,\partial_{\tau}p_{0}
+\rho^{3}\,p_{2}\right)+\mathcal{O}(l^{3},\rho^{4})\,,
\\
A&=&\frac{1}{\rho^2}-\rho^2+l^{2}\left(-\frac{1}{6}p_{0}^{2}
+\rho^{2}\,a_{2}\right)+\mathcal{O}(l^{4},\rho^{2}\ln\rho)\,,
\\
\Sigma_{d}&=&\frac{1}{\rho}+l^{2}\left(-\rho^{2}\,\frac{p^{2}_{0}}{12}
-\rho^{3}\,\frac{p_{0}\partial_{\tau}p_{0}}{9}
+\rho^{4}\,d_{4}\right)+\mathcal{O}(l^{4},\rho^{4}\ln\rho)\,,
\\
\Sigma_{b}&=&\frac{1}{\rho}+l^{2}\left(-\rho^{2}\,\frac{p^{2}_{0}}{12}
-\rho^{3}\,\frac{p_{0}\partial_{\tau}p_{0}}{9}
+\rho^{4}\,b_{4}\right)+\mathcal{O}(l^{4},\rho^{4}\ln\rho)\,,
\\
\Xi&=&l^{2}\left(-\rho\,\frac{p_{0}\partial_{x}p_{0}}{9}
+\rho^{2}\,f_{2}\right)+\mathcal{O}(l^{3},\rho^{2}\ln\rho)\,,
\eea
where in the above $p_{0}$, $p_{2}$, $a_{2}$, $b_{4}$, $d_{4}$ and $f_{2}$ depend on $(\tau,x)$.  Note that from the AdS/CFT dictionary $m_{f}=p_{0}$.
 These functions will satisfy Einstein equations that are coupled second order partial differential equations.
 To solve them numerically, we will apply spectral methods and techniques developed by Chesler and Yaffe \cite{Chesler:2013lia} and use Dirichlet boundary
 condition for the longitudinal direction. The accuracy of our physical results are certainly  limited to our computational resources. While we could quantify
 the effect of the numerical artifacts to be of a few percents to our knowledge none of physical conclusions that are deduced are affected by them.

In this paper, we study various observables already in the literature such as apparent horizon, two-point
Wightman functions and entanglement entropy (EE). Our goal is  to study the thermalization under the  quench in \Eq(p0-intro) for various parameters  with a special emphasize on the study of EE. In section 2, we look into these different nonlocal observables as a measure of the thermalization and different aspects of them will be studied in detail. In section 3, we recap the conclusions and the physical picture deduced from the simulations in previous sections. Section 4 is dedicated to a discussion on fast quenches and section 6 will be our appendix with a through derivation of the equations of motion and numerics.

\section{Thermalization observables}

\subsection{Apparent horizon}
One of the most important quantities in the description of the thermodynamics of a black hole is its statistical entropy as a measure of the number of quantum states. Hawking's famous area relation, $S=\frac{A_{h}}{4G_{5}}$,  makes a connection between this entropy and the area of the black hole's horizon. 
The radius of the former area is determined by the position of the horizon and in our scenario 
 as the scalar field falls into the black hole and radiates,
black hole will expand and its rate is directly related to behavior of the radius.

We consider the metric in \Eq(fullmetric1)
with a simplifying feature of setting a cutoff in the backreaction at second order, explicitly  assuming\footnote{Since the metric is invariant under the residual diffeomorphism
$r\rightarrow r+f(\tau)$ with $r\equiv 1/\rho$, we use this property to fix the expansion of $A(\tau,\rho,x)$ not
to have any linear term in $r$.} 
\bea
\label{backvar}A(\tau,\rho,x)&\!=\!&\frac{1}{\rho^2}-\rho^2+l^2\hat{A}(\tau,\rho,x)+\mathcal{O}(l^4)\,,\\
\label{backvar2}\Sigma(\tau,\rho,x)&\!=\!&\frac{1}{\rho}e^{l^2\hat{\Sigma}(\tau,\rho,x)}+\mathcal{O}(l^4)\,,\\
\label{backvar3}\Xi(\tau,\rho,x)&\!=\!&l^{2}\hat{\Xi}+\mathcal{O}(l^3)\,,
\eea
where $\Sigma$ in the above notation can be either  of $\Sigma_{b}$ and $\Sigma_{d}$ and the expansion parameter  is determined by  
$l\sim m_{f}/T$. Basically, the argument is that we look at the variations of $\phi$ at the order of $l$ and neglect the backreaction on itself. Implementing 
this assumption in the Einstein equations allows us to truncate the series at $\mathcal{O}(l^{3})$ or $\mathcal{O}(l^{4})$ on  different metric components. For an interesting discussion of the thermodynamics of the model refer to \cite{ALMN}. In the following, we use the above components to study the behavior of the apparent horizon of the black hole deep in the bulk.

In a much simpler case where $\Sigma_{d}=\Sigma_{b}\equiv \Sigma$ (the homogeneous spacetime), the equation for the position of the trapping surface
follows from $d_{+}\Sigma=0$ with $d_{+}\equiv\partial_{\tau}-\frac{A\rho^{2}}{2}\partial_{\rho}$. In the general case \cite{YaffeLong}, this equation is modified to\footnote{The $\nabla$ and the dot product are defined according to  $\hat{g}$ with spatial components given by $\hat{g}_{11}=\left(\frac{\Sigma_{d}}{\Sigma_{b}}\right)^{4/3}$ and $\hat{g}_{22}=\hat{g}_{33}=\left(\frac{\Sigma_{b}}{\Sigma_{d}}\right)^{2/3}$.}
\be
d_{+}\Sigma=-\half \partial_{\rho}\Sigma\,\Xi^{2}+\frac{1}{3}\Sigma\, \nabla\cdot\Xi
\ee
with $\Sigma$ now given by $\Sigma\equiv (\Sigma_{d}\Sigma^{2}_{b})^{1/3}$. Applying the expansions in \Eq(backvar)-\Eq(backvar3) gives the position of the trapping surface
\be
\rho_{h}(\tau,x)=\left[\frac{\hat{A}(\tau,\rho,x)}{4}
+\frac{\partial_{\tau}\hat{\Sigma}_{d}(\tau,\rho,x)}{6}+\frac{\partial_{\tau}\hat{\Sigma}_{b}(\tau,\rho,x)}{3}-\frac{\partial_{x}\hat{\Xi}(\tau,\rho,x)}{6}\right]_{\rho=1}\,.
\ee
Knowing the position of the apparent horizon, $\rho_{h}$, the natural quantity to calculate is  the volume of the horizon. 
The volume density of the entropy given by $\mathcal{S}=\frac{V_{h}}{4G_{5}}$
 corresponds to the explicit expression for the perturbation of the volume element
\be\label{volume}
V_{h}=\Sigma_{d}\,\Sigma_{b}^{2}\equiv 1+l^{2}\delta V_{h}\,,
\ee 
 where it has to be calculated at  $\left(\tau,1+l^{2}\rho_{h},x\right)$. This gives the final expression for variation in
the volume element of the apparent horizon
\be
\delta V_{h}=\Big[-\frac{3}{4}\hat{A}-\frac{\partial_{\tau}\hat{\Sigma}_{d}}{2}+\frac{\partial_{x}\hat{\Xi}}{2}-\partial_{\tau}\hat{\Sigma}_{b}
+\hat{\Sigma}_{d}+2\hat{\Sigma}_{b}\Big]_{\rho=1}\,.
\ee
From the above expression, we can see that the introduction of the inhomogeneity directly
 changes the location of the apparent horizon in comparison with the previous calculations  in \cite{Alex2014} and 
\cite{Auzzi}.

\begin{figure}
\begin{subfigure}{.5\textwidth}
{\includegraphics[width=7.0cm]{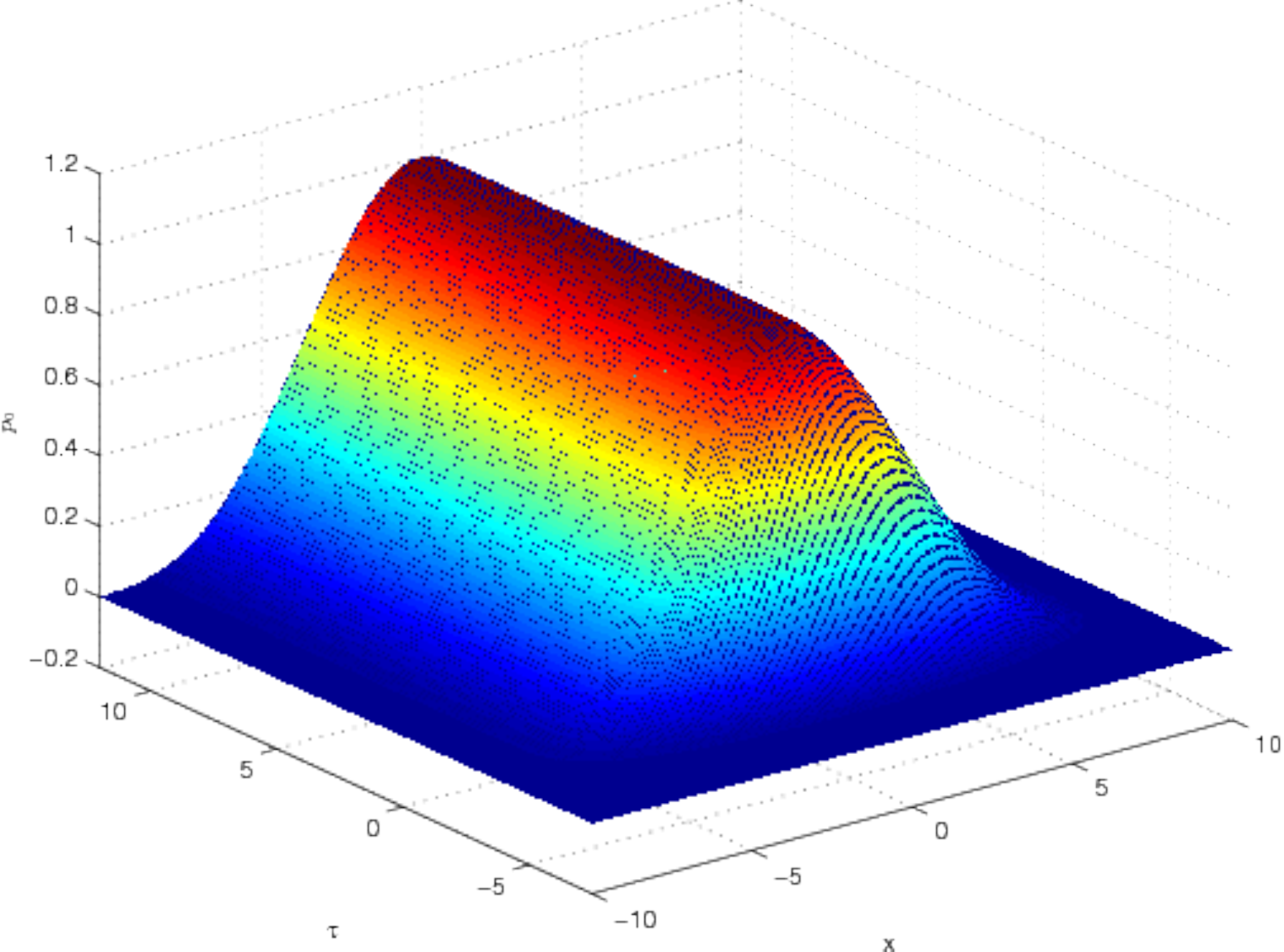}}
  \caption{}
  \label{fig:ref1}
\end{subfigure}%
\begin{subfigure}{.5\textwidth}
{\includegraphics[width=7.0cm]{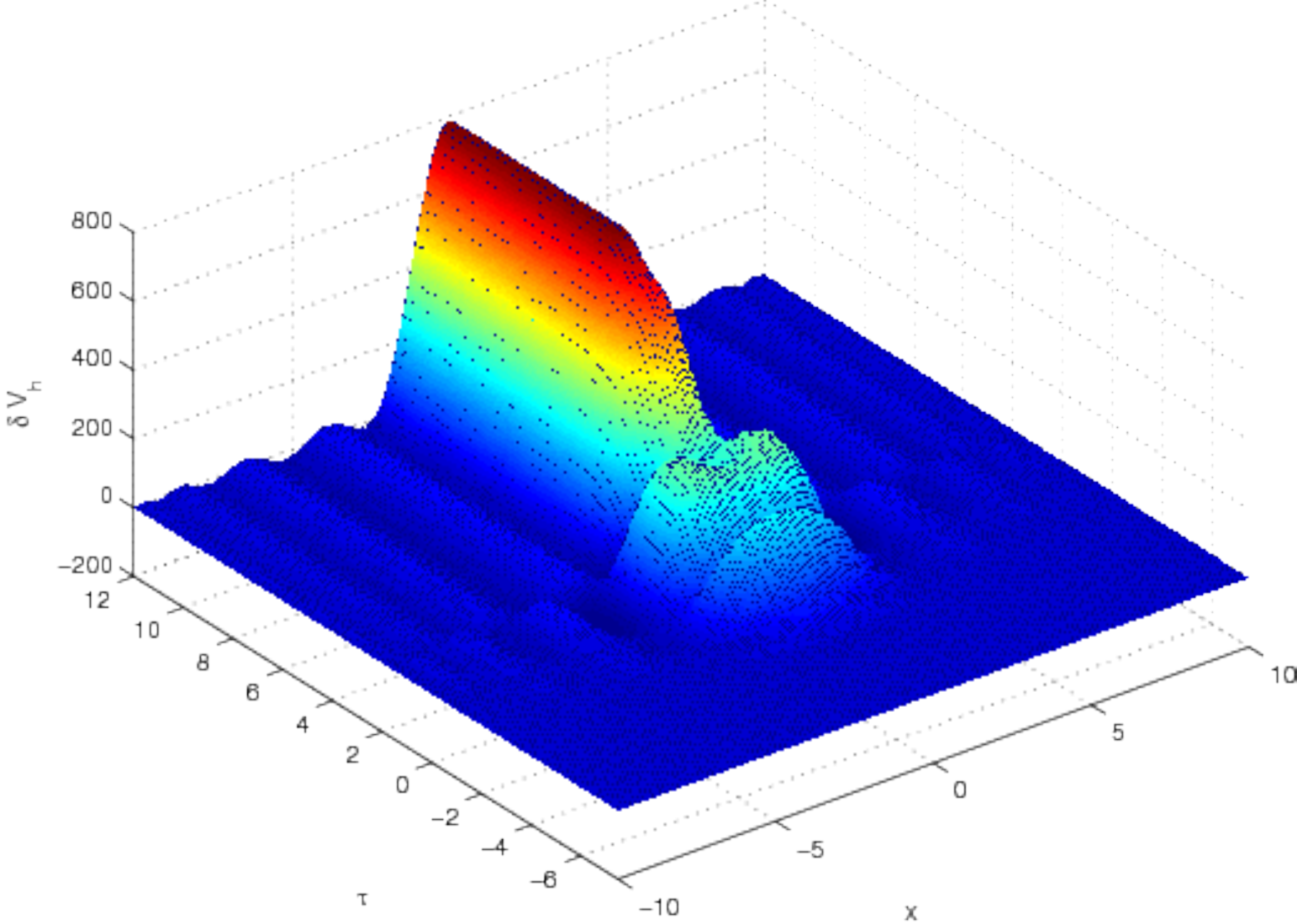}}
  \caption{}
  \label{fig:Vh2cc}
\end{subfigure}
\begin{subfigure}{.5\textwidth}
{\includegraphics[width=7.0cm]{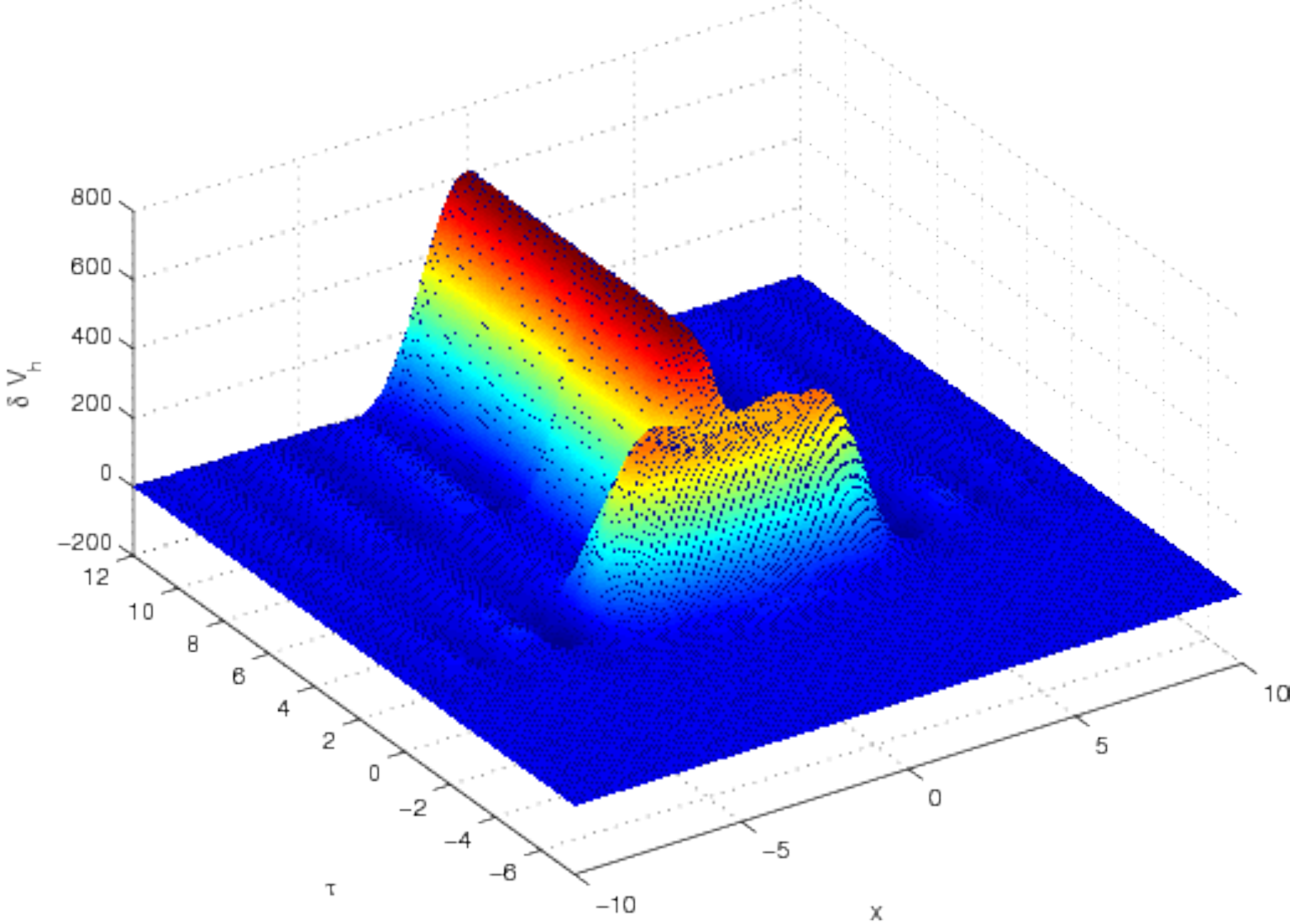}}
  \caption{}
  \label{fig:Vh3a2}
\end{subfigure}  
\begin{subfigure}{.5\textwidth}
{\includegraphics[width=7.0cm]{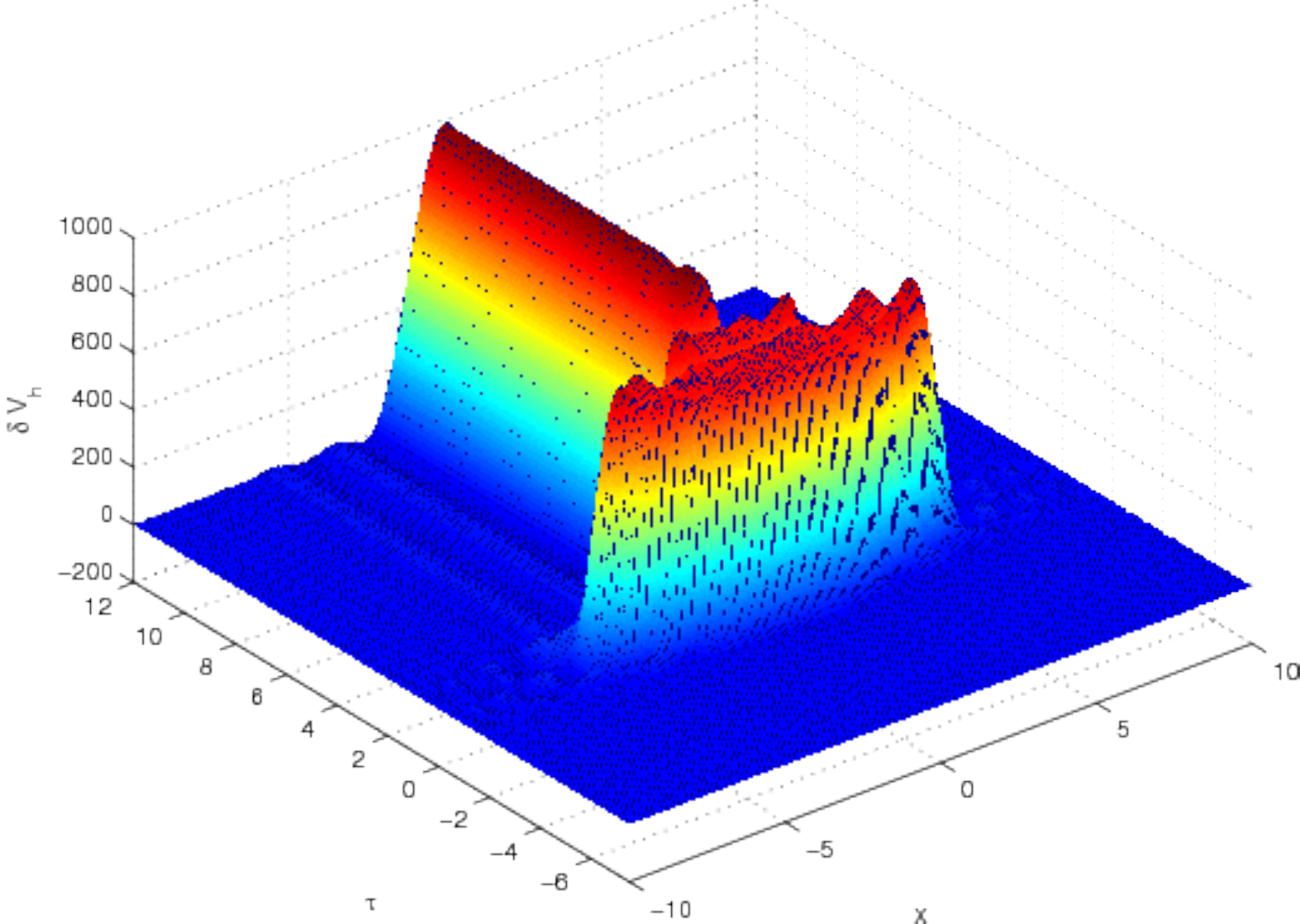}}
  \caption{}
  \label{fig:Vh3b2}
\end{subfigure}%

\begin{subfigure}{.5\textwidth}
{\includegraphics[width=7.0cm]{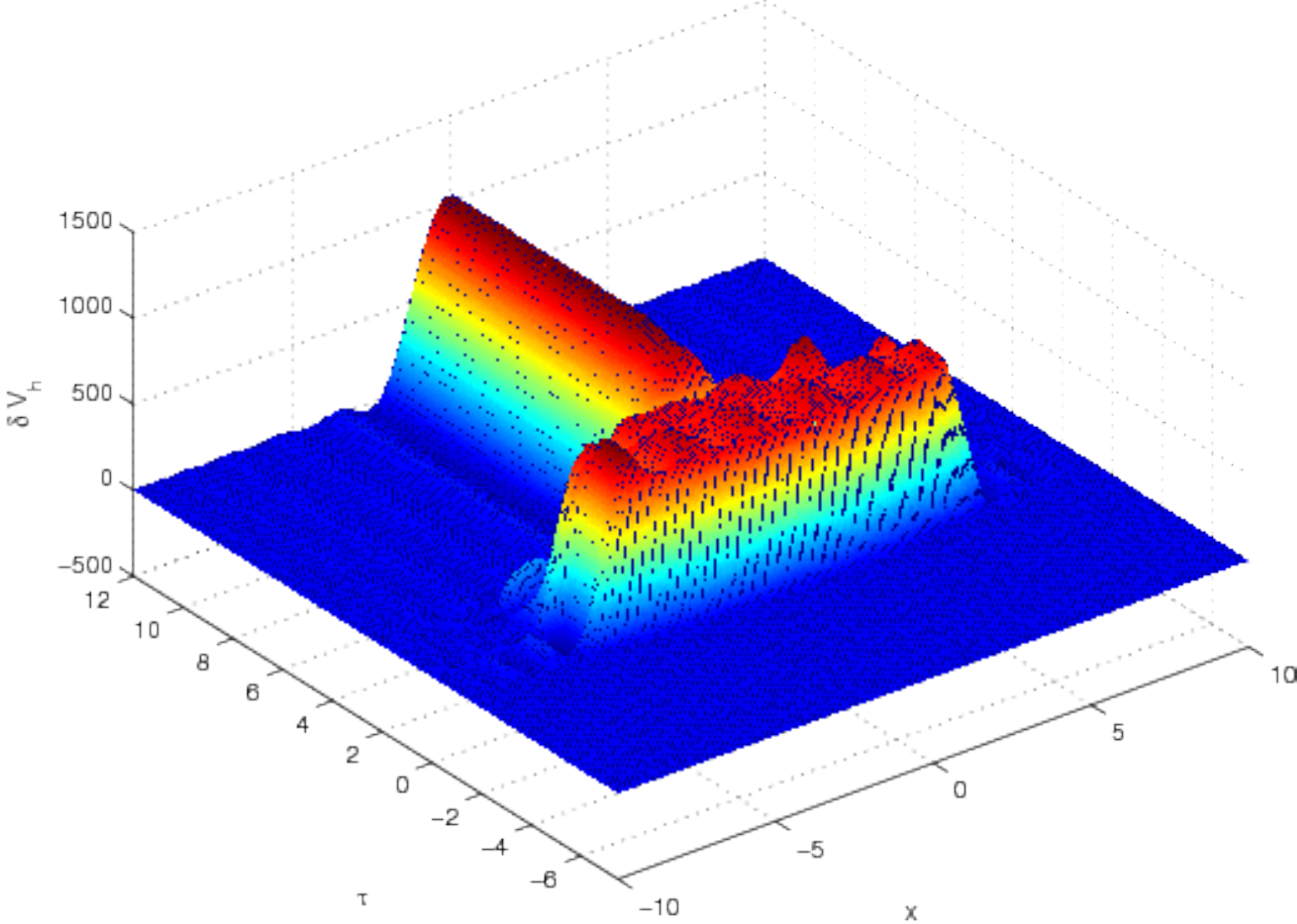}}
  \caption{}
  \label{fig:Vh3c2}
\end{subfigure}
\begin{subfigure}{.5\textwidth}
{\includegraphics[width=7.0cm]{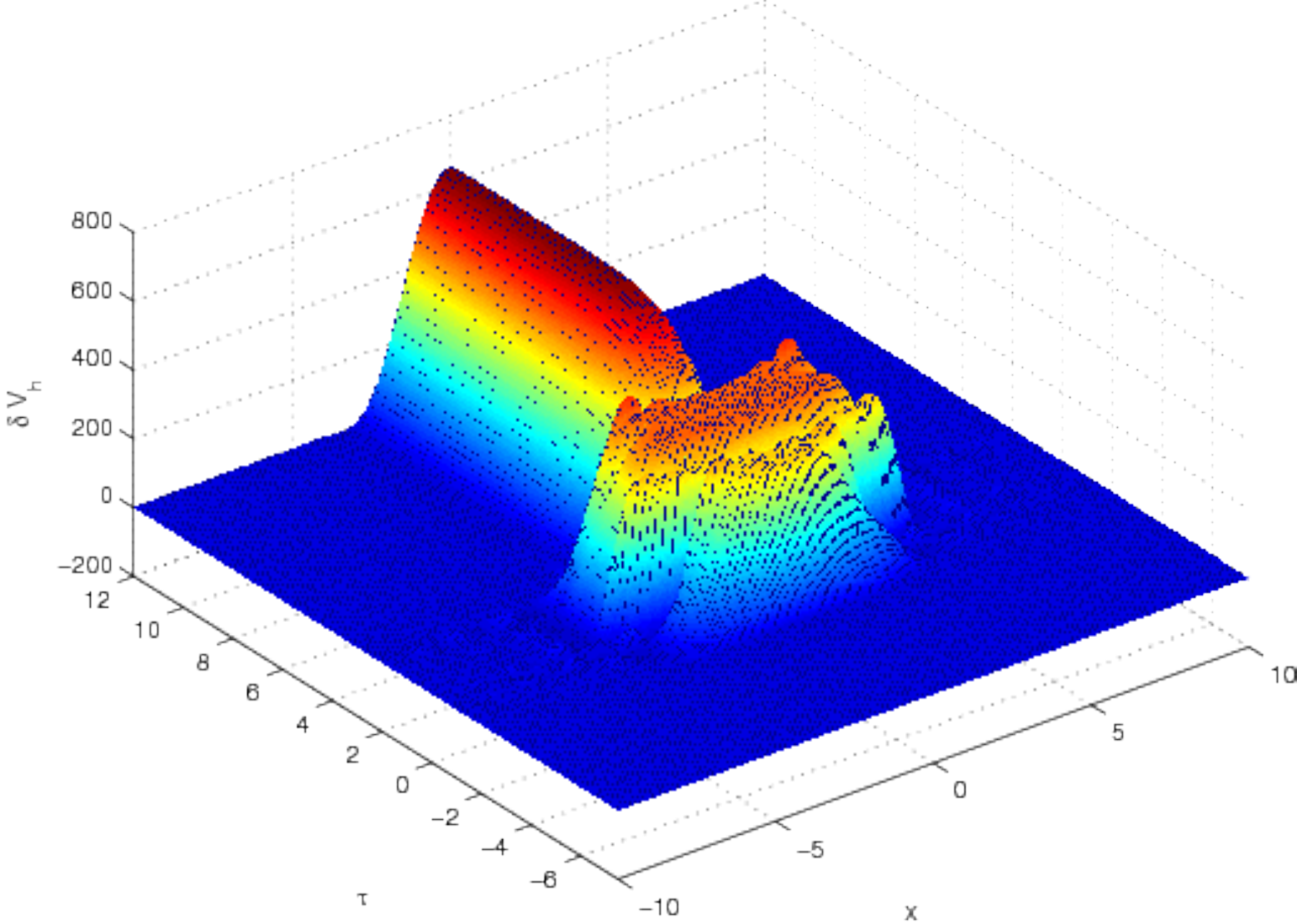}}
  \caption{}
  \label{fig:Vh4a2}
\end{subfigure}%
\caption{Figure (a) is the profile of $p_{0}$ that is being sent into the black hole. The rest of the plots are  time evolution of variations in the radial position of the horizon. In (b), (c), (d) and (e)  plots are drawn for fixed value of $\sigma=\sqrt{L_{x}}$ with $L_{x}=10$ the length of the domain in the $x$ direction. The varying parameters are correspondingly $\alpha\in\left\{1,\half,\frac{1}{4},\frac{1}{8} \right\}$. In (f),  these parameters are $\alpha=1$ and $\sigma=\sqrt{1.5 L_{x}}$.
The interpolation  are based on $N_{x}=N_{\rho}=20-30$ the number of Chebyshev points along the inhomogeneous direction $x$ and radial direction $\rho$. The number of time steps used for the fourth order Runge-Kutta  varies between $7810-17560$.  }
\end{figure}

As a reference, Figure \ref{fig:ref1} shows the plot for $p_{0}(\tau,x)$, read it $m_{f}$, as a function of real  time $\tau$ and inhomogeneous direction  $x$. This is equivalent to the profile of the scalar field that is falling into the black hole from the boundary and the effect of this infall can be seen in the fluctuations of the apparent horizon in Figures \ref{fig:Vh2cc}-\ref{fig:Vh4a2} in $x-\tau$ coordinates.
These plots that match those of \cite{Alex2014}, have been specifically chosen as they show different physics as we vary the tuning parameters. One first clear point is that they all roughly imitate behaviors of their sources. Choosing $x=0$ in $p_{0}(\tau,x)$ will reduce our problem to  \cite{Alex2014}. As it is clear from Figures \ref{fig:Vh2cc}-\ref{fig:Vh4a2}, their behaviors along $x=0$ is very similar. They all follow the profile of $p_{0}(\tau,x=0)$. But they follow different patterns along the inhomogeneous direction. In $p_{0}(\tau,x)$, there are Gaussian profiles in the $x$ direction with amplitudes that are almost constant far away from $\tau=0$, either $\tau>0$ or $\tau<0$. Close to $\tau=0$, the amplitude of the Gaussian distribution increases linearly. This is when the quench has been turned on  and in the vacuum of the QFT a mass gap has been formed. This is evident in Figures \ref{fig:Vh2cc}, \ref{fig:Vh3a2} and \ref{fig:Vh4a2}  for $\tau=0$. It is an interesting fact that at this moment excitations occupy a length equal to the width of the initial Gaussian profile and their amplitudes seem to follow a universal behavior, occupying the whole available space.

As we reduce the value of $\alpha$ in $p_{0}(\tau,x)$, excitations will not only occupy the available space at the $\tau=0$ but they also overrun the original profile of $p_{0}(\tau,x)$ for all $\tau>0$ as seen in Figures \ref{fig:Vh2cc}-\ref{fig:Vh3c2}. In fact it's very hard to distinguish between Figure \ref{fig:Vh3b2} and Figure \ref{fig:Vh3c2} although they physically belong to different sizes of the mass gaps. This is the universal behavior associated to the abrupt quenches that has been discovered in \cite{Buchel,AMN}. 

An interesting feature is captured in Figure \ref{fig:Vh4a2}. By increasing $\sigma$, the tuning parameter corresponding to the width of the Gaussian distribution, mass gap excitations will fill up the available space. 

Some of the features in the plots below should not be confused by physics. They are discretization artifacts and one can in principle
factor them out by improving the computational resources. For instance, the amplitude of the corrugations in the flat areas surrounding the bump to
the highest peak is at maximum $5\%$.  Similarly, the local peaks on top of the bumps at the time of switching the quench is at maximum $9\%$. A short 
discussion about the size of the numerical artifacts and their effects on the thermalization is given in sections \ref{Discretization} and \ref{Thermalization}. 

\clearpage

\subsection{Two-point correlator}

\begin{figure}[h!]
\centering
{\includegraphics[width=8cm]{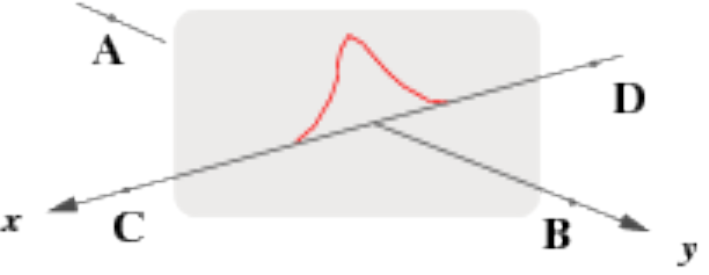}}
{\caption{The disturbance drawn in red pen, is that of a  Gaussian function, representing the inhomogeneity.
We are interested in the correlation of points off this plane i.e. points A and B in Case I. Similarly, in Case II, the correlation
between C and D will be studied. Note the resemblance of the setup to the elliptic flow in heavy-ion collisions. }
\label{fig:Two-point}}
\end{figure}

Two-point Wightman functions are good candidates of probing thermalization. For operators with large masses, the correlation functions
will have a simple interpretation in term of spacelike geodesics that connect two sample points on the boundary of the CFT through the bulk space. Since we have a special direction which is the
direction of the inhomogeneity, we can categorize our setup   into two groups. Case I, will be  the situation where this special
 direction is orthogonal to the axis of observation and Case II, refers to the situation where the points chosen are along the axis of the 
inhomogeneity. This is explained in Figure \ref{fig:Two-point}.

Similar categorization also applies to our discussion in the next section where we extend this setup and study  thermalization
of the  quenches by the entanglement entropy.

\addsubsubsection{Case I: plane A-B }\label{CaseI-two-point}
 To see the effect of the quenches, we are
interested in the length of a geodesic that stretches  along one of the spatial directions.  The other 
simplifying assumption here is that similar to \cite{Alex2014}, we look into correlator of operators
with  large conformal dimensions\footnote{This limit omits the possibility of studying the  correlator of the quenching operator itself.}. 
Then the two-point Wightman function will be proportional to the length of the
 boundary-to-boundary spacelike geodesic \cite{Ross}.
 
For simplicity,  our choice is the curve that satisfies boundary conditions, $\tau_{1}=\tau_{\ast}$, $y_{1}=-y_{m}$, $x_{1}=z_{1}=0$
and $\tau_{2}=\tau_{\ast}$, $y_{2}=y_{m}$, $x_{2}=z_{2}=0$. In other words, not the specific direction that the inhomogeneity 
will act on.
In this setup, the geodesic connects points A and B through their extension in the bulk. The inhomogeneity appears at 
$\mathcal{O}(l^{2})$  along the axis where points C and D are positioned.
To see  how the quench affects the geodesic as we mentioned before, we choose a cut off for 
the backreaction at $\mathcal{O}(l^{2})$.
The effect of this backreaction on the coordinates will be parametrized by
\be
\tau=\tau_{0}+l^{2}\tau_{2}\,,\quad \rho=\rho_{0}+l^{2}\rho_{2}\,,\quad x=l^{2}x_{2}\,.
\ee 
Our former boundary condition imposes $\tau_{0}=\tau_{\ast}$. It's instructive to compute
 the geodesic first, to see explicitly the effect of the inhomogeneity. 
Since the geodesic equations follow from 
$\frac{d^{2}x^{\kappa}}{d\lambda^{2}}+\Gamma^{\kappa}_{\mu\nu}\frac{dx^{\mu}}{d\lambda}
\frac{dx^{\nu}}{d\lambda}\!=\!0$ in some general affine parametrization $\lambda$ 
in Case I and II, different equations of motion will be derived.
It is also interesting to see how the inhomogeneity affects the geodesic beyond  our approximation
for the backreaction. The equations of motion  in this case are cumbersome and  it suffices to mention that the above parametrization will still work out to solve the equations of motion.

\textit{The geodesic equation for $\boldsymbol\tau$.--} At the zeroth order, the equation is trivially satisfied, 
when $l=0$, one can see that
\be\label{tau-geodesic}
\ddot{\tau}_{0}-\frac{1}{\rho_{0}}\left[1-(\dot{\tau}_{0})^{2}\left(1+\rho^{4}_{0}\right)\right]=0\,,
\ee
and at the second order, we get
\bea
\ddot{\tau_{2}}+2\frac{\dot{\tau}_{0}\dot{\tau}_{2}}{\rho_{0}}\left(1+\rho_{0}^{4}\right)
+\frac{\rho_{2}}{\rho_{0}^{2}}\left[\dot{y}^{2}_{0}-\dot{\tau}^{2}_{0}\left(1-3\rho^{4}_{0}\right)\right]
-\frac{2\dot{y}^{2}_{0}\hat{\Sigma_{b}}}{\rho_{0}}-\frac{1}{2}\dot{\tau}^{2}_{0}\rho_{0}^{2}\partial_{\rho}\hat{A}+\partial_{\rho}\hat{\Sigma}_{b}=0\,,
\eea
where in the above we have constraint the geodesic by $\dot{x}_{0}=\dot{z}_{0}=0$. Also note that
the metric components depend on $\left(\tau_{0},\rho_{0},x_{0},y_{0}\right)$ with $\tau_{0}(y_{0})$ and $\rho_{0}(y_{0})$. 
This means that we are looking at constant intervals on the geodesic along the $x$ axis.  

\textit{The  geodesic equation for $\boldsymbol\rho$.--} At zero order reads
\be
\ddot{\rho}_{0}+\frac{1}{\rho_{0}}\left[\dot{y}^{2}_{0}-\dot{\tau}^{2}_{0}-2\dot{\tau}_{0}\dot{\rho}_{0}-2\dot{\rho}_{0}^{2}\right]
-\rho^{3}_{0}\left(\dot{y}^{2}_{0}+2\dot{\tau}_{0}\dot{\rho}_{0}-\rho^{4}_{0}\dot{\tau}^{2}_{0}\right)=0\,,
\ee
and for $\mathcal{O}(l^{2})$,
\bea
&&\ddot{\rho}_{2}
+\dot{\rho}_{2}\left(-2\frac{\dot{\tau}_{0}}{\rho_{0}}
-4\frac{\dot{\rho}_{0}}{\rho_{0}}-2\dot{\tau}_{0}\rho^{3}_{0}\right)
+\dot{\tau}_{2}\left(-2\frac{\dot{\tau}_{0}}{\rho_{0}}
-2\frac{\dot{\rho}_{0}}{\rho_{0}}-2\dot{\rho}_{0}\rho^{3}_{0}+2\dot{\tau}_{0}\rho^{7}_{0}\right)
\nonumber\\&&
+\rho_{2}\left(-\frac{\dot{y}^{2}_{0}}{\rho^{2}_{0}}
+\frac{\dot{\tau}^{2}_{0}}{\rho^{2}_{0}}+2\frac{\dot{\tau}_{0}\dot{\rho}_{0}}{\rho^{2}_{0}}
+2\frac{\dot{\rho}_{0}^{2}}{\rho^{2}_{0}}-3\dot{y}^{2}_{0}\rho^{2}_{0}
-6\dot{\tau}_{0}\dot{\rho}_{0}\rho^{2}_{0}+7\dot{\tau}^{2}_{0}\rho^{6}_{0}\right)
+\hat{A}\left(\dot{y}^{2}_{0}\rho_{0}-\dot{\tau}^{2}_{0}\rho_{0}
-\dot{\tau}^{2}_{0}\rho^{5}_{0}\right)
\nonumber\\&&
+\frac{\partial_{\rho}\hat{A}}{2\rho^{2}_{0}}\left(\dot{\tau}^{2}_{0}\rho^{2}_{0}
+2\dot{\tau}_{0}\dot{\rho}_{0}\rho_{0}^{2}-\dot{\tau}^{2}_{0}\rho^{6}_{0}\right)
+\frac{2\dot{y}^{2}_{0}\hat{\Sigma}_{b}}{\rho_{0}}(1-\rho^{4}_{0})
+\dot{y}^{2}_{0}\partial_{\rho}\hat{\Sigma}_{b}(-1+\rho^{4}_{0})
\nonumber\\&&
+\half \dot{\tau}^{2}_{0}\rho^{2}_{0}\partial_{\tau}\hat{A}+\dot{y}^{2}_{0}\partial_{\tau}\hat{\Sigma}_{b}=0\,.
\eea

\textit{The inhomogeneous direction $\boldsymbol x_{2}$.--}  Simplifying the equation will yield

\bea\label{xgeocaseI}
&&\ddot{x}_{2}-2\frac{\dot{\rho}_{0}\dot{x}_{2}}{\rho_{0}}
+\rho_{0}\Xi\left[\dot{y}^{2}_{0}-(\dot{\tau}_{0})^{2}\left(1+\rho^{4}_{0}\right)\right]
+\half \dot{\tau}^{2}_{0}\rho^{2}_{0}\partial_{x}\hat{A}
-\dot{y}^{2}_{0}\partial_{x}\hat{\Sigma}_{b}
\nonumber\\&&
+\rho^{2}_{0}\dot{\tau}_{0}\left(\dot{\rho}_{0}\partial_{\rho}\hat{\Xi}_{f}
+\dot{\tau}_{0}\partial_{\tau}\hat{\Xi}_{f}\right)=0
\,.
\eea 
As we said before, we are looking at constant intervals along the $x$ axis and by varying the affine parameter that causes the geodesic to go deeper in the bulk, a non-zero value for $x_{2}$ will be produced.
 Note the $\partial_{x}$ in \Eq(xgeocaseI) which produce a distance of the order of $l^{2}$ between constant intervals.

From the metric compatibility condition, $\epsilon=-g_{\mu\nu}\frac{dx^{\mu}}{d\lambda}\frac{dx^{\nu}}{d\lambda}$
and the condition on  spacelike geodesics,  $\epsilon=-1$ at zeroth order  in $l$, one obtains
\be\label{compatibility-zero}
-\eta_{\mu\nu}\dot{x}^{\mu}_{0}\dot{x}^{\nu}_{0}+
2\dot{\tau}_{0}\dot{\rho}_{0}-\dot{\tau}_{0}^{2}\rho_{0}^{4}=-\rho^{2}_{0}\,,
\ee
in which we have to impose $\dot{x}_{0}=\dot{z}_{0}=0$ and $\dot{y}_{0}=1$.
 After expanding to $\mathcal{O}(l^{2})$, the corresponding equation simplifies to
\bea\label{compatibility2}
\frac{1}{\rho^{2}_{0}}\left(\dot{\tau}_{0}\dot{\tau}_{2}+\dot{\tau}_{2}\dot{\rho}_{0}
+\dot{\tau}_{0}\dot{\rho}_{2}-\dot{y}^{2}_{0}\hat{\Sigma}_{b}\right)
-\dot{\tau}_{0}\dot{\tau}_{2}\rho^{2}_{0}
+\frac{\rho_{2}}{\rho_{0}^{3}}\left(\dot{y}^{2}_{0}-\dot{\tau}^{2}_{0}-2\dot{\tau}_{0}\dot{\rho}_{0}\right)
-\dot{\tau}^{2}_{0}\rho_{0}\rho_{2}
+\half\dot{\tau}^{2}_{0}\hat{A}=0\,.\nonumber\\
\eea
Similar expansion to the order of $\mathcal{O}(l^{2})$ for the geodesic equations in the direction of $y$ and $z$, will produce

\be
-\frac{\dot{\rho}_{2}}{\rho_{0}}+\frac{\rho_{2}}{\rho^{2}_{0}}\dot{\rho}_{0}
+\dot{\rho}_{0}\partial_{\rho}\Sigma_{b}+\dot{\tau}_{0}\partial_{\tau}\Sigma_{b}=0\,.
\ee
The Killing vector in $y$ direction satisfies $\Sigma_{b}^{2}\,\dot{y}_{0}=const.$, expanding to zero
 order will yield $\dot{y}_{0}=\rho_{0}^{2}\times conts.$ and this will fix the value of $\dot{y}_{0}$ in
 \Eq(tau-geodesic)-\Eq(compatibility2).

After this short study of the behavior of the geodesics under the quench, we can compute the length of geodesics of interest.
The length of the geodesic connecting operators inserted at $(\tau_{1}=\tau_{\ast}$, $y_{1}=-y_{m}$, $x_{1}=z_{1}=0)$
and $(\tau_{2}=\tau_{\ast}$, $y_{2}=y_{m}$, $x_{2}=z_{2}=0)$ evaluates to
\be
\mathcal{L}=\int_{-y_m}^{y_m}dy_{0}
\sqrt{\Sigma_{b}^{2}+\Sigma_{d}\dot{x}^{2}-A\dot{\tau}^{2}+2\Xi\dot{x}\dot{\tau}
-2\frac{\dot{\rho}\dot{\tau}}{\rho^{2}}}\,,
\ee
with all the metric components as a function of $(\tau,\rho,x,y)$. After expanding
to the first order of $l^{2}$, we get a correction for the length of the geodesic that has the form of
$\mathcal{L}=\mathcal{L}_{0}+l^{2}\mathcal{L}_{2}$, with
\be\label{I-zerolength}
\mathcal{L}_{0}=\int_{-y_{m}}^{y_{m}}dy_{0}\frac{\sqrt{D(\tau_{0},\rho_{0},x_{\ast})}}{\rho_{0}}\,,
\ee
here $x_{\ast}$ is the boundary coordinate in the 
inhomogeneous direction. The second order  correction given by
\be\label{twopoint-length}
\mathcal{L}_{2}=\int_{-y_{m}}^{y_{m}}dy_{0}\left[
\frac{\hat{\Sigma}_{b}-\dot{\tau}_{0}^{2}\rho^{2}_{0}\hat{A}/2}{\rho_{0}\sqrt{D}}
 -\frac{\dot{\rho}_{0}+\dot{\tau}_{0}(1-\rho^{4}_{0})}{\rho_{0}\sqrt{D}}\dot{\tau}_{2}
-\frac{D-2\dot{\tau}_{0}\rho^{4}_{0}}{\rho^{2}_{0}\sqrt{D}}\rho_{2}
-\frac{\dot{\tau}_{0}}{\rho_{0}\sqrt{D}}\dot{\rho}_{2}
\right]\,,
\ee
with 
\be\label{Dd}
D=1-2\dot{\tau}_{0}\dot{\rho}_{0}+\dot{\tau}_{0}^{2}\left(-1+\rho^{4}_{0}\right)\,.
\ee
Note that if we were assuming $\dot{x}_{0}\neq0$ then there would be a term
 proportional to $x_{2}$ in \Eq(twopoint-length).
It  is convenient to use the equations of motion for the geodesics for the last three terms in \Eq(twopoint-length) to show
that the total contribution is zero after a partial integration. This is a  consequence of perturbation around the extremal trajectory as  it was noticed in \cite{Alex2014}.

Constraint on the static geodesics come from $K_{\mu}\frac{dx^{\mu}}{d\lambda}=const.$ In the absence of the quench, time
 is a Killing vector.
 With $K_{\tau}=g_{\tau\tau}$, $K_{\rho}=g_{\rho\tau}$ and
$K_{x}=g_{x\tau}$, the zeroth-order  equation is given by
\be\label{twopoint-tdot}
\left(\rho^{4}_{0}-1\right)\dot{\tau}_{0}-\dot{\rho}_{0}=const.
\ee
Another way of seeing this is from the zeroth order geodesic equation for $y$. 
At the horizon $\rho_{0}=1$ and $\dot{\rho}_{0}=0$, this fixes the constant coefficient to zero.
 The general solution is \cite{Alex2014},
\be\label{Twopoint-t0}
\frac{d\tau_{0}}{d\rho_{0}}=-\frac{1}{1-\rho^{4}_{0}}\,,\quad\text{or}\quad
 \tau_{0}(\rho_{0})=\tau_{\ast}-\tan^{-1}(\rho_{0})-\tanh^{-1}(\rho_{0})\,,
\ee
here $\tau_{\ast}$ is the  time on the boundary as  an observer in the bulk reaches the boundary at $\rho\rightarrow 0$.
From the compatibility condition of the metric, \Eq(compatibility-zero), we have
\be\label{Twopoint-u0}
\left[1+(\rho^{4}_{0}-1)\dot{\tau}_{0}^{2}-2\dot{\tau}_{0}\dot{\rho}_{0}\right]\rho^{2}_{0}=\rho^{2}_{m}\,,
\ee
where the constant $\rho_{m}$ is the maximum value for the radius of the arc that attaches the two points on the 
boundary.
Thus   \Eq(twopoint-length) reduces to
\be
\mathcal{L}_{2}=-\frac{2}{\rho_{m}}\int_{0}^{\rho_{m}}d\rho_{0}
\frac{\hat{\Sigma}_{b}-\dot{\tau}_{0}^{2}\rho^{2}_{0}\hat{A}/2}{\dot{\rho}_{0}}\,,
\ee
where in the above, the metric components of $\hat{\Sigma}_{b}$, $\hat{\Sigma}_{d}$ and $\hat{A}$ depend on 
$(\tau_{0},\rho_{0},x_{\ast})$ with $\tau_{0}(y_{0})$ and $\rho_{0}(y_{0})$. This is exactly the result in \cite{Alex2014} with
 the exception that now the profile
of the geodesic is nonlinearly a function of the $x_{\ast}$.  To prepare the integral for
 numerics following \cite{Alex2014},
after a change of variable such as $\rho_{0}=\rho_{m}\left(1-q^{2}\right)$, the former expression takes the following  form
\be\label{L2}
\mathcal{L}_{2}=2\int_{0}^{1}\left(1-q^{2}\right)dq \left[
\frac{2\hat{\Sigma}_{b}}{\sqrt{\left(2-q^{2}\right)\left(1-\left(1-q^{2}\right)^{4}\rho_{m}^4\right)}}
-q\rho^{2}_{m}\hat{A}\frac{\sqrt{1-\left(1-q^{2}\right)^{2}}}{\left(1-\left(1-q^{2}\right)^{4}\rho^{4}_{m}\right)^{3/2}}
\right]\,,
\ee
where again the components of the metric in the above expression are functions of
$(\tau_{0}$, $\rho_{0}$, $x_{\ast})$ with $\tau_{0}(q)$ and $\rho_{0}(q)$.

\begin{figure}
\begin{subfigure}{.5\textwidth}
{\includegraphics[width=7.0cm]{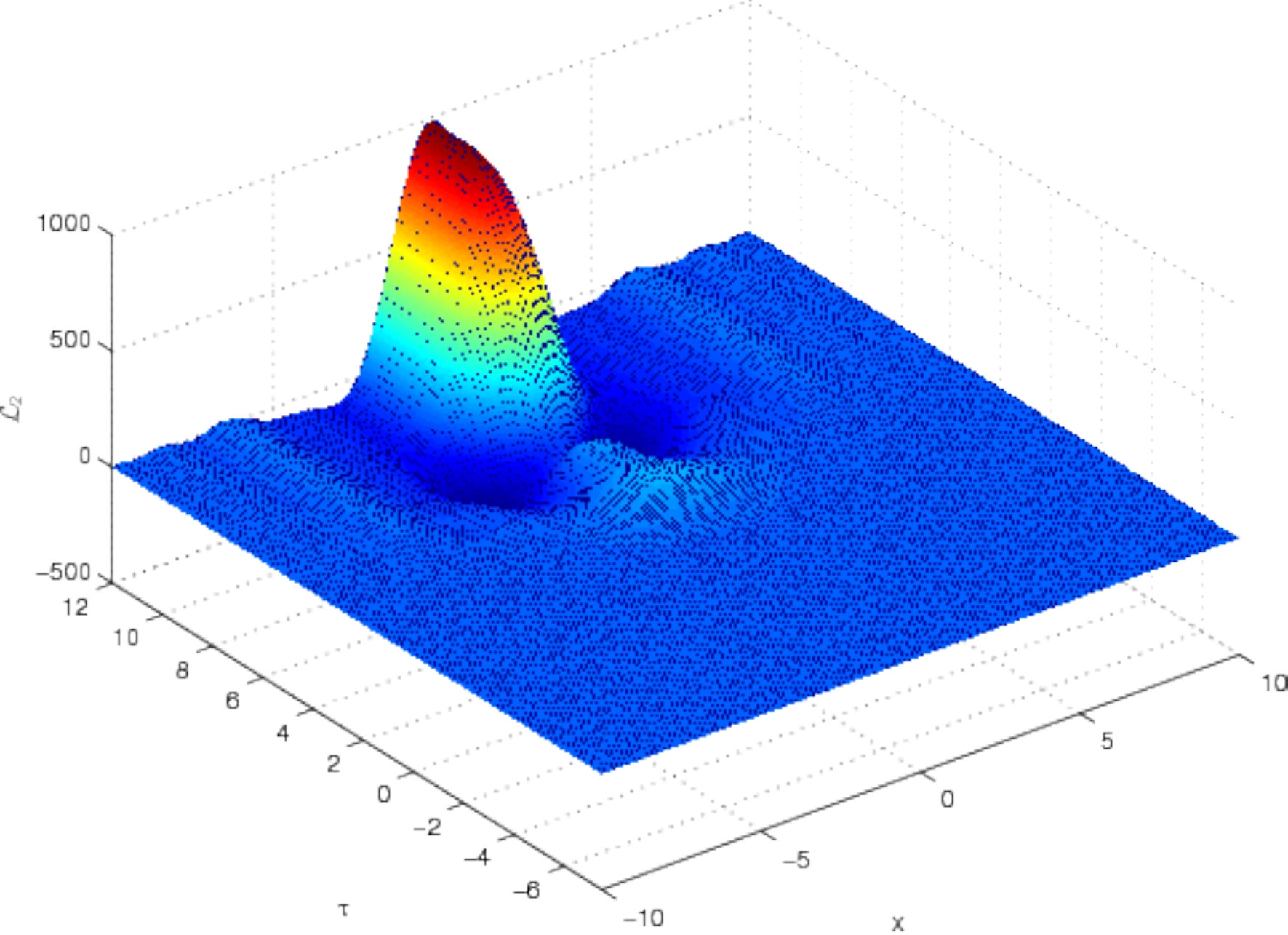}}
  \caption{}\label{fig:TwoPointA2cc999a}
\end{subfigure}
\begin{subfigure}{.5\textwidth}
{\includegraphics[width=7.0cm]{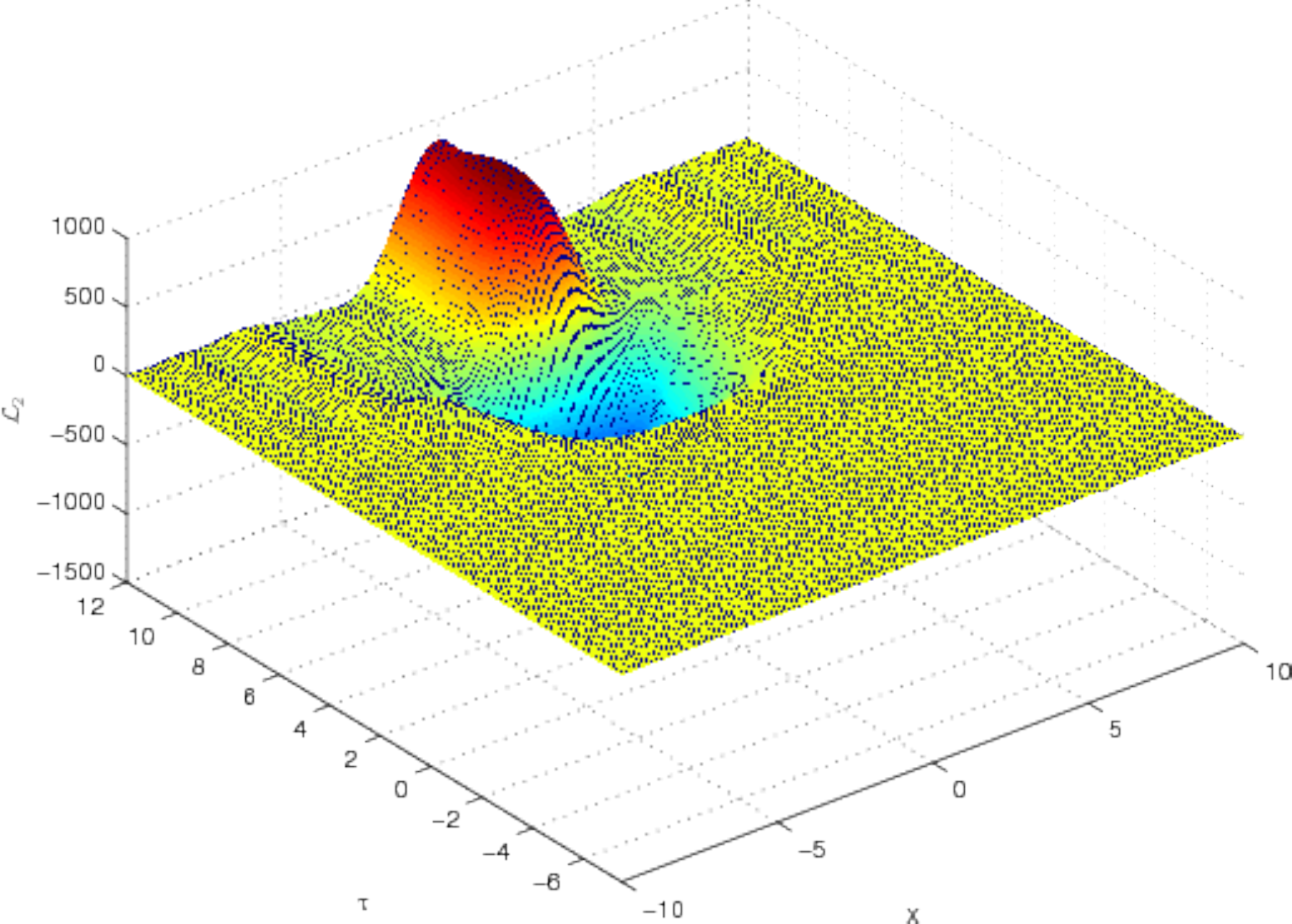}}
  \caption{}\label{fig:TwoPointA3a2999}
\end{subfigure}

\begin{subfigure}{.5\textwidth}
{\includegraphics[width=7.0cm]{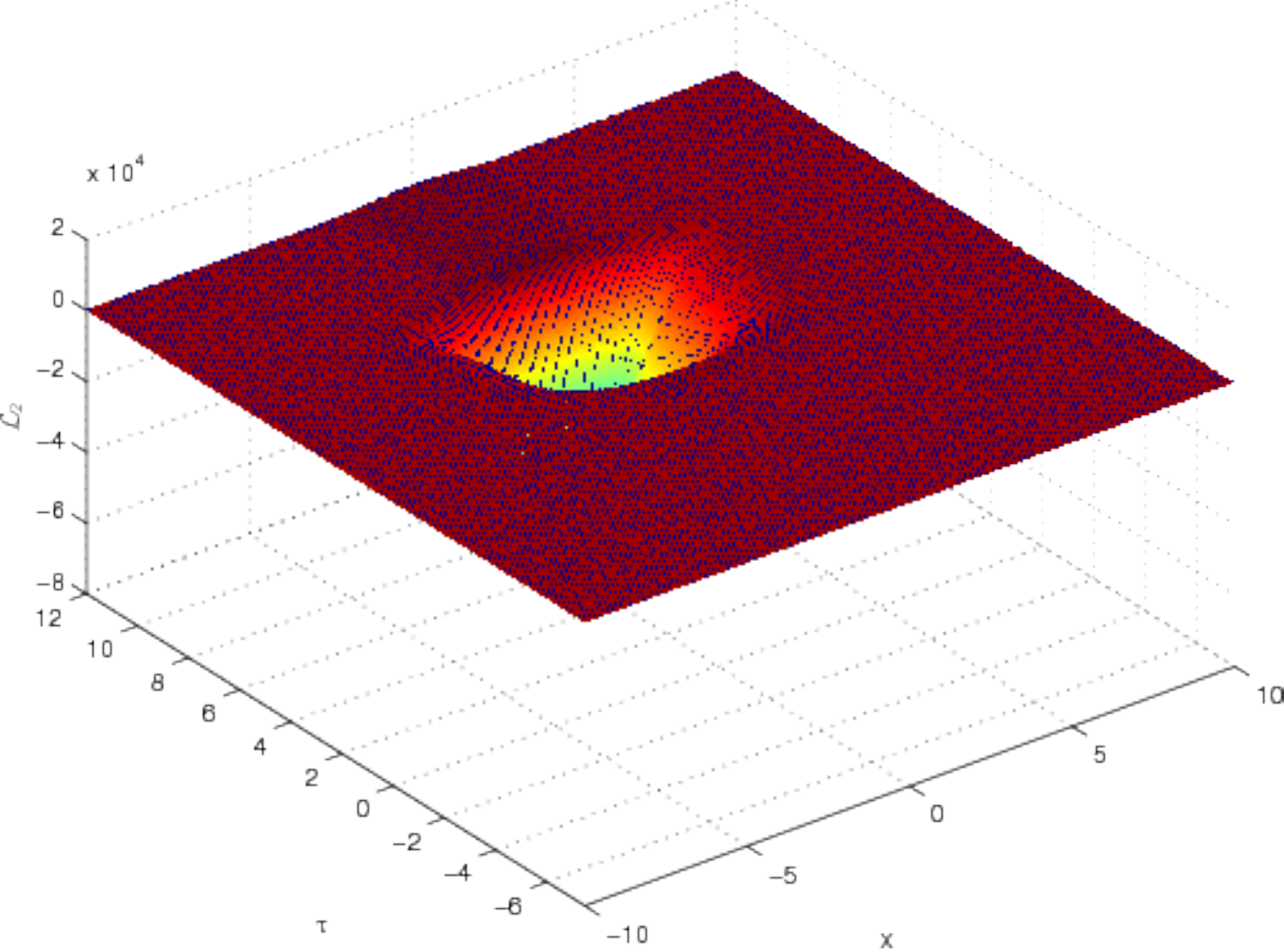}}
  \caption{}\label{fig:TwoPointA3b2999}
\end{subfigure}
\begin{subfigure}{.5\textwidth}
{\includegraphics[width=7.0cm]{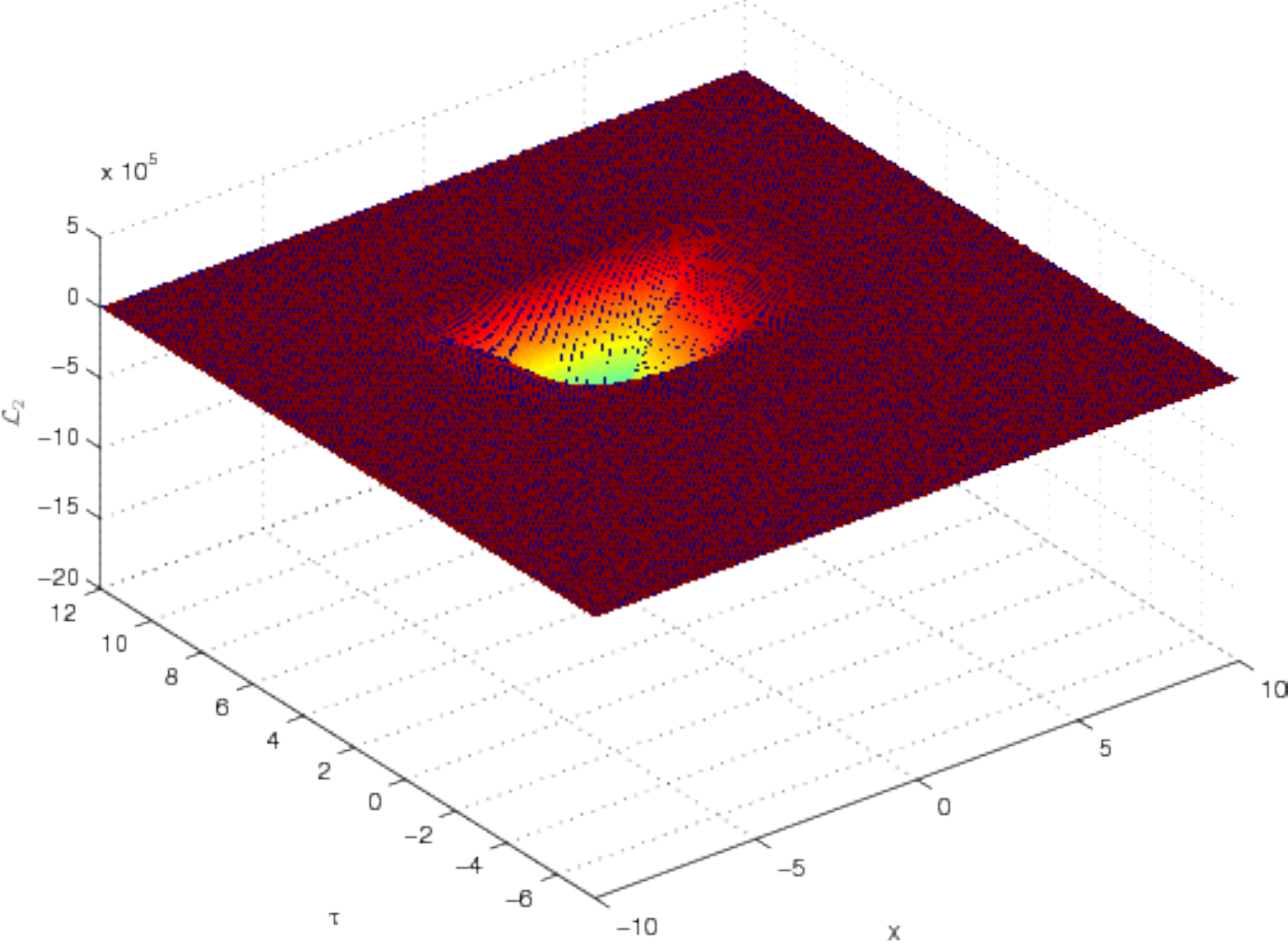}}
  \caption{}\label{fig:TwoPointA3c2999}
\end{subfigure}
\caption{Time evolution  of the two-point Wightman functions for operators with large conformal dimensions. 
Figures in (a), (b), (c) and (d) are plotted for $\alpha\in\left\{1,\half,\frac{1}{4},\frac{1}{8}\right\}$ and $\sigma=\sqrt{L_{x}}$. 
The interpolation of points are based on $N_{x}=N_{\rho}=20-30$ along the inhomogeneity direction $x$ and radial direction $\rho$. The number of time steps for RK4  are $7810-17560$.   }
\end{figure}

We can interpret the final Gaussian distribution that is produced at late times as a signal of a successful thermalization. Among the different simulations that have been performed in this section for parameters in the range of $\rho_{m}\in \left\{0.1\rho_{h},0.5\rho_{h},0.9\rho_{h},0.999\rho_{h}\right\}$\footnote{For the rest of the simulations in the paper, 
we fixed $\rho_{h}=1$.}, those that correspond to $\rho_{m}=0.9\rho_{h}-0.999\rho_{h}$ could be verified  to have reached the thermalization. Figures \ref{fig:TwoPointA2cc999a}-\ref{fig:TwoPointA3c2999} show the correlation between two fixed points in the $y$ axis for different $\alpha\in\left\{1,\half,\frac{1}{4},\frac{1}{8}\right\}$  while a scalar field that has a Gaussian profile as a function of $x$ is falling into the black hole in the bulk space. In these figures different observers stationed on the $x$ axis will measure the correlation between the two specific points on the $y$ axis differently. The maximum correlation is measured on the $x=0$ axis and other measurements are symmetric around this axis as the original profile for $p_{0}(\tau,x)$ has this symmetry. As the quench is triggered, there appears  a ``phase transition'' in a sense that  the sign of the correlation function changes sign; from  zero in the ground state, goes to a minimum negative value and  undo itself  and reaches a final saturated maximum. The rather simple form of \Eq(L2) shows that this transition is due to the interplay between $\hat{\Sigma}_{b}$ and the warp factor $\hat{A}$. The first term is always positive while the sign of the second term varies depending on the sign of   $\hat{A}$.  Reduction of the value of $\alpha$ makes the late time Gaussian-like distribution to disappear, signaling a fully thermalized equilibrium state measured by the observable in the universal (abrupt quench) limit.

\begin{figure}[!ht]
\begin{subfigure}{.5\textwidth}
{\includegraphics[width=7.0cm]{L2A20points2cc-crop}}
  \caption{}\label{fig:TwoPointA2cc999b}
\end{subfigure}
\begin{subfigure}{.5\textwidth}
{\includegraphics[width=7.0cm]{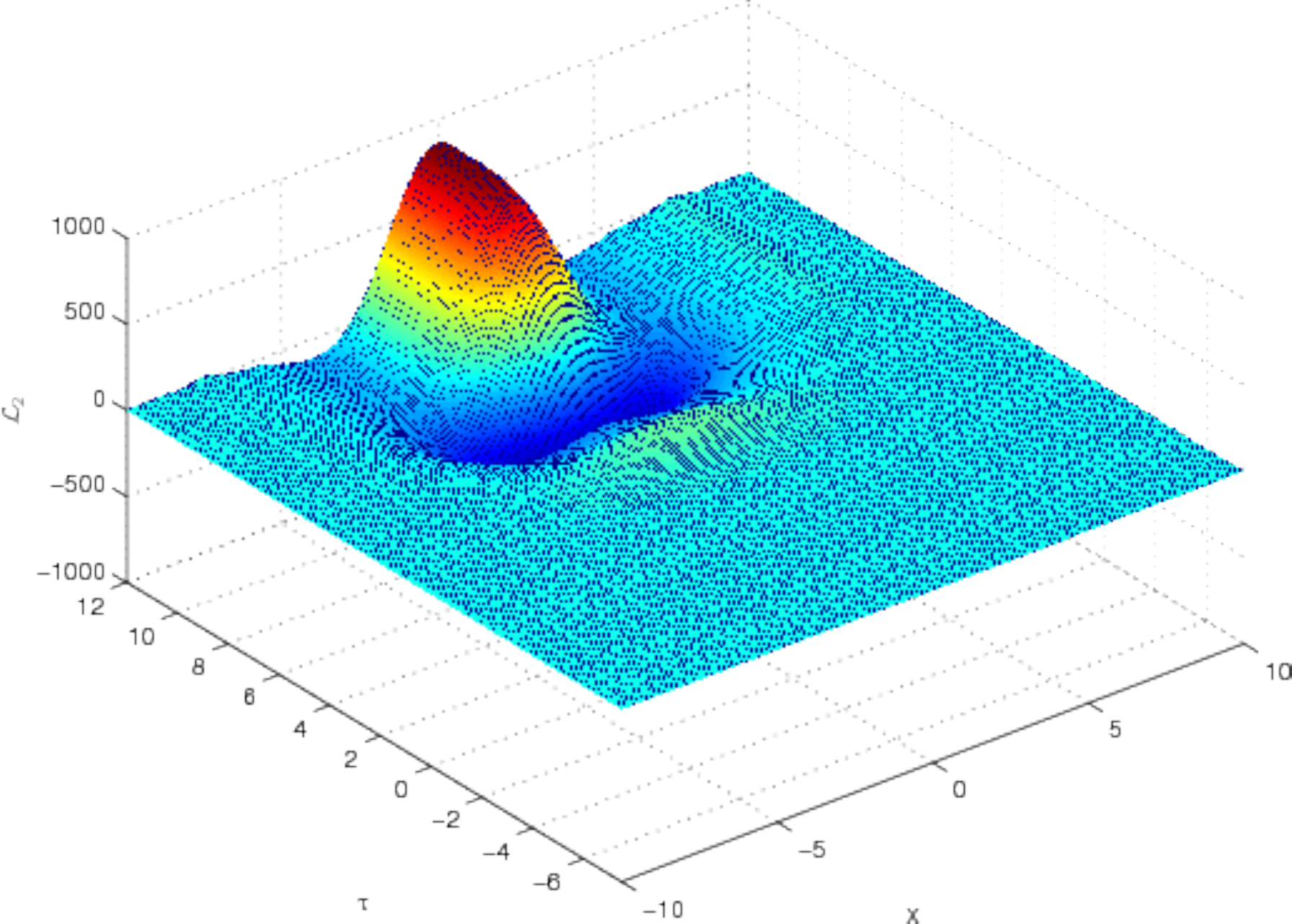}}
  \caption{}
  \label{fig:TwoPointA4a2999}
\end{subfigure}
\caption{Changing the value of $\sigma$ from $\sqrt{L_{x}}$ to  $\sqrt{1.5L_{x}}$ from left to right causes the distributions to rescale. This factor must be a nontrivial function of the dynamics under study.
On the left-hand side $N_{x}=N_{\rho}=20$ and on the right-hand side $N_{x}=N_{\rho}=30$ Chebyshev points have been used.}
\end{figure}

In Figures \ref{fig:TwoPointA2cc999b}-\ref{fig:TwoPointA4a2999}, we compare the effect of changing $\sigma$ in the range $\sqrt{L_{x}}-\sqrt{1.5L_{x}}$. In the next section, we will compare these results with those of Case II.

\addsubsubsection{Case II: plane C-D}\label{caseII}
In this section, we consider  two-point correlations again, while  we  measure the inhomogeneity in a plane perpendicular to the one in the previous section. For an illustration refer to Figure \ref{fig:Two-point} and the comments at the begging of that section.
The relative geometry of the setup here is more important as it has a resemblance to the setup of the elliptic flow in heavy-ion collisions. In both cases, there are distributions that are localized in the transverse directions. Of course, the physics of the two cases are not directly related.

The effect of the backreaction on the coordinates will be parametrized by
\be
\tau=\tau_{0}+l^{2}\tau_{2}\,,\quad \rho=\rho_{0}+l^{2}\rho_{2}\,,\quad x=x_{0}+l^{2}x_{2}\,.
\ee 
In what follows, we will use $x_{0}$ to parametrize the geodesic. Expansion
in terms of the above series will then  yield,

\textit{The geodesic equation for $\boldsymbol \tau_{2}$.--} 
\bea
&&\ddot{\tau}_{2}+2\frac{\dot{\tau}_{0}\dot{\tau}_{2}}{\rho_{0}}\left(1+\rho^{4}_{0}\right)
-2\frac{\dot{x}_{2}}{\rho_{0}}+\frac{\rho_{2}}{\rho^{2}_{0}}
\left(1-\dot{\tau}^{2}_{0}+3\dot{\tau}^{2}_{0}\rho^{4}_{0}\right)
-2\frac{\hat{\Sigma}_{d}}{\rho_{0}}-\half \dot{\tau}^{2}_{0}\rho^{2}_{0}\partial_{\rho}\hat{A}
+\dot{\tau}_{0}\rho^{2}_{0}\partial_{\rho}\hat{\Xi}_{f}
\nonumber\\&&
+\partial_{\rho}\hat{\Sigma}_{d}=0\,.
\eea
\textit{The geodesic equation for $\boldsymbol \rho_{2}$.--} 
\bea\label{eqrho}
&&\ddot{\rho}_{2}+2\frac{\dot{x}_{2}}{\rho_{0}}\left(1-\rho^{4}_{0}\right)
-2\dot{\rho}_{2}\left(\frac{\dot{\tau}_{0}}{\rho_{0}}+2\frac{\dot{\rho}_{0}}{\rho_{0}}
+\dot{\tau}_{0}\rho^{3}_{0}\right)
-2\dot{\tau}_{2}\left(\frac{\dot{\tau}_{0}}{\rho_{0}}+\frac{\dot{\rho}_{0}}{\rho_{0}}
+\dot{\rho}_{0}\rho^{3}_{0}-\dot{\tau}_{0}\rho^{7}_{0}\right)
\nonumber\\&&
+\left(-\frac{1}{\rho^{2}_{0}}+\frac{\dot{\tau}^{2}_{0}+2\dot{\rho}^{2}_{0}
+2\dot{\tau}_{0}\dot{\rho}_{0}}{\rho^{2}_{0}}-3\rho^{2}_{0}-6\dot{\tau}_{0}\dot{\rho}_{0}\rho^{2}_{0}
+7\dot{\tau}^{2}_{0}\rho^{6}_{0}\right)\rho_{2}
+\rho_{0}\left(1-\dot{\tau}^{2}_{0}\left(1+\rho^{4}_{0}\right)\right)\hat{A}
\nonumber\\&&
-2\dot{\rho}_{0}\rho_{0}\Xi+\frac{2}{\rho_{0}}\left(1-\rho^{4}_{0}\right)\hat{\Sigma}_{d}
+\dot{\tau}_{0}\rho^{2}_{0}\partial_{x}A-\rho^{2}_{0}\partial_{x}\hat{\Xi}_{f}
+\frac{\dot{\tau}_{0}\rho^{2}_{0}}{2}
\left(\dot{\tau}_{0}+2\dot{\rho}_{0}-\dot{\tau}_{0}\rho^{4}_{0}\right)\partial_{\rho}\hat{A}
\nonumber\\&&
+\rho^{2}_{0}\left(-\dot{\tau}_{0}-\dot{\rho}_{0}+\dot{\tau}_{0}\rho^{4}_{0}\right)
\partial_{\rho}\hat{\Xi}_{f}+\left(-1+\rho^{4}_{0}\right)\partial_{\rho}\hat{\Sigma}_{d}
+\half\dot{\tau}^{2}_{0}\rho^{2}_{0}\partial_{\tau}\hat{A}+\partial_{\tau}\hat{\Sigma}_{d}=0\,.
\eea
\textit{The geodesic equation for $\boldsymbol x_{2}$.--} 
\bea\label{ddotxII}
&&\ddot{x}_{2}-2\frac{\dot{x}_{2}\dot{\rho}_{0}}{\rho_{0}}-2\frac{\dot{\rho}_{2}}{\rho_{0}}
+2\frac{\dot{\rho}_{0}\rho_{2}}{\rho^{2}_{0}}
+\rho_{0}\left(1-\dot{\tau}^{2}_{0}\left(1+\rho^{4}_{0}\right)\right)\hat{\Xi}_{f}
+\half\dot{\tau}^{2}_{0}\rho^{2}_{0}\partial_{x}\hat{A}+\partial_{x}\hat{\Sigma}_{d}+
\dot{\tau}_{0}\dot{\rho}_{0}\rho^{2}_{0}\partial_{\rho}\hat{\Xi}_{f}
\nonumber\\&&
+2\dot{\rho}_{0}\partial_{\rho}\hat{\Sigma}_{d}+\dot{\tau}^{2}_{0}\rho^{2}_{0}\partial_{\tau}\hat{\Xi}_{f}
+2\dot{\tau}_{0}\partial_{\tau}\hat{\Sigma}_{d}=0\,,
\eea
and we can verify that the geodesics on the $y$ and $z$ axis are not affected at $\mathcal{O}(l^{2})$. The metric compatibility condition will subsequently change to
\bea\label{caseII-compatibility}
&&\dot{x}_{2}-\dot{\tau}_{0}\dot{\rho}_{2}
+\left(-\dot{\tau}_{0}-\dot{\rho}_{0}+\dot{\tau}_{0}\rho^{4}_{0}\right)\dot{\tau}_{2}+
\frac{\rho_{0}}{\rho_{2}}\left(-1+\dot{\tau}^{2}_{0}+2\dot{\tau}_{0}\dot{\rho}_{0}+\dot{\tau}^{2}_{0}\rho^{4}_{0}\right)
-\half\dot{\tau}^{2}_{0}\rho^{2}_{0}A+\dot{\tau}_{0}\rho^{2}_{0}\Xi
\nonumber\\&&
+\Sigma_{d}=0\,.
\nonumber\\
\eea
Note the appearance of the disturbances in \Eq(eqrho) for the bulk radius and compare  it to the previous case. This completes  the list of the required geodesics which   could have been driven otherwise from the action principle.

The length of the spacelike geodesic that connects   points  C and D on
 $(x_{1}=-x_{m}, y_{1}=0, z_{1}=0, \tau_{1}=\tau_{\ast})$ and $(x_{2}=x_{m}, y_{2}=0, z_{2}=0, \tau_{2}=\tau_{\ast})$ is given by
\be
\mathcal{L}=\int_{-x_{m}}^{x_{m}}dx_{0}
\sqrt{-A\dot{\tau}^{2}+\Sigma_{d}^{2}\left(1+\dot{x}_{2}\right)^{2}
+2\Xi_{f}\dot{\tau}(1+\dot{x}_{2})
-2\frac{\dot{\rho}\dot{\tau}}{\rho^{2}}}\,,
\ee
where in the above $\dot{\tau}=\dot{\tau}_{0}+l^{2}\dot{\tau}_{2}$, and we are assuming a similar
 expression for $\dot{\rho}$ too. In addition to $\rho(x_{0})$, the metric components $\Sigma_{d}$, $A$ and $\Xi$
are functions of $(\tau,\rho,x_{0})$ with  $\tau(x_{0})$ and $\rho(x_{0})$.
Expanding to $\mathcal{O}(l^{2})$, at zero order, we find \Eq(I-zerolength) and to the second order it simplifies to
\bea\label{caseII-length-original}
\mathcal{L}_{2}&=&
\int_{-x_{m}}^{x_{m}}\frac{dx_{0}}{\rho_{0}\sqrt{D}}
\left(\hat{\Sigma}_{d}-\half\dot{\tau}^{2}_{0}\hat{A}+\dot{\tau}_{0}\rho^{2}_{0}\hat{\Xi}_{f}\right)
\nonumber\\&&
+\int_{-x_{m}}^{x_{m}}\frac{dx_{0}}{\rho_{0}\sqrt{D}}
\left[\dot{x}_{2}-\dot{\tau}_{0}\dot{\rho}_{2}
+\left(-\dot{\tau}_{0}-\dot{\rho}_{0}+\dot{\tau}_{0}\rho^{4}_{0}\right)
\dot{\tau}_{2}+\frac{-D+2\dot{\tau}^{2}_{0}\rho^{4}_{0}}{\rho_{0}}\rho_{2}
\right]\,,
\eea
with $D$ defined in \Eq(Dd).
Similar to  Case I, the  equations of motion at zero order will allow us to simplify the above expression.
 The term proportional to $\dot{\tau}_{2}$ and the combination of the coefficients
that multiply $\rho_{2}$ and $\dot{\rho}_{2}$ will cancel out. The only non-zero 
contribution from the second line of \Eq(caseII-length-original) comes from $\dot{x}_{2}$.
The interpretation of this term is the following;  we have chosen $x_{0}$ as a parameter 
that covers the geodesic between the two fixed points on the boundary but 
this coordinate is also along the axis that the inhomogeneity is sourced accordingly  by the profile of the  scalar field.
Therefore this term compensates for the  fact that we are constraining  the geodesic in a fixed interval.

By partial integration and equations of motion, we can reduce the contribution to
\be\label{caseII-length}
\mathcal{L}_{2}=
\int_{-x_{m}}^{x_{m}}\frac{dx_{0}}{\rho_{0}\sqrt{D}}
\left(\hat{\Sigma}_{d}-\half\dot{\tau}^{2}_{0}\hat{A}+\dot{\tau}_{0}\rho^{2}_{0}\hat{\Xi}_{f}+\dot{y}_{0}\hat{\Sigma}_{b}\right)
+\left.\frac{x_{2}}{\rho_{0}\sqrt{D}}\right|_{-x_{m}}^{x_{m}}\,.
\ee
Now, if we assume  $2x_{m}\gg1$. This means  $x_{2}=0$ at $\pm x_{m}$. In this case, there is no
contribution from the second term in \Eq(caseII-length). While this is an interesting scenario, we pursue
the general case and therefore do not impose this latter boundary condition. Notice that splitting the integral into
$\int_{0}^{x_{m}}$, wouldn't help at all since in order to know the value of $x_{2}$ at $x_{0}=0$, we have to
solve the geodesic equations all the way  from the boundary down to the maximum value of the bulk radius.

\begin{figure}[!ht]
\begin{subfigure}{.5\textwidth}
{\includegraphics[width=7.0cm]{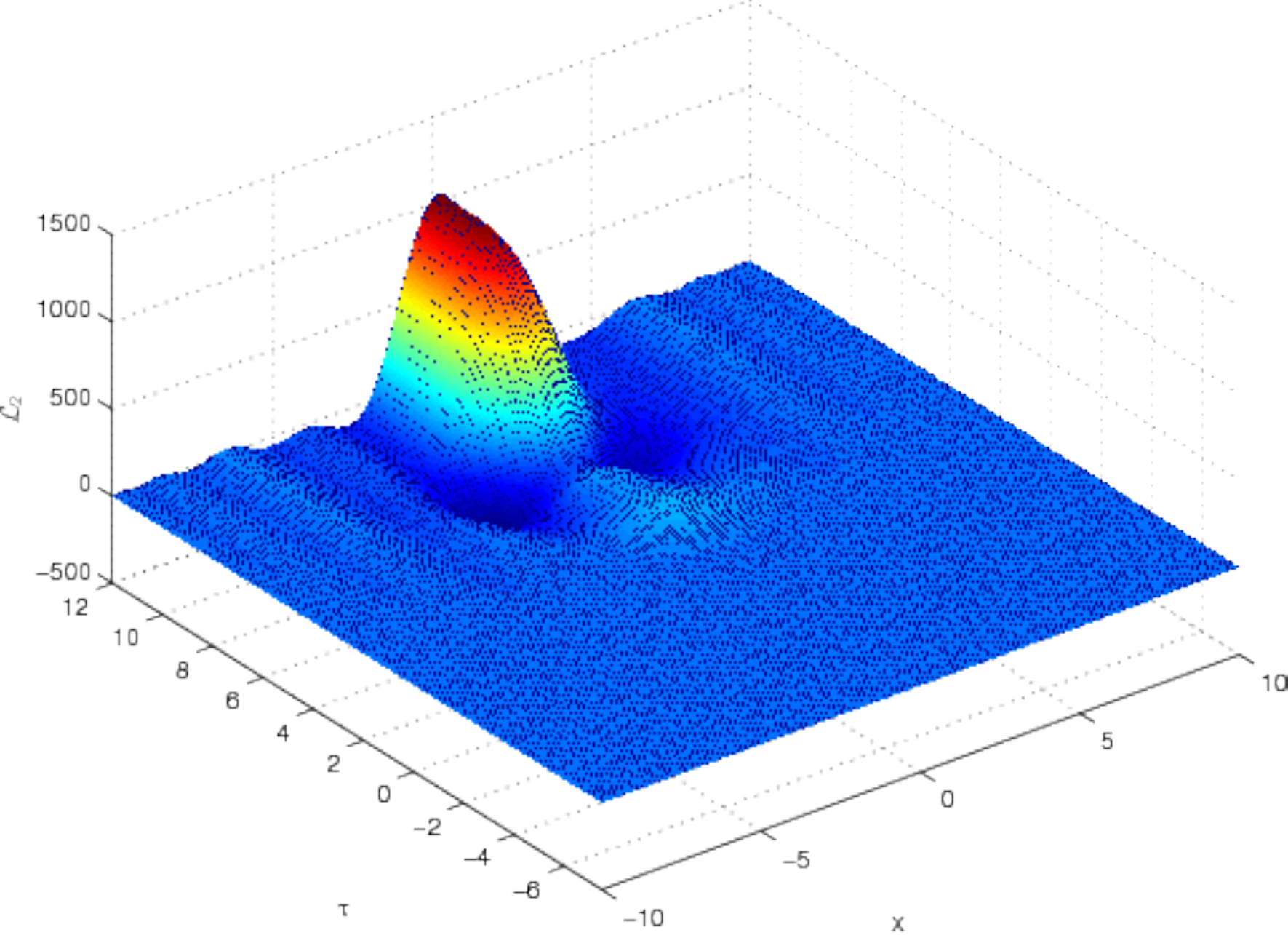}}
  \caption{}
  \label{fig:TwoPointB2cc999}
\end{subfigure}
\begin{subfigure}{.5\textwidth}
{\includegraphics[width=7.0cm]{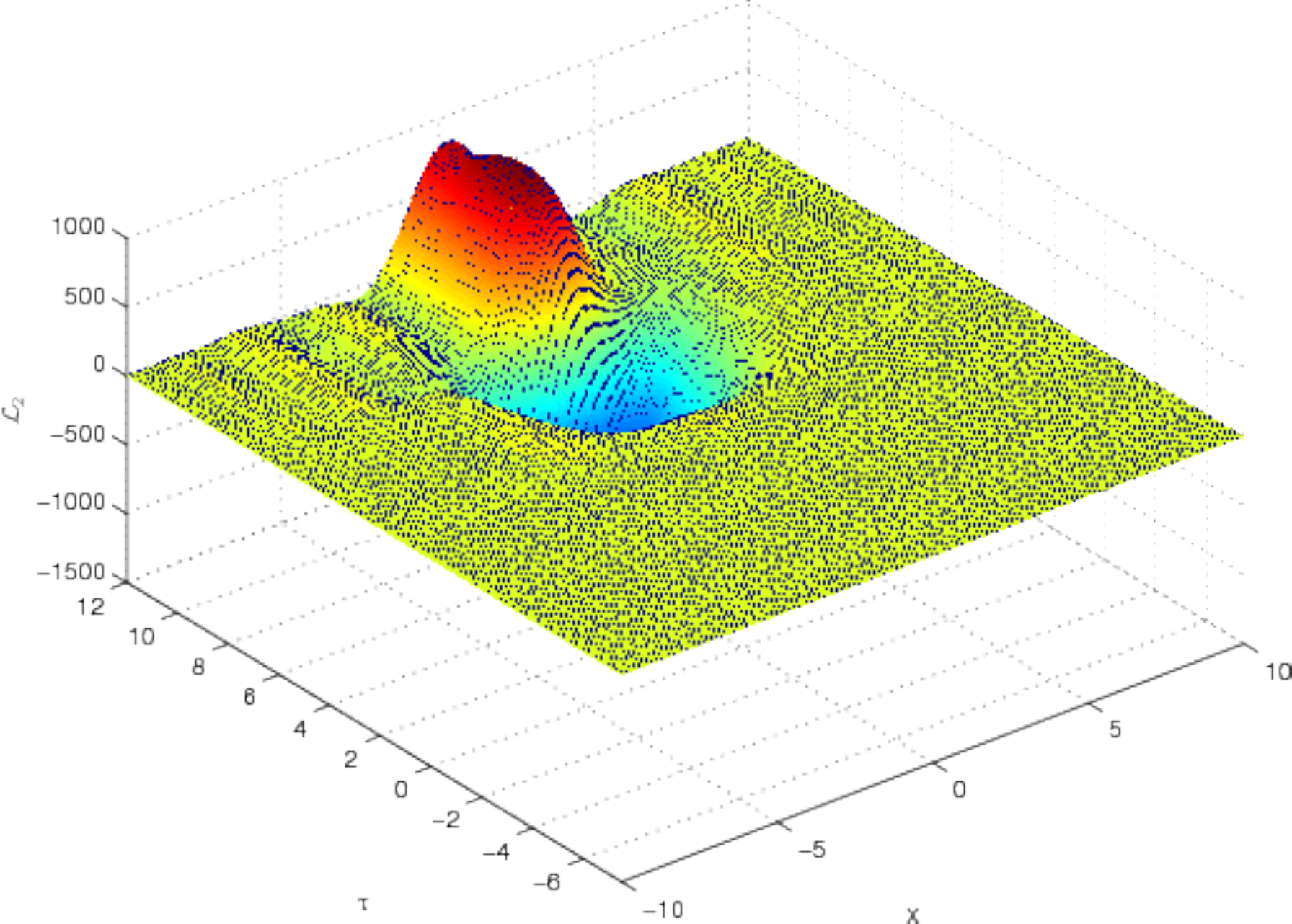}}
  \caption{}
  \label{fig:TwoPointB3a2999}
\end{subfigure}

\begin{subfigure}{.5\textwidth}
{\includegraphics[width=7.0cm]{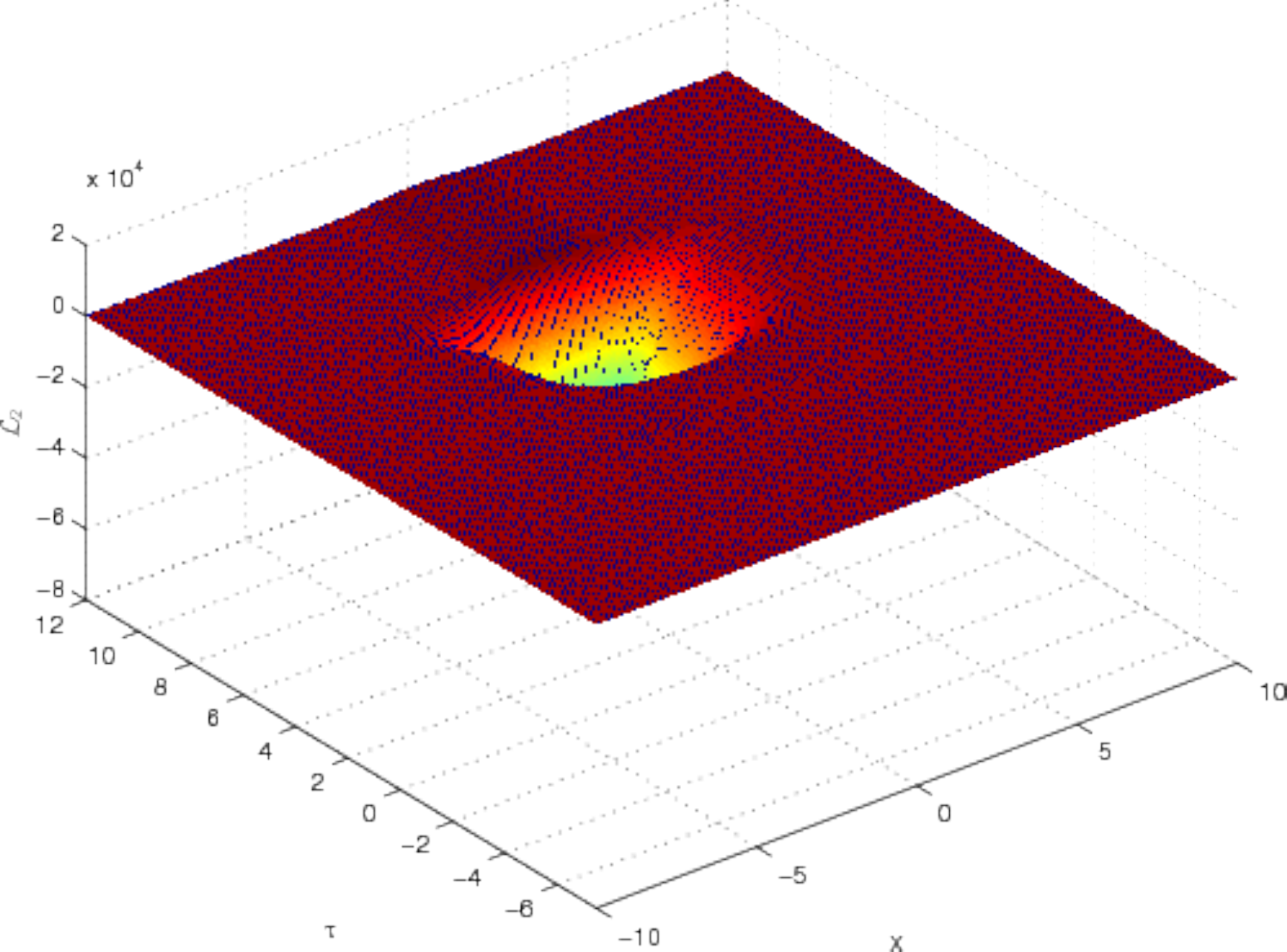}}
  \caption{}
  \label{fig:TwoPointB3b2999}
\end{subfigure}
\begin{subfigure}{.5\textwidth}
{\includegraphics[width=7.0cm]{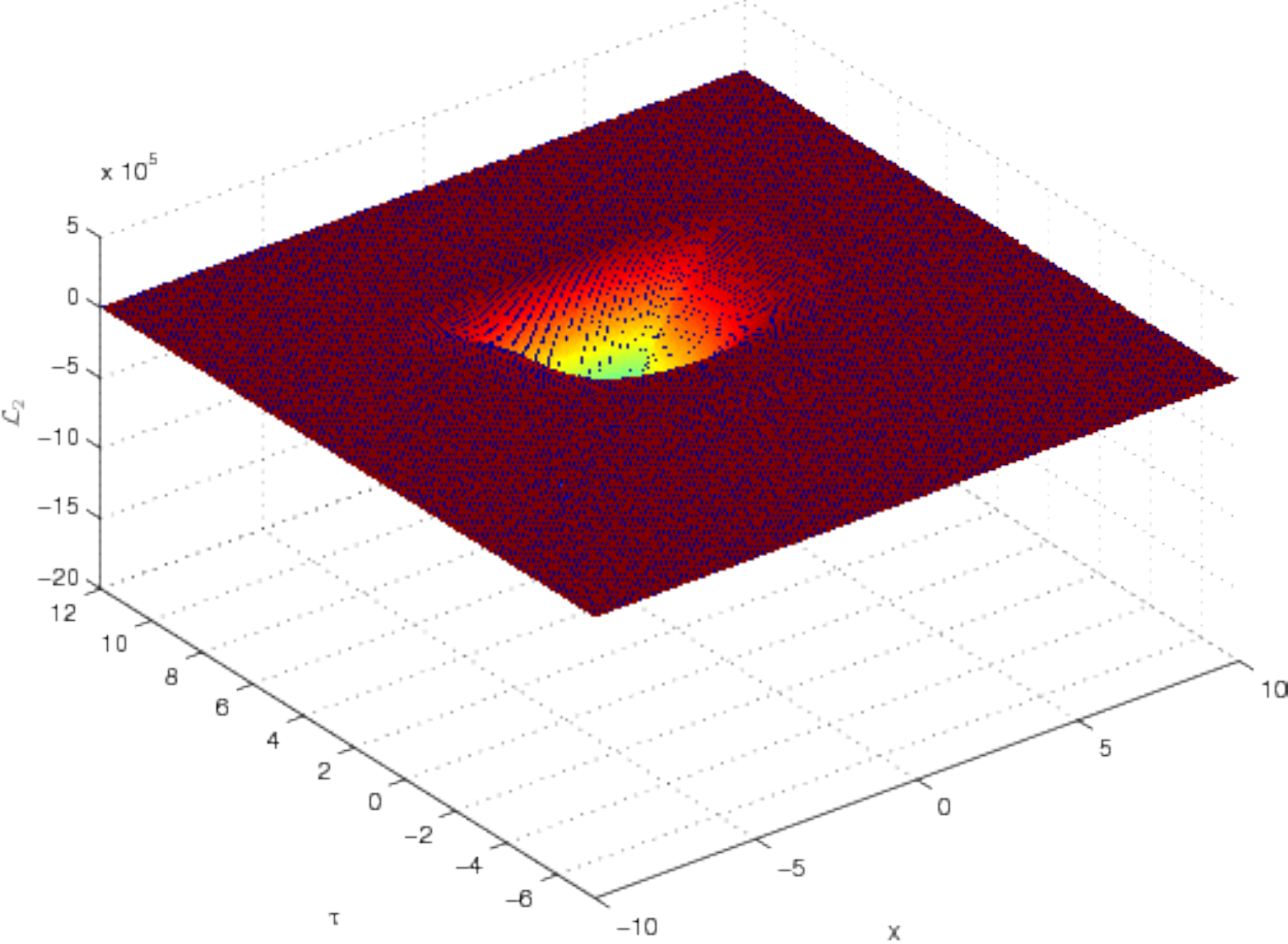}}
  \caption{}
  \label{fig:TwoPointB3c2999}
\end{subfigure}

\centering
\begin{subfigure}{.5\textwidth}
{\includegraphics[width=7.0cm]{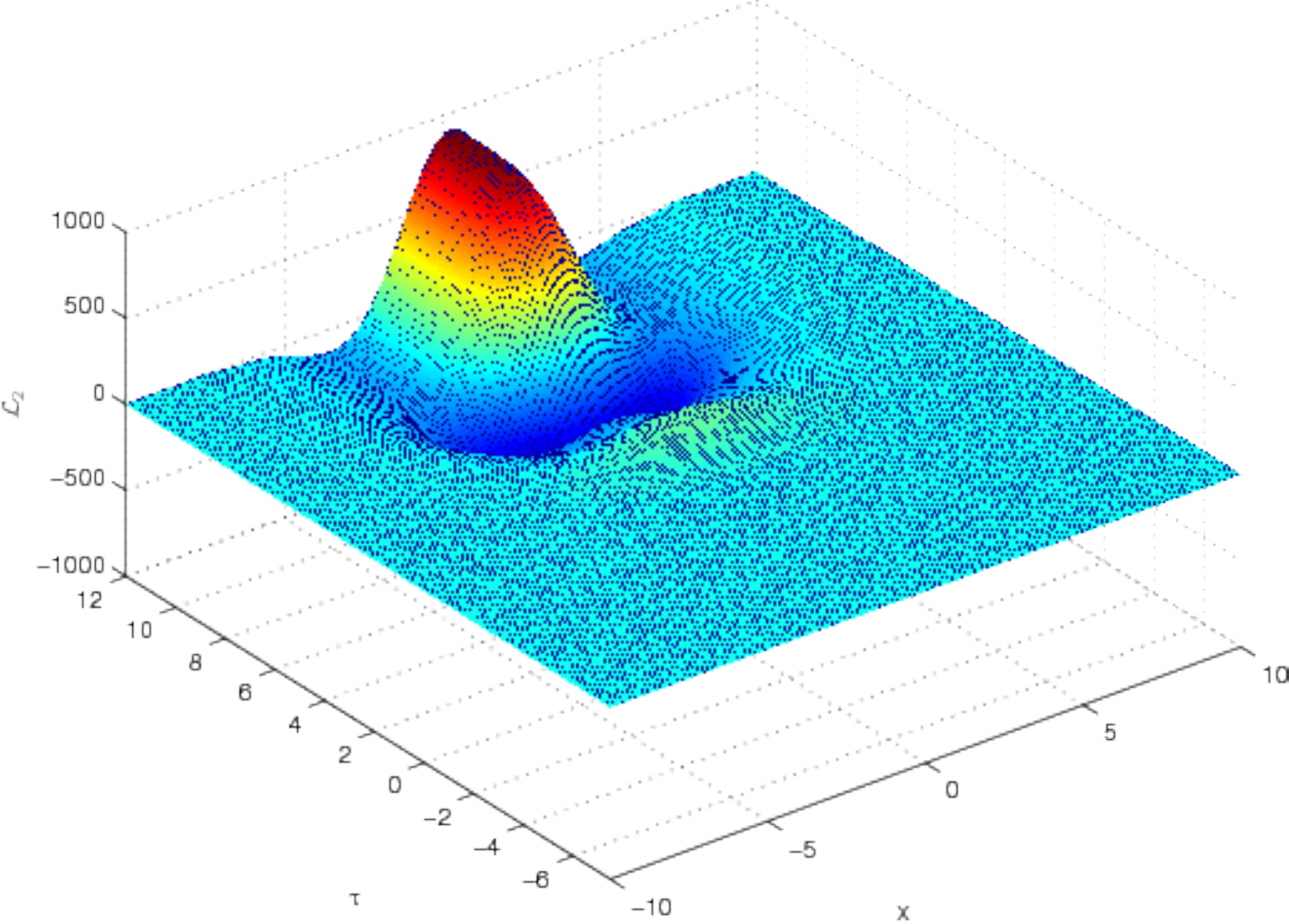}}
  \caption{}
  \label{fig:TwoPointB4a2999}
\end{subfigure}
\caption{Time evolution  of the two-point Wightman function for operators with large conformal dimension. In Case II,
the correlations are measured by an observer along the plane of reactions.
Plots in (a), (b), (c) and (d) are for fixed value of $\sigma=\sqrt{L_{x}}$ while varying  $\alpha\in\left\{1,\half,\frac{1}{4},\frac{1}{8}\right\}$. Instead in (e), $\alpha=1$ with $\sigma=\sqrt{1.5L_{x}}$. All these figures are deduced for geodesics with the deepest bulk penetration which is given by the choice $\rho_{m}=0.999\rho_{h}$ in our setup.  Other parameters of the simulations are similar to ones used in the previous sections.}
\label{fig:23}
\end{figure}

First, we have to solve the equations of motion for $\tau_{0}$ and $\rho_{0}$ in terms of $x_{0}$. 
They are already mentioned in 
\Eq(Twopoint-t0) and \Eq(Twopoint-u0). Choosing the positive root, the solution is given by
\be
\frac{d\rho_{0}}{d\tilde{x}_{0}}=
\frac{\sqrt{\left(1-\rho^{4}_{0}\right)\left(\rho^{2}_{m}-\rho^{2}_{0}\right)}}{\rho_{0}}\,,
\ee
with the change of variable $\tilde{x}_{0}\equiv x_{m}-x_{0}$. Solving the above equation for $\tilde{x}_{0}$, 
in the limit of $\rho_{0}\rightarrow0$, we find 
$\rho_{0}=\sqrt{2\rho_{m}\tilde{x}_{0}}$. From \Eq(Twopoint-u0), 
we find $D=\frac{\rho_{m}}{2\tilde{x}_{0}}$ and therefore the denominator of
the last term in \Eq(caseII-length) behaves as
\be
\frac{1}{\rho_{0}\sqrt{D}}\sim\frac{1}{\rho_{m}}\,,
\ee
which has a finite value. This means that imposing the boundary condition $x_{2}=0$ at $\pm x_{m}$ is 
safe and its contribution vanishes
as the profile is symmetric around $x_{0}=0$. To write it in the final form, 
we use 
$\rho_{0}=\rho_{m}\left(1-q^{2}\right)$ and solve for $\dot{\tau}_{0}$ 
from \Eq(twopoint-tdot) to obtain
\bea\label{L2-ref}
\mathcal{L}_{2}&=&\int_{0}^{1}\frac{4\left(1-q^{2}\right)dq}{\sqrt{1-\rho^{4}_{m}\left(1-q^{2}\right)^{4}}}
\left(\frac{\hat{\Sigma}_{d}}{\sqrt{2-q^{2}}}
-\frac{q^{2}\sqrt{2-q^{2}}}{2\left(1-q^{2}\right)^{2}\left(1-\rho^{4}_{m}\left(1-q^{2}\right)^{4}\right)}\hat{A}
\right.\nonumber\\&&\left.\hspace{2cm}
+\frac{\rho^{2}_{m} q\left(1-q^{2}\right)}{\sqrt{1-\rho^{4}_{m}\left(1-q^{2}\right)^{4}}}\,\hat{\Xi}_{f}\right)\,.
\eea
Similar to the last section, plots for the above expression are shown in Figures \ref{fig:TwoPointB2cc999}-\ref{fig:TwoPointB4a2999} for various tuning parameters $\alpha$ and $\sigma$ in \Eq(p0-intro). In Figures \ref{fig:TwoPointB2cc999}-\ref{fig:TwoPointB3c2999}, plots for $\alpha\in\left\{1,\half,\frac{1}{4},\frac{1}{8}\right\}$ are shown and in Figure \ref{fig:TwoPointB4a2999}, $\sigma=\sqrt{1.5L_{x}}$. An important observation is made by comparing our plots to those of the last section. In fact they look very identical. Let us remind ourself  about the difference between Case I in Figures \ref{fig:TwoPointA2cc999a}-\ref{fig:TwoPointA4a2999} and Case II with the figures listed below. In the first scenario, correlation between two points on the $y$ axis is measured while a scalar field with a Gaussian profile falls into the black hole. The correlation between the points is found by computing the geodesic connecting these pair of points through the bulk. This means that as the scalar field $\phi$ is falling into the bulk, the excitations that are produced by the form of the profile will affect the length of the geodesic.  The plane of the flow of these excitations  are orthogonal to the plane where the geodesic is drawn. In Case II, both the excitations of the scalar field and the geodesics are on the same plane. 
The resemblance of the two scenarios is very nontrivial  although we also have to remember
that our results are valid for correlations of operators with large mass dimensions.  A rough explanation is that in $\mathcal{L}_{2}$ in both cases apart from the geometrical factors that
parametrize the geodesics,  in case I, the functional dependence is given by $\mathcal{L}_{2}(\hat{\Sigma_{d}},\hat{A},\hat{\Xi}_{f})$, 
while in case II, the explicit dependence is clear from \Eq(L2-ref). From our simulations, it was
clear that $\hat{\Sigma}_{d,b}$ were roughly at the same order while $\hat{\Xi}_{f}\ll1$. 
Notice also that  $\hat{\Xi}_{f}$ is an odd function of $x$, this means that the plots in 
Figures \ref{fig:TwoPointB2cc999}-\ref{fig:TwoPointB4a2999} are not completely symmetric 
along $x=0$ compared to those mentioned in Figures \ref{fig:TwoPointA2cc999a}-\ref{fig:TwoPointA4a2999} 
of Case I. For a similar conclusion on the connection between  inhomogeneity and appearance of odd
functionalities in the correlation functions refer to \cite{Aarts:1999zn}.

In the next section, we study  entanglement entropies and they show that they are more distinctive when it comes to different setups for thermalization.

\clearpage

\subsection{Entanglement entropy}

In this section, we generalize our previous arguments on two-point functions. Among  different options for the 
minimal surfaces that one can use, we restrict ourself to the strip geometry. Then rather than probing the bulk by
a single geodesic, we will measure the thermalization by a minimal surface that satisfies the boundary of a strip. 
We will follow  Ryu and Takayanagi \cite{Ryu:2006} prescription for calculating the entanglement entropy (EE) for
holographic theories which is based on extremizing bulk surfaces. For related works on EE refer to \cite{EE-related}.

\addsubsubsection{Case I: plane A-B}
One natural way for parameterizing the boundary is to use the set of coordinates $(x,y,z)$. Let's
parametrize the direction  that forms an arc by going through the bulk to be $y_{0}$.
Then the geometry is extended indefinitely along  $x$ and $z$ axes. The situation that these two coordinates 
are cyclic has been considered recently in \cite{Alex2014}. As before, we assume that the inhomogeneity 
backreacts along the $x$ direction while leaving  $\partial_z$ as the Killing vector. The reader who is familiar with the derivations can skip to the discussion at the end of this subsection.

The surface area will be evaluated from the induced metric using
coordinates $(x,y,z)$. The induced metric  to the hypersurface is conveniently derived by confining  
line elements to displacements confined to the hypersurface.
Doing so we find that
\be\label{EE-action}
S_{\Sigma}=\int_{-\infty}^{\infty}dx_{0}\int_{-\infty}^{\infty}dz_{0}\int_{-y_{m}}^{y_{m}}dy_{0}\,\sqrt{\gamma_{ind}} \,\,\Sigma_{b}\,,
\ee
with tangent vectors of the curves on the hypersurface defined by $e^{\alpha}_{a}\equiv\frac{dx^{\alpha}}{dy^{a}}$ and
\bea
\gamma_{ind}&=&-2\frac{A}{\rho^{2}}e^{\tau}_{x}e^{\tau}_{y}\left(e^{\tau}_{y}e^{\rho}_{x}
+e^{\tau}_{x}e^{\rho}_{y}\right)
-3A^{2}\left(e^{\tau}_{x}e^{\tau}_{y}\right)^{2}
-\frac{2}{\rho^{2}}\left(e^{\tau}_{x}e^{\rho}_{x}\Sigma^{2}_{b}+e^{\tau}_{y}e^{\rho}_{y}\Sigma^{2}_{d}\right)
 +\Sigma_{b}^{2}
\Sigma_{d}^{2}
\nonumber\\&&
+2\Xi\left(
3A e^{\tau}_{x}\left(e^{\tau}_{y}\right)^{2}
+2\frac{\left(e^{\tau}_{y}\right)^{2}e^{\rho}_{x}}{\rho^{2}}+e^{\tau}_{x}\Sigma^{2}_{b}
\right)
+4\frac{e^{\tau}_{x}e^{\tau}_{y}e^{\rho}_{x}e^{\rho}_{y}}{\rho^{4}}-4\left(e^{\tau}_{y}\right)^{2}\Xi^{2}
\nonumber\\&&
-A\left[\left(e^{\tau}_{y}\Sigma_{d}\right)^{2}+\left(e^{\tau}_{x}\Sigma_{b}\right)^{2}\right]\,.
\eea
The equations of motion follow by varying the action
\bea
\partial_{i}\frac{\partial S_{\Sigma}}{\partial(\partial_{i}\tau)}-\frac{\partial S_{\Sigma}}{\partial \tau}=0
\,,\quad
\partial_{i}\frac{\partial S_{\Sigma}}{\partial(\partial_{i}\rho)}-\frac{\partial S_{\Sigma}}{\partial \rho}=0\,,
\eea
with $i\in\{x,y\}$. 
Expanding the coordinates to $\mathcal{O}(l^{2})$, the EE similar to the two-point Wightman functions, will
have an expansion of the form $S_{\Sigma}=S_{\Sigma(0)}+l^{2}S_{\Sigma(2)}+l^{2}\delta S_{\Sigma(0)}$. 
To zeroth-order in the perturbation  one gets for the hypersurface
\be\label{EE-zerolength}
S_{\Sigma(0)}=2K^{2}\int_{0}^{y_{m}}dy_{0}\frac{\sqrt{D}}{\rho^{3}_{0}}\,,
\ee
where since the effect of the inhomogeneity comes from the backreaction of the metric and hence it's a 
$\mathcal{O}(l^{2})$ effect, it will consequently be absent here and the integral over $x$ will be done
 trivially. The  cut off $K$  has been  introduced for trivial integrations. 

To second order, we have
\bea
S_{\Sigma(2)}=2K\int_{-\infty}^{\infty}dx_{0}\int_{0}^{y_{m}}dy_{0}\frac{1}{2\rho^{3}_{0}\sqrt{D}}
\left[2\hat{\Sigma}_{b}+2\left(\hat{\Sigma}_{b}+\hat{\Sigma}_{d}\right)D
-\dot{\tau}_{0}^{2}\rho^{2}_{0}\hat{A}\right]\,,
\eea
also note that in the above expression, the integral over the coordinate $x$ is now nontrivial as all the metric
components $\hat{\Sigma}_{b}$, $\hat{\Sigma}_{d}$ and $\hat{A}$ are the backreacted corrections. The next contribution changes 
the boundary volume since it depends on $\tau_{2}$, $\rho_{2}$ and $x_{2}$ according to
\bea\label{EE-surface}
\hspace{-.4cm}\delta S_{\Sigma(0)}&\!=\!&2K\int_{-\infty}^{\infty}\!\!dx_{0}\int_{0}^{y_{m}}\!\!dy_{0}
\left[\frac{2\dot{\tau}_{0}^{2}\rho^{4}_{0}-3D}{\rho^{4}_{0}\sqrt{D}}\rho_{2}
-\frac{\dot{\tau}_{0}}{\rho^{3}_{0}\sqrt{D}}\dot{\rho}_{2}-\frac{\dot{\tau}_{0}\left(1-\rho^{4}_{0}\right)+\dot{\rho}_{0}}{\rho^{3}_{0}\sqrt{D}}
\dot{\tau}_{2}\right]\,.
\eea
It should be pointed out that if we assume $\dot{x}_{0}\neq0$ then a term proportional to $x_{2}$ will appear in
the EE contribution. Similar to the previous case, looking at the geodesics will provide us the 
following equations for the profiles
of $\rho_{0}(y)$ \cite{Alex2014},
\be\label{Alex-EQ}
(1-\rho_{0}^{4})\dot{\tau}_{0}+\dot{\rho}_{0}=0\,,\quad D\rho^{6}_{0}=\rho^{6}_{m}\,,
\ee
which reduce to
\be\label{EE-motion}
\frac{d\rho_{0}}{dy}=-\frac{\sqrt{\left(1-\rho^{4}_{0}\right)\left(\rho^{6}_{m}-\rho^{6}_{0}\right)}}{\rho^{3}_{0}}\,.
\ee
Although a full analytic solution to the above equation will be desirable, it suffices to find an
asymptotic solution which will be required in the subsequent section,
\be
y_{0}=y_{\ast}-\frac{\rho^{4}_{0}}{4\rho^{3}_{m}}+\mathcal{O}(\rho_{0}^{8})\,,
\ee
this is the boundary coordinate as seen from an observer falling deep in the bulk. The straight
substitution from \Eq(Alex-EQ) and \Eq(EE-motion) has shown that \cite{Alex2014},
\bea
S_{\Sigma(0)}&=&2K^{2}\int_{\epsilon}^{\rho_{m}}d\rho_{0}\frac{\rho^{3}_{m}}{\rho^{3}\sqrt{(1-\rho^{4})(\rho_{m}^{6}-\rho^{6})}}\,,
\\\label{EE-Entropy}
S_{\Sigma(2)}&=&2K\int_{-\infty}^{\infty}dx_{0}\int_{0}^{\rho_{m}}d\rho_{0}\left[
\frac{2\rho^{6}_{0}(1-\rho^{4}_{0})\hat{\Sigma}_{b}+2\rho^{6}_{m}(1-\rho^{4}_{0})\left(\hat{\Sigma}_{b}+\hat{\Sigma}_{d}\right)
-\rho^{2}_{0}(\rho^{6}_{m}-\rho^{6}_{0})\hat{A}}{2\rho^{3}_{0}\rho^{3}_{m}(1-\rho^{4}_{0})^{3/2}
\sqrt{\rho^{6}_{m}-\rho^{6}_{0}}}\right]\,.\nonumber\\
\eea 

From \Eq(EE-surface) it is evident that we can simplify the
expression using the equations of motion . The coefficients of $\dot{\tau}_{0}$ cancel out. The
derivative over $\dot{\rho}_{2}$ can be  rewritten using the partial derivative in terms of $\rho_{2}$
which will be again proportional to the equations of motion. The only contribution emerging
from the surface term is
\be
\delta S_{\Sigma(0)}=2K^{2}\left.
\frac{\dot{\tau}_{0}}{\rho^{3}_{0}\sqrt{D}}\rho_{2}\right|_{0}^{y_{m}}\,.
\ee
It's easiest first to evaluate the coefficient of $\rho_{2}$  because it is at zero order  in the backreaction rather than calculating the whole expression.
Since only the quantities such as $\dot{\tau}_{0}$ and $\dot{\rho}_{0}$ are required, we can expand
around $y=0$ which is equivalent to the top of the arc in the bulk where it gets its maximum 
value $\rho_{m}$. Perturbatively solving the equation of motion in \Eq(EE-motion), we obtain the following
solutions 
\be
\rho_{0}(y)=\rho_{m}+\frac{3}{2}\left(\frac{-1+\rho^{4}_{m}}{\rho_{m}}\right)y^{2}\,,
\quad 
\tau_{0}(y)=\frac{3}{2}\frac{y^{2}}{\rho_{m}}\,.
\ee
There is also a non-physical solution $\rho_{0}(y)=\rho_{m}$ and $\tau_{0}(y)=\frac{3}{2}\frac{y^{2}}{\rho_{m}}$, this
 solution can be discarded
as it takes an infinite time for the geodesic to satisfy the boundary condition. Nonetheless, both solutions
give a vanishing contribution to the value of the expression that we are interested. 

The value of the expression at  $y_{0}=y_{m}$ requires more work. Since the boundary time $\tau_{\ast}$ will
be the time at which $\rho_{0}\rightarrow 0$, we can solve the differential equation in \Eq(EE-motion) to obtain 
$\rho_{0}\sim (y_{m}-y_{0})^{1/4}$.
Putting everything together \cite{Alex2014}, we obtain  the coefficient of $\rho_{2}(y_{0})$, 
\be\label{EE-singularity}
-2K\frac{1}{2\sqrt{2}\rho_{m}^{9/4}\delta^{3/4}}\,,
\ee
where in the above $\delta$ is a regulator to avoid the singularity of the upper limit of $y=y_{m}$.
As it has been argued, one needs to evaluate the behavior of $\rho_{2}(y_{0})$ to find the finite
contribution to the entanglement entropy. Following the method described in \cite{Alex2014}, we
vary the action in \Eq(EE-action) for  $\tau_{2}(y_{0})$ and $\rho_{2}(y_{0})$ as it's 
not clear from the beginning whether or not there will be a modification from terms that  depend on the inhomogeneity
  in the  action of \Eq(EE-action). From the Euler-Lagrange equations
\be
\delta_{\rho_{2}} S_{\Sigma}-\frac{d}{dy_{0}}\left(\delta_{\dot{\rho}_{2}}S_{\Sigma}\right)=0\,,\quad
\delta_{\tau_{2}} S_{\Sigma}-\frac{d}{dy_{0}}\left(\delta_{\dot{\tau}_{2}}S_{\Sigma}\right)=0\,,
\ee
at $\mathcal{O}(l^{2})$, naturally, we recover the equations of motion for the unperturbed variables $\rho_{0}$
and $\tau_{0}$. 
Along the same line, at  $\mathcal{O}(l^{4})$, we find the equations of motion for $\tau_{2}$ and $\rho_{2}$.
These are ab initio nonlinear equations involving components of metric $A$, $\Sigma_{b}$,
$\Sigma_{d}$ and $\Xi$  on one hand and $\tau_{0}$, $\rho_{0}$, $\tau_{2}$ and $\rho_{2}$ on the other. 
As the singularity in \Eq(EE-singularity) originates from the limit of $\rho\rightarrow 0$, we can replace
the components of the metric with their leading  values in \Eq(B_A)-\Eq(B_Xi) from the appendix. Using the asymptotic
expansions for $\tau_{0}$ and $\rho_{0}$ as mentioned in the paragraph above  \Eq(EE-singularity), at leading order,
we find
\bea
\ddot{\rho}_{2}+\ddot{\tau}_{2}=\frac{1}{24\sqrt{2}}\frac{\rho^{9/2}_{m}p_{0}^{2}(\tau_{\ast},x_{\ast})}{\tilde{y}^{5/4}_{0}}
+\mathcal{O}(1/\tilde{y}_{0})\,,
\eea
where in the above $\tilde{y}_{0}=(y_{m}-y)$.  In the limit of 
$\tilde{y}_{0}\rightarrow0$, assuming the derivatives of $p_{0}$ are suppressed by extra factors of $\tilde{y}_{0}$, 
the former degenerate equation  \cite{Alex2014} 
yields 
\be
\rho_{2}+\tau_{2}=-\frac{\sqrt{2}}{9}\,p^{2}_{0}(\tau_{\ast},x_{\ast})\,\rho_{m}^{9/2}\,\delta^{3/4}\,.
\ee
Since there is no modification from the other components of the metric, this is  identical to the homogeneous case in \cite{Alex2014}. Finding
 the coefficient will result in
\be
\delta S_{\Sigma(0)}=K^{2}\frac{5}{36}\,p^{2}_{0}(\tau_{\ast},x_{\ast})\,.
\ee

The integral in \Eq(EE-Entropy)  is singular at $\rho_{0}=0$ and we have to regularize it.
To do so as before, we make use of the asymptotic expansions of the metric components for
$\rho_{0}\rightarrow0$ in \Eq(B_A)-\Eq(B_Sigmab),
\bea\label{expand1}
\hat{A}&=&-\frac{1}{6}p_{0}^{2}+\rho_{0}^{2}\,a_{2}+\mathcal{O}(\rho_{0}^{2}\ln\rho_{0})\,,\\
\hat{\Sigma}_{d}&=&-\frac{1}{12}\rho_{0}^{2}\,p^{2}_{0}+\rho_{0}^{4}\,d_{4}+\mathcal{O}(\rho^{4}_{0}\ln\rho_{0})\,,\\
\label{expand2}
\hat{\Sigma}_{b}&=&-\frac{1}{12}\rho_{0}^{2}\,p^{2}_{0}+\rho_{0}^{4}\,b_{4}+\mathcal{O}(\rho^{4}_{0}\ln\rho_{0})\,,
\eea
then from the expansion around the singularity, a counter term can be formed
\be
S_{counter}=\frac{K^{2}}{6}\,p^{2}_{0}(\tau_{\ast},x_{\ast})\int_{\epsilon}^{\rho_{m}}
\,\frac{d\rho_{0}}{ \rho_{0}}\,,
\ee
where $\epsilon$ is a regulator for the integral. Substituting from \Eq(expand1)-\Eq(expand2), the finite part of \Eq(EE-Entropy) reads
\be
S^{fin}_{\Sigma(2)}=2K\int_{-\infty}^{\infty}dx_{0}\int_{0}^{\rho_{m}}\rho_{0}d\rho_{0}\left[
\frac{2\rho^{6}_{0}(1-\rho^{4}_{0})b_{4}+2\rho^{6}_{m}(1-\rho^{4}_{0})\left(b_{4}+d_{4}\right)
-(\rho^{6}_{m}-\rho^{6}_{0})a_{2}}{2\rho^{3}_{m}(1-\rho^{4}_{0})^{3/2}
\sqrt{\rho^{6}_{m}-\rho^{6}_{0}}}\right]\,,
\nonumber\\
\ee
with $a_{2}$, $b_{4}$ and $d_{4}$ function of $(\tau_{0},x_{0})$ with $\tau_{0}(\rho_{0})$. The corresponding divergent part evaluates to
\be
S^{div}_{\Sigma(2)}=-2K\int_{-\infty}^{\infty}dx_{0}\int_{\epsilon}^{\rho_{m}}d\rho_{0}
\frac{p^{2}_{0}(\tau_{0},x_{0})}{12}\left[
\frac{2\rho^{6}_{0}(1-\rho^{4}_{0})+4\rho^{6}_{m}(1-\rho^{4}_{0})
-2(\rho^{6}_{m}-\rho^{6}_{0})}{2\rho_{0}\rho^{3}_{m}(1-\rho^{4}_{0})^{3/2}
\sqrt{\rho^{6}_{m}-\rho^{6}_{0}}}\right]\,.
\nonumber\\
\ee
Now, it is convenient to make the process of regularization scheme independent by  adding
\be
S_{cor}=-\frac{K^{2}}{6}p^{2}_{0}(\tau_{\ast},x_{\ast})\log\rho_{m}\,.
\ee
Finally, the total entanglement entropy for the strip geometry, including the inhomogeneity implicitly, will be
\bea\label{EE-I}
\hspace{-2cm}&&S_{\Sigma(2)}=S^{fin}_{\Sigma(2)}+S^{div}_{\Sigma(2)}+S_{counter}+S_{cor}+\delta S_{\Sigma(0)}
\nonumber\\
&&\!=\!4K\!\!\int_{-\infty}^{\infty}\!\!dx_{0}\!\int_{0}^{1}\!\!qdq\!\left[
\frac{\rho^{2}_{m}(1-q^{2})^{7}\,b_{4}}{\sqrt{(1-\!\rho^{4}_{m}(1\!-\!q^{2})^{4})(1\!-\!(1\!-\!q^{2})^{6})}}
+\frac{\rho^{2}_{m}(1-q^{2})(b_{4}+d_{4})}{\sqrt{(1-\rho_{m}^{4}(1-q^{2})^{4})(1-(1-q^{2})^{6})}}
\right.
\nonumber\\&&
\left.
-\frac{\rho^{2}_{m}(1-q^{2})\sqrt{1-\left(1-q^{2}\right)^{6}}\,a_{2}}{2\left(1-\rho^{4}_{m}\left(1-q^{2}\right)^{4}\right)^{3/2}}
-\frac{\left(1-q^{2}\right)^{5}p^{2}_{0}(\tau_{0},x_{0})}{12\rho_{m}\sqrt{\left(1-\rho^{4}_{m}\left(1-q^{2}\right)^{4}\right)
\left(1-(1-q^{2})^{6}\right)}}
\right.
\nonumber\\&&
\left.
-\frac{p^{2}_{0}(\tau_{0},x_{0})}{6\rho_{m}\left(1\!-\!q^{2}\right)\sqrt{\left(1\!-\!\rho^{4}_{m}\left(1\!-\!q^{2}\right)^{4}\right)
\left(1\!-(1\!-\!q^{2})^{6}\right)}}
+\frac{p^{2}_{0}(\tau_{0},x_{0})\sqrt{1-\left(1-q^{2}\right)^{6}}}{12\rho_{m}\left(1\!-\!q^{2}\right)\left(1\!-\!\rho^{4}_{m}
\left(1\!-\!q^{2}\right)^{4}\right)^{3/2}}
\right]
\nonumber\\&&
+\frac{K^{2}}{6}p^{2}_{0}(\tau_{\ast},x_{\ast})\left[\int_{0}^{1}\frac{2q dq}{1-q^{2}}-\log\rho_{m}
+\frac{5}{6}\right]\,.
\eea
Note the difference between $p_{0}(\tau_{0},x_{0})$ and $p_{0}(\tau_{\ast},x_{\ast})$. They  will have some overlap in their values when they
cover the spacetime with $\tau_{0}(q)$ but in general are independent. The fact that the metric components
$a_{2}(\tau_{0},x_{0})$,  $b_{4}(\tau_{0},x_{0})$ and $d_{4}(\tau_{0},x_{0})$ are  nonlinear functions of the
inhomogeneity makes \Eq(EE-I) a nontrivial generalization of the result in \cite{Alex2014}.

\begin{figure}[!ht]
\begin{subfigure}{.5\textwidth}
{\includegraphics[width=7.0cm]{p0figureastime-crop}}
  \caption{}
  \label{fig:p0figureastime2}
\end{subfigure}%
\begin{subfigure}{.5\textwidth}
{\includegraphics[width=7.0cm]{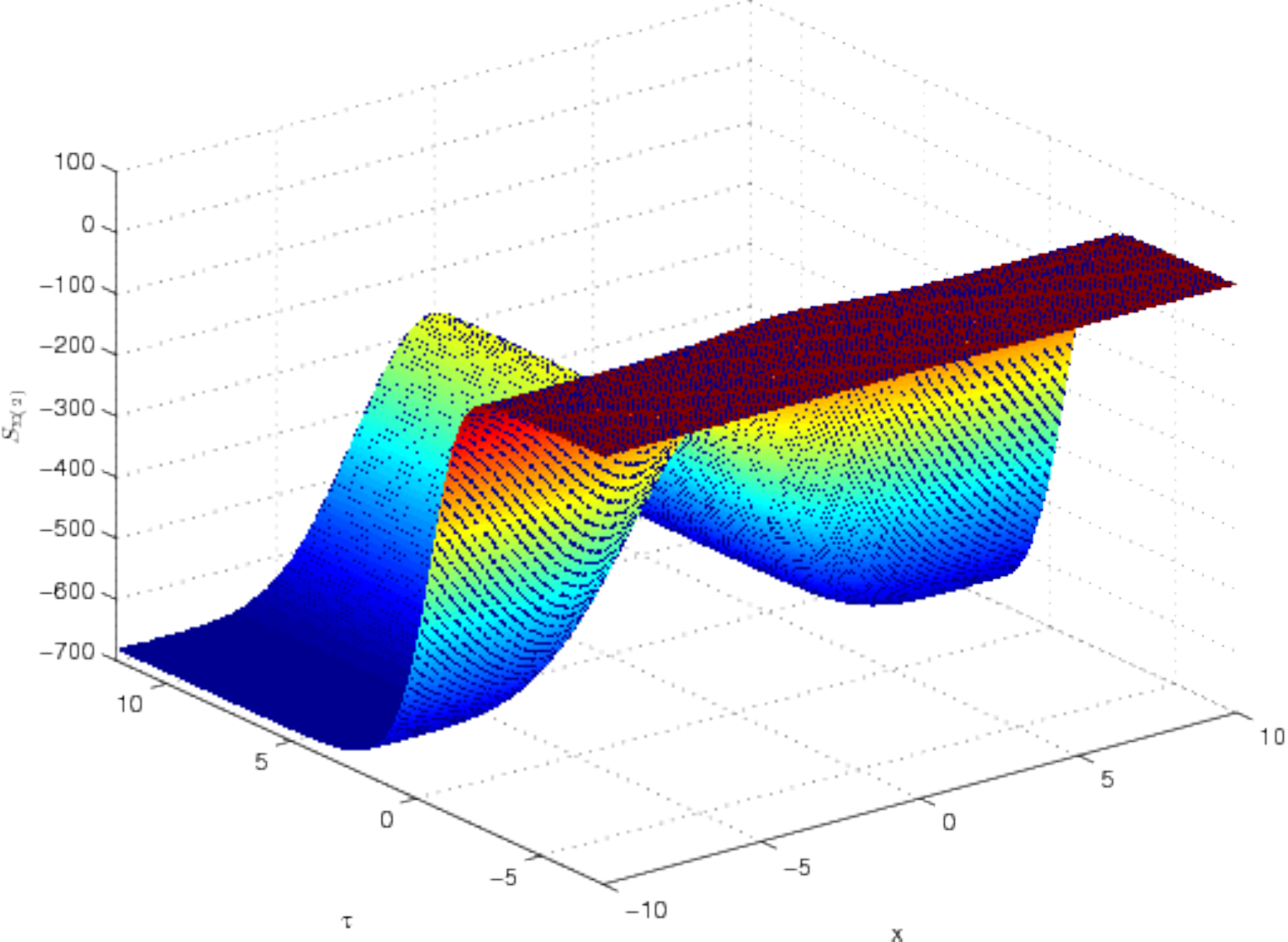}}
  \caption{}
  \label{fig:EEA2cc1}
\end{subfigure}%

\begin{subfigure}{.5\textwidth}
{\includegraphics[width=7.0cm]{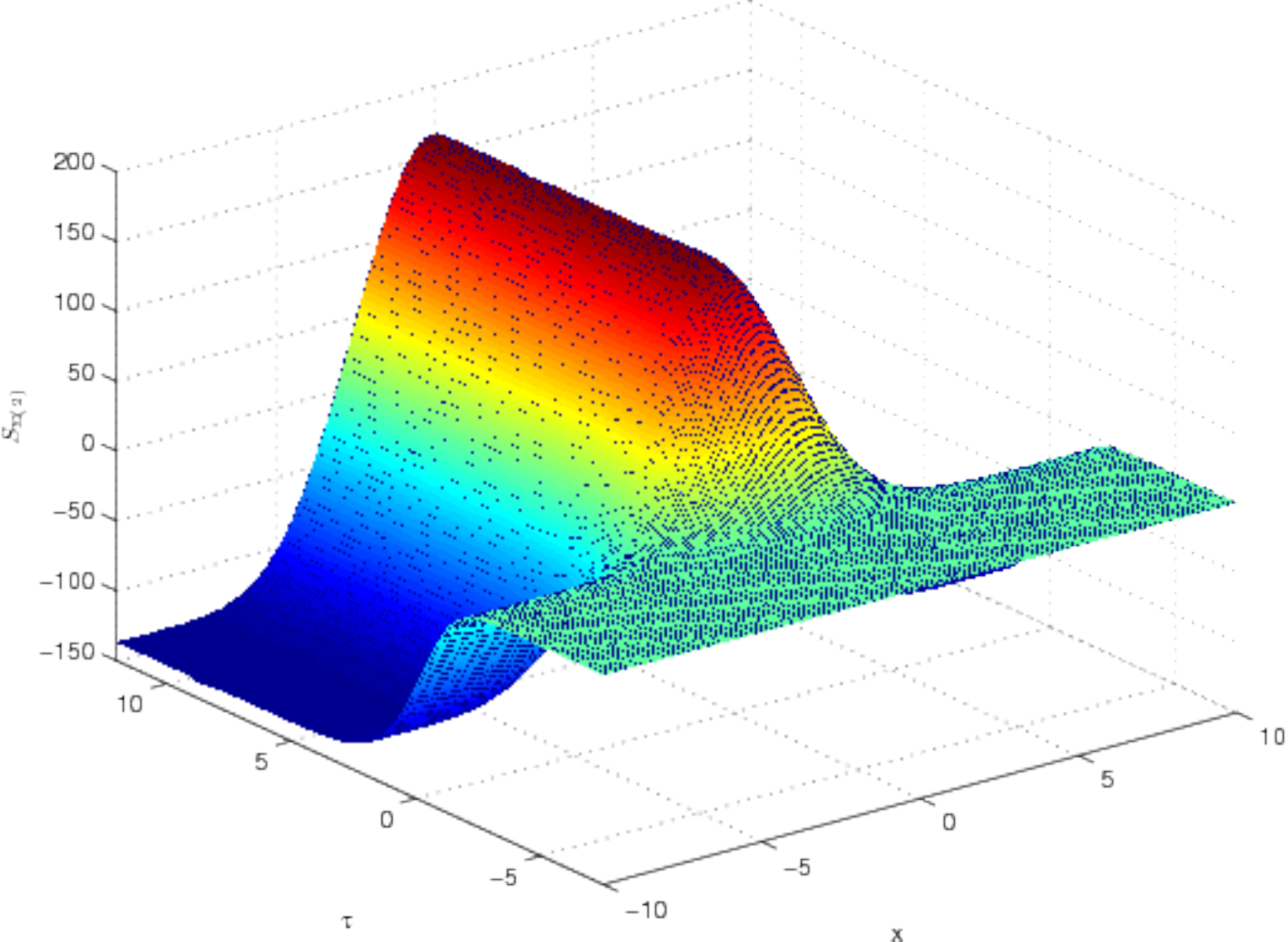}}
  \caption{}
  \label{fig:EEA2cc5}
\end{subfigure}
\begin{subfigure}{.5\textwidth}
{\includegraphics[width=7.0cm]{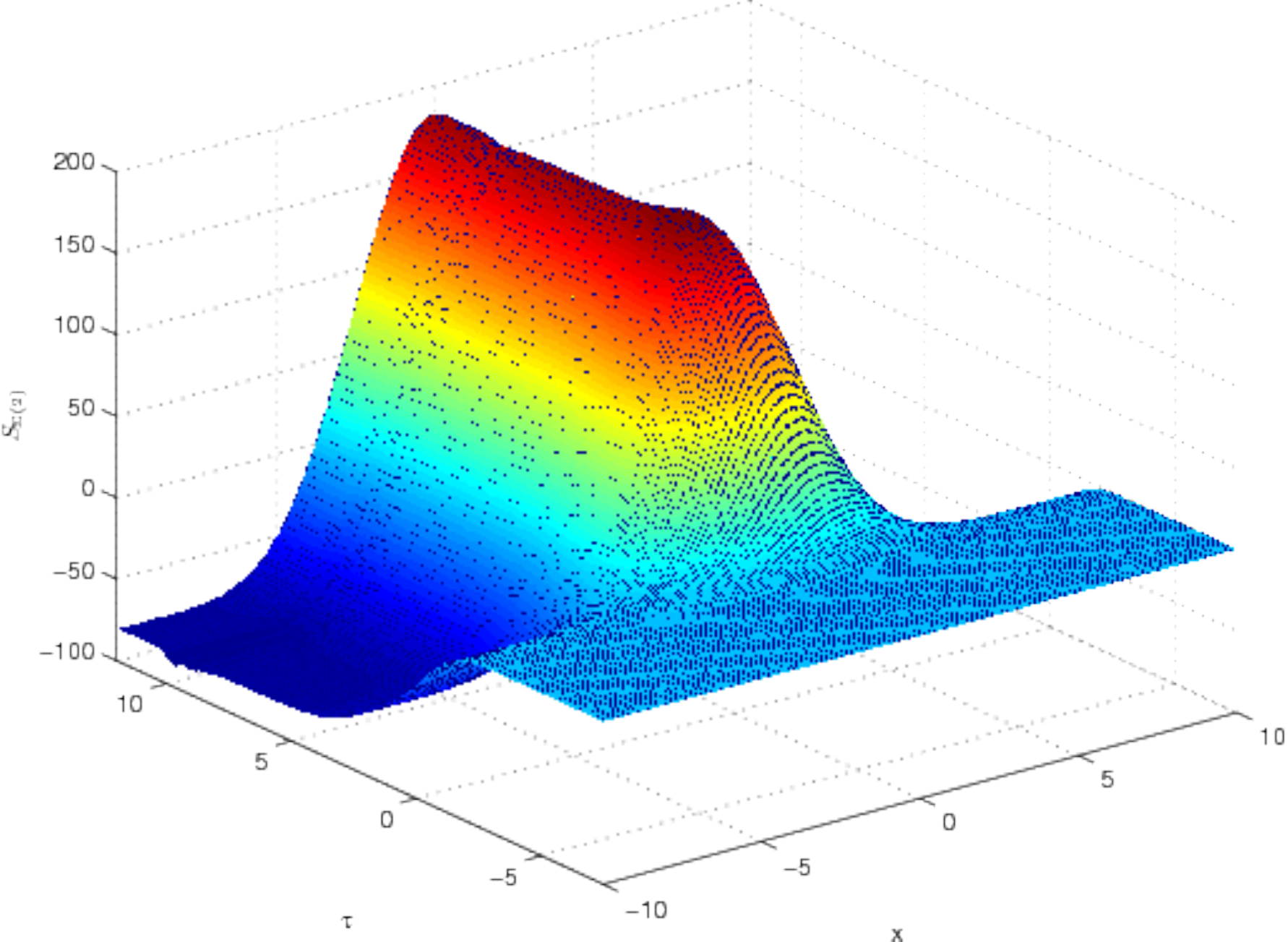}}
  \caption{}
  \label{fig:EEA2cc}
\end{subfigure}%

\centering
\begin{subfigure}{.5\textwidth}
{\includegraphics[width=7.0cm]{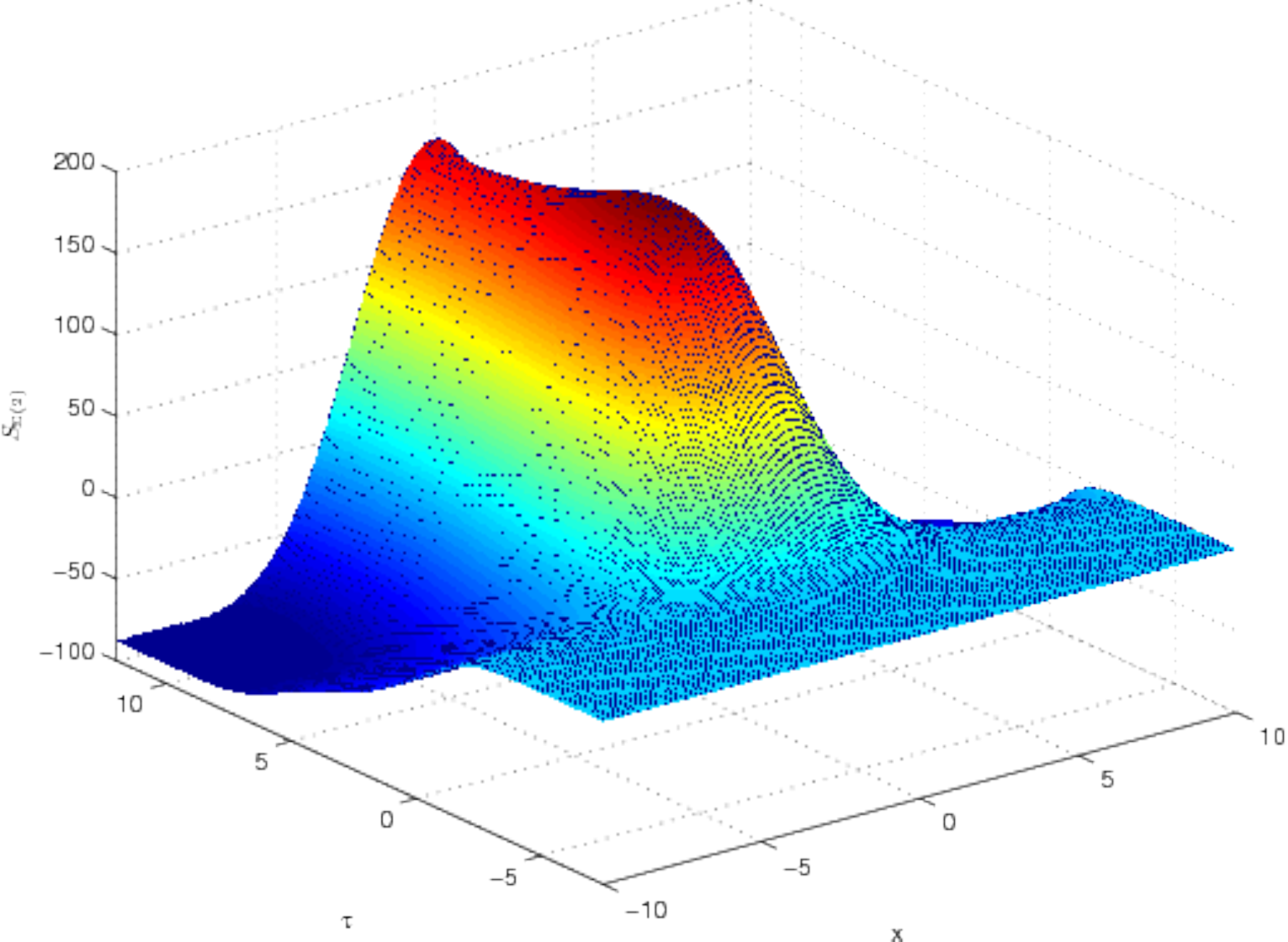}}
  \caption{}
  \label{fig:EEA2cc999}
\end{subfigure}
\caption{Plots of the time evolution  of the variation of the entanglement entropy at $\mathcal{O}(l^{2})$. In Case I,
 the correlating region is orthogonal to the plane of reaction. (a) is the source on the boundary and (b)-(e) are the corresponding plots for the EE as we vary $\rho_{m}$ for fixed $\alpha=1$ and $\sigma=L_{x}$. The numerical setup is identical to the previous sections.}
\label{fig:37}
\end{figure}

EE as a local observable  provides more detailed information for thermalization compared to  other observables that we have studied so far. First, we plan to study its dependence on the cut off $\rho_{m}$ that we have chosen in our analysis.
Figure \ref{fig:p0figureastime2} is  the profile of $p_{0}(\tau,x)$, the  non-normalizable mode 
of the scalar field, which is falling into the black hole.
Figures \ref{fig:EEA2cc1}-\ref{fig:EEA2cc999} are the corresponding variation of the EE as a function of the coordinates $x-\tau$ as we increase the value of the maximum depth of the entangling surface into the bulk from $0.1\rho_{m}$ to $0.999\rho_{m}$. This has the effect of shifting the amplitudes toward more positive values. 
It is easy to see from \Eq(EE-I) that the dynamics of EE for $\rho_{m}\ll 1$ is dominated by the original profile of $p_{0}(\tau,x)$ in addition to a constant offset contribution for $\tau<0$. At $\rho_{m}\sim 1$, this dynamics will be dominated by the backreacted components of the metric instead. This also explains why in Figure \ref{fig:EEA2cc999} the early Gaussian peak that appears at $\tau\simeq0$ is wider than the same Gaussian peak at late times due to the sudden appearance of the mass gap and plethora of excitations that follow. Figure \ref{fig:EEA2cc999} is the closest configuration to a realistic thermalization.

\clearpage

\begin{figure}[!ht]
\begin{subfigure}{.5\textwidth}
{\includegraphics[width=7.0cm]{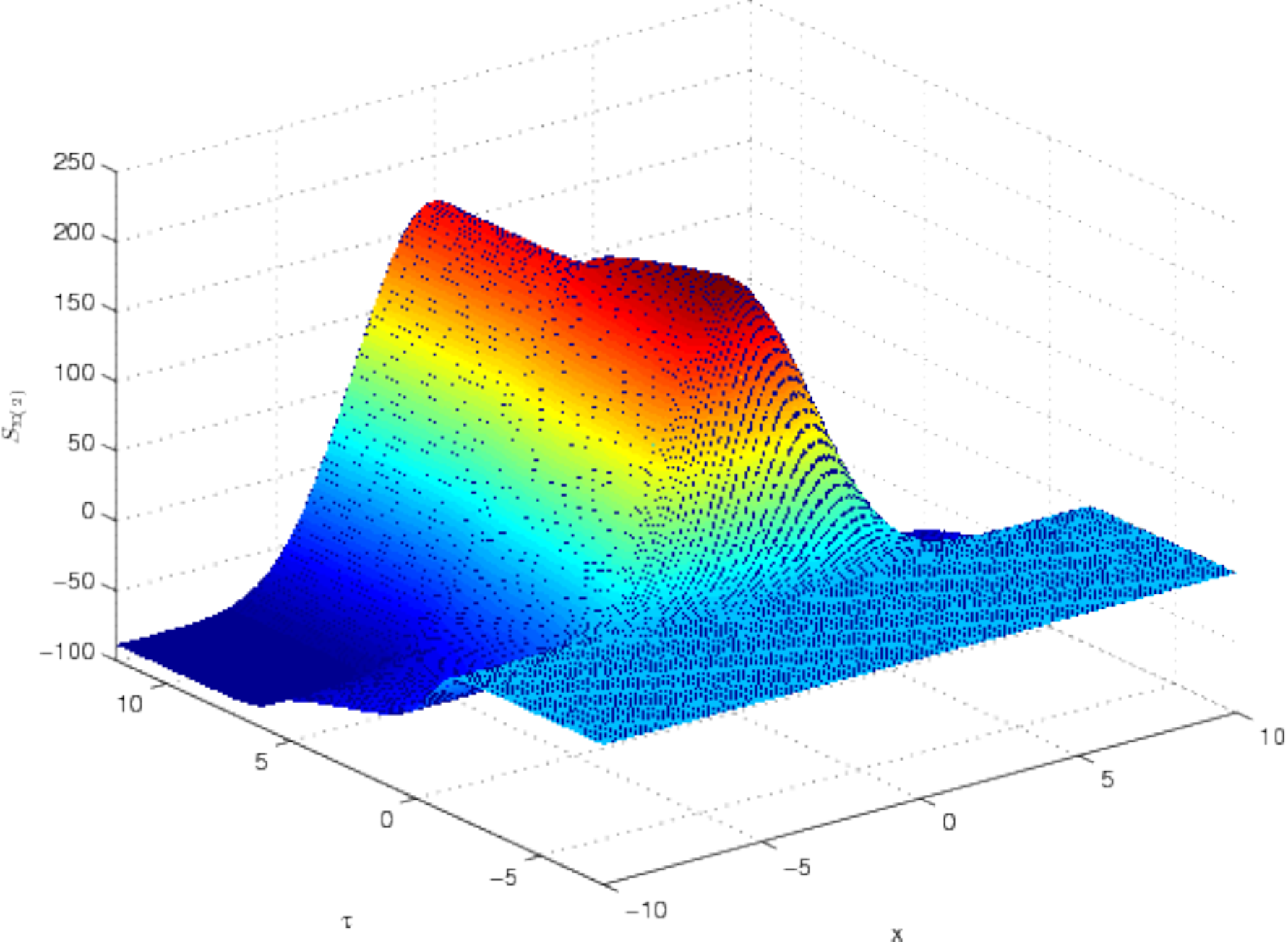}}
  \caption{}
  \label{fig:EEA3a2999}
\end{subfigure}
\begin{subfigure}{.5\textwidth}
{\includegraphics[width=7.0cm]{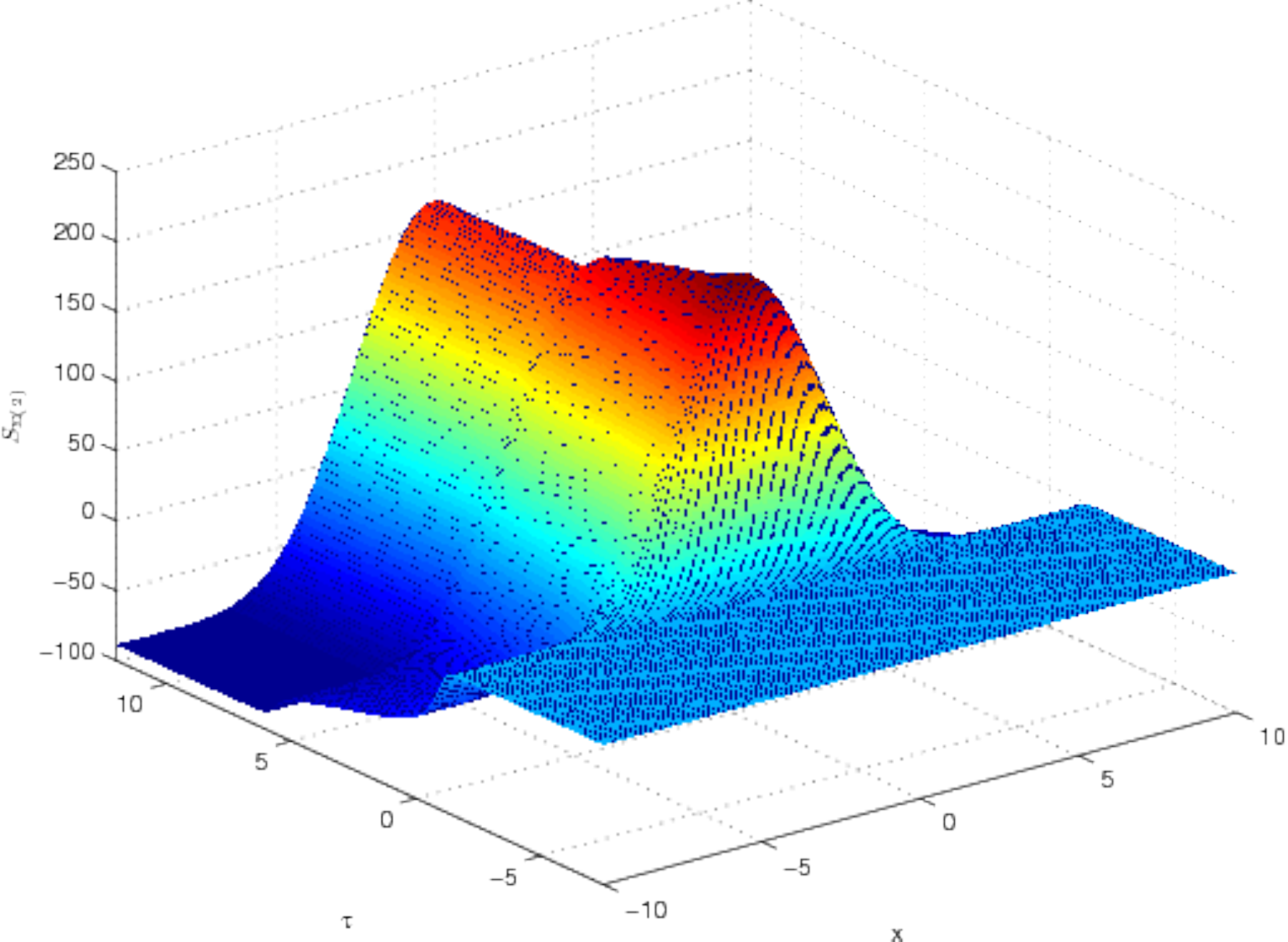}}
  \caption{}
  \label{fig:EEA3b2999}
\end{subfigure}

\begin{subfigure}{.5\textwidth}
{\includegraphics[width=7.0cm]{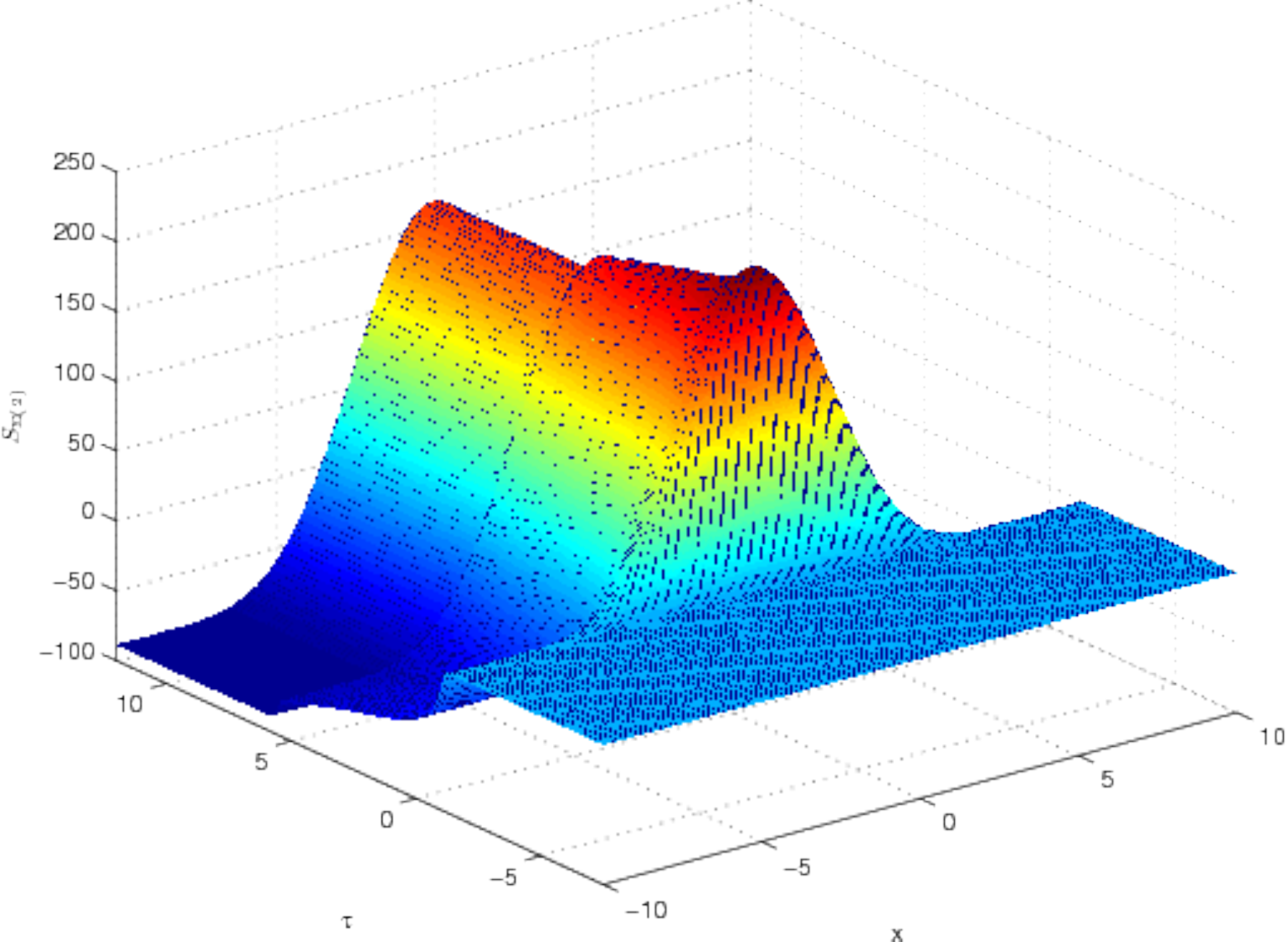}}
  \caption{}
  \label{fig:EEA3c2999}
\end{subfigure}
\begin{subfigure}{.5\textwidth}
{\includegraphics[width=7.0cm]{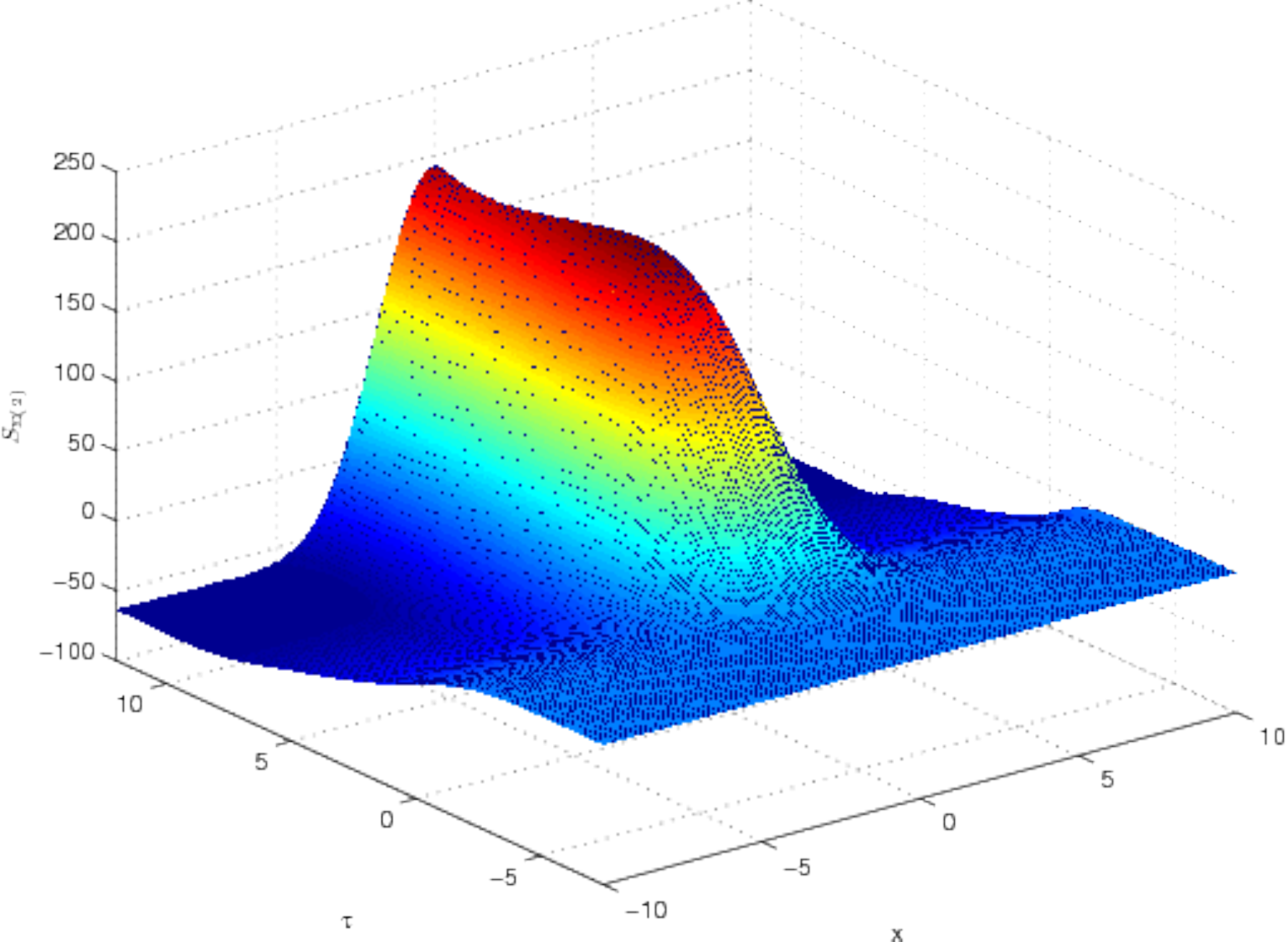}}
  \caption{}
  \label{fig:EEA4a2999}
\end{subfigure}

\begin{subfigure}{.5\textwidth}
{\includegraphics[width=7.0cm]{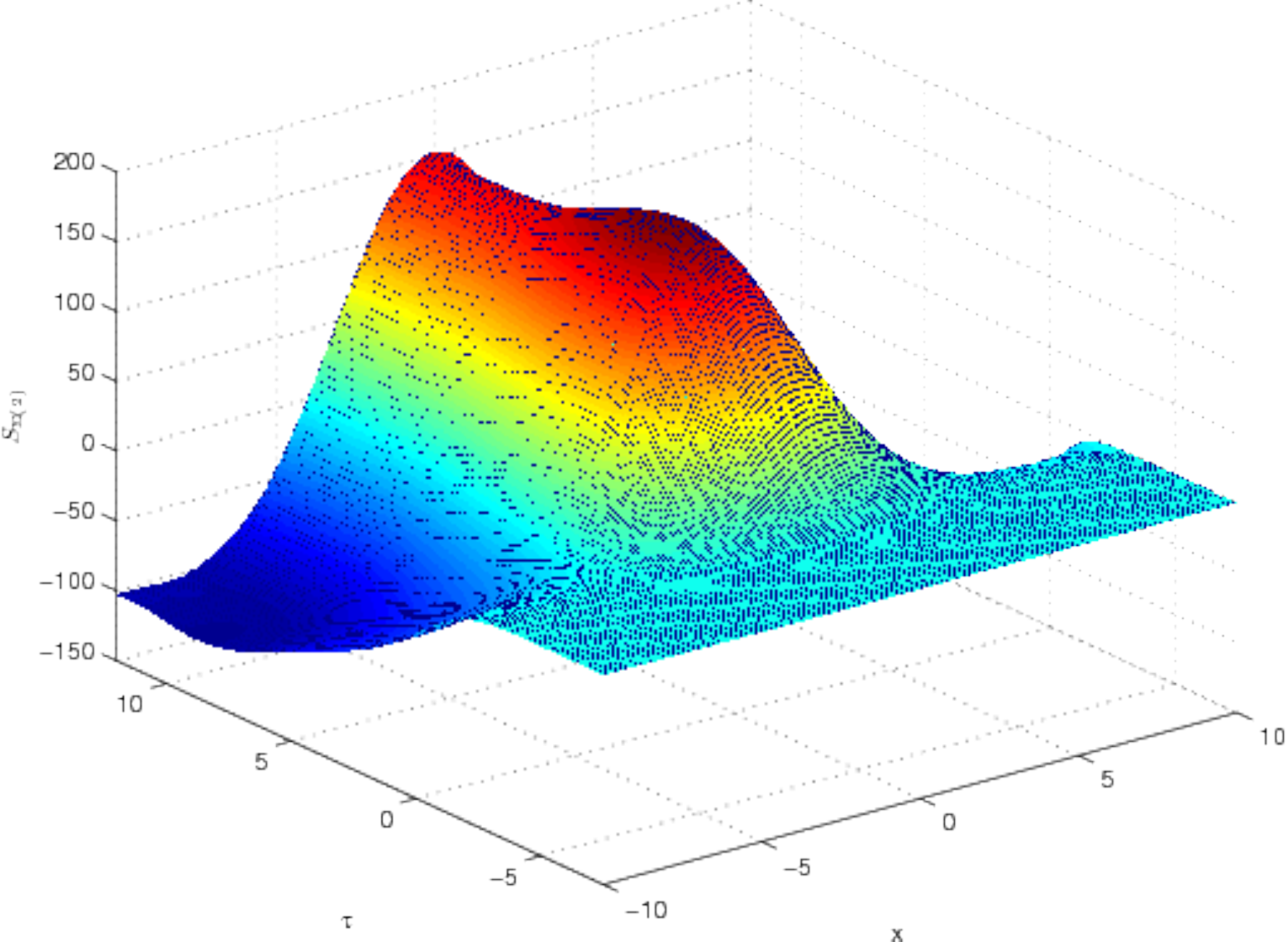}}
  \caption{}
  \label{fig:EEA4b2999}
\end{subfigure}
\begin{subfigure}{.5\textwidth}
{\includegraphics[width=7.0cm]{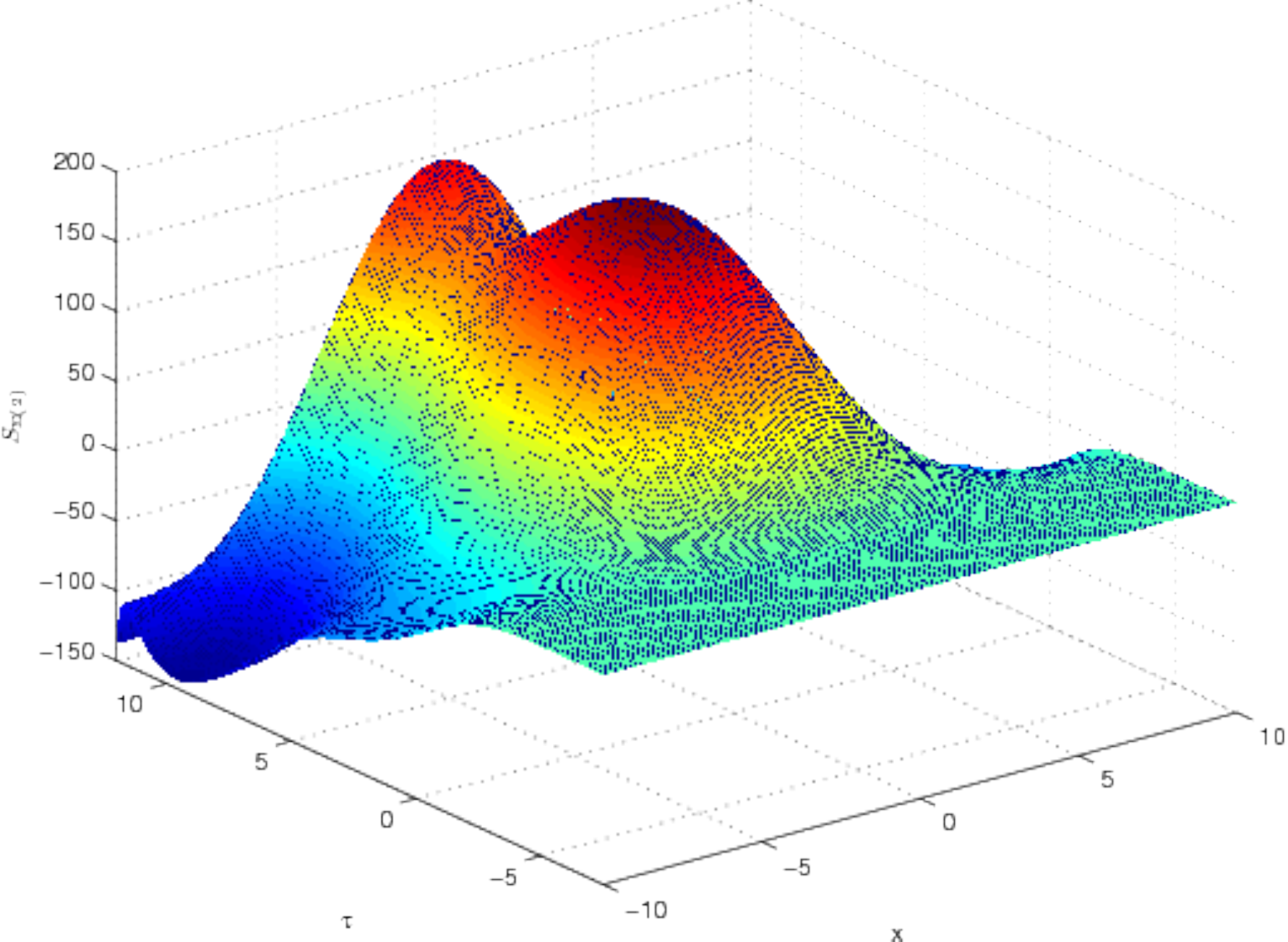}}
  \caption{}
  \label{fig:EEA4c2999}
\end{subfigure}
\caption{Corresponding plots for the EE as we reduce  $\alpha$ in (a)-(c) for $\sigma=\sqrt{L_{x}}$. In (d)-(f), we  increase $\sigma$  with fixed $\alpha=1$. We are also assuming $\rho_{m}=0.999\rho_{h}$ and $L_{x}=10$ in the above plots.}
\end{figure}

Our EE expressions are complicated and they don't  show the simple quasi-particle picture proposed by Cardy et al.  \cite{Calabrese:2005in,0808.0116}. Nevertheless, we can still connect to this idea. As it is shown in Figures \ref{fig:EEA3a2999}-\ref{fig:EEA3c2999}, we vary the tuning parameter  $\alpha\in\{\half,\frac{1}{4},\frac{1}{8}\}$. While we reduce the values of $\alpha$,  the mass gap production  will have a steep slope. This in part causes more excitations per volume. These ``quasiparticles'' are constrained by causality and from a given Cauchy surface at $\tau=0$, it will take them $\tau=x/v_{max}$ to reach to their ``horizon''. This effect can be seen in Figure \ref{fig:EEA3c2999} in a very pronounced way as it makes a slight wiggle on the surface  at $\tau\sim 5$.   

In Figures \ref{fig:EEA4a2999}, \ref{fig:EEA4b2999} and  \ref{fig:EEA4c2999}, we are gradually increasing the width of the  Gaussian profile for $p_{0}(\tau,x)$. This causes the blue region (in color), surrounding the bump, to shift toward the negative values and to expand the width of the peak at $\tau=0$. Curiously, this latter effect doesn't exceed a circular-shape region obeying radius $\tau=x/v_{max}$. We want to point out that this is not trivial.

\clearpage

\addsubsubsection{Case II: plane C-D}
Similar to the case considered in Section \ref{caseII}, for the two-point function, we reconsider
similar problem assuming
that the direction of the inhomogeneity is orthogonal to the boundaries of the entangling region. Let's call this region $A$. The geometry of $A$ is that of a strip and we parametrize it with  $(x_{0},y_{0},z_{0})$. The extremal surface that bounds $A$ throughout the bulk is derived from: 
\be\label{action-caseII}
S_{\Sigma}=\int_{-\infty}^{\infty}dy_{0}\int_{-\infty}^{\infty}dz_{0}\int_{-x_{m}}^{x_{m}}dx_{0}
\,\sqrt{\gamma_{ind}}\,\,\Sigma^{2}_{b}\,,
\ee
with
\be
\gamma_{ind}=-A\dot{\tau}^{2}+2\dot{\tau}\left(1+\dot{x}_{2}\right)\Xi
+(1+\dot{x}_{2})^{2}\Sigma^{2}_{d}-2\frac{\dot{\tau}\dot{\rho}}{\rho^{2}}\,,
\ee
and the boundary for the hypersurface (strip) is from $-x_{m}$ to $x_{m}$ and it's indefinitely extended along $y$ 
and $z$ directions. Note that in writing \Eq(action-caseII), we relied on the lessons learned from the geodesic equations
mentioned at the beginning such as \Eq(ddotxII).
Expansion has the general form
$S_{\Sigma}=S_{\Sigma(0)}+l^{2}S_{\Sigma(2)}+l^{2}\delta S_{\Sigma(0)}$. The first term has already been
calculated in \Eq(EE-zerolength). For $S_{\Sigma(2)}$, we get
\be\label{EE-caseII-S_sigma}
S_{\Sigma(2)}=2K^{2}\int_{0}^{x_{m}}\frac{dx_{0}}{2\rho^{3}_{0}\sqrt{D}}
\Big(
2\hat{\Xi}_{f}\,\dot{\tau_{0}}\,\rho^{2}_{0}-\dot{\tau_{0}}^{2}\rho^{2}_{0}\,\hat{A}+4D\,\hat{\Sigma}_{b}+2\hat{\Sigma}_{d}
\Big)\,,
\ee
with $D=1-\dot{\tau_{0}}^{2}+\dot{\tau_{0}}^{2}\rho^{4}_{0}-2\dot{\tau_{0}}\dot{\rho_{0}}$. Similar 
expansion for the dynamical variables  such as $\tau_{2}$, $\rho_{2}$ and $x_{2}$ gives
\bea
\delta S_{\Sigma(0)}&=&2K^{2}\int_{0}^{x_{m}}\!\!\frac{dx_{0}}{2\rho^{4}_{0}\sqrt{D}}
\left[2\rho_{0}\dot{x}_{2}+\left(-6D+4\dot{\tau}^{2}_{0}\rho^{4}_{0}\right)\rho_{2}
-2\rho_{0}\left(\dot{\tau}(1-\rho^{4}_{0})+\dot{\rho}\right)\dot{\tau}_{2}
-2\rho_{0}\dot{\tau}_{0}\dot{\rho}_{2}
\right]\,.\nonumber\\
\eea
As it was noticed in the last section the coefficient of  $\dot{\tau}_{2}$ is zero if we use the equations of motion 
at zero order. Again, the coefficient of the terms $\rho_{2}$ and $\dot{\rho}_{2}$ group 
together by partial integrations, yielding 
\be\label{EE-caseII-infty}
\delta S_{\Sigma(0)}=2K^{2}l^{2}\left.\frac{\dot{\tau}_{0}\,\rho_{2}}{\rho^{3}_{0}
\sqrt{D}}\right|^{x_{m}}_{0}
+K^{2}l^{2}\left.\frac{x_{2}}{\rho^{3}_{0}\sqrt{D}}\right|^{x_{m}}_{-x_{m}}\,,
\ee
where in the above, we applied  the equations of motion such as \Eq(Alex-EQ). In addition, we have changed the lower
 bound of the second term as we explained below \Eq(caseII-length). They are both
 diverging with $\delta^{-3/4}$ where $\delta$ is the cutoff in the $x_{0}$ direction when 
$\rho_{0}$ approaches the boundary. The first term is identical to the contribution from the surface term in Case 
I, but the second term is new and is due to the effect of the inhomogeneity. It's also challenging since 
if we want to enforce the boundary  condition of $x_{2}=0$ at $\pm x_{m}$ the coefficient must be finite. To find the exact value of the 
coefficient, we have to solve for the equations of motion for $x_{2}$ close to the boundary.

Using the fact that $\rho_{0}\sim (x_{m}-x_{0})^{1/4}$ and the boundary expansions to leading order for the metric
coefficients, such as
\bea
\hat{\Sigma}_{b}&=&-\frac{\rho_{0}^{2}}{12}\,p^{2}_{0}(\tau_{0},x_{0})+\mathcal{O}(\rho^{4}_{0})\,,\\
\hat{\Sigma}_{d}&=&-\frac{\rho_{0}^{2}}{12}\,p^{2}_{0}(\tau_{0},x_{0})+\mathcal{O}(\rho^{4}_{0})\,,\\
\hat{\Xi}_{f}&=&-\frac{\rho_{0}}{9}\,p_{0}(\tau_{0},x_{0})\,\frac{\partial p_{0}(\tau_{0},x_{0})}{\partial x_{0}}+\mathcal{O}(\rho^{2}_{0})\,,\\
\hat{A}&=&-\frac{1}{6}\,p^{2}_{0}(\tau_{0},x_{0})+\mathcal{O}(\rho^{2}_{0})\,,
\eea
together with the equations of motion derived from the Euler-Lagrange equations
\bea
&&\delta_{\rho_{2}} S_{\Sigma}-\frac{d}{dx}(\delta_{\dot{\rho}_{2}}S_{\Sigma})=0\,,\\
&&\frac{d}{dx}(\delta_{\dot{\tau}_{2}}S_{\Sigma})=0\,,\\
&&\frac{d}{dx}(\delta_{\dot{x}_{2}}S_{\Sigma})=0\,,
\eea
we find the 
following geodesic equations around the boundary surface\footnote{We assume the branch in the solutions that satisfies $x_{m}>x_{0}$.},
\bea\label{cof}
&&\ddot{\rho}_{2}+\ddot{\tau}_{2}=\frac{1}{24\sqrt{2}}\frac{\rho^{9/2}_{m}p_{0}^{2}(\tau_{\ast},x_{\ast})}{\tilde{y}^{5/4}_{0}}\,,
\\
&&4\ddot{x}_{2}-\frac{2\sqrt{2}}{\rho^{3/4}_{m}}\tilde{x}^{3/4}_{0}\ddot{\rho}_{2}
-\frac{3\sqrt{2}}{\rho^{3/4}_{m}}\frac{\dot{\rho}_{2}}{\tilde{x}^{1/4}_{0}}
+\frac{3}{4\sqrt{2}}\frac{\rho_{2}}{\tilde{x}^{5/4}_{0}}=
\frac{5\rho^{3/2}_{m}}{12}\frac{p^{2}_{0}(\tau_{\ast},x_{\ast})}{\tilde{x}^{1/2}_{0}}\,.
\eea
Therefore in this case, we recover the degenerate equations of motion for $\ddot{\rho}_{2}$ and $\ddot{\tau}_{2}$ and
an extra  equation of motion for $\ddot{x}_{2}$. The same coefficients that have been obtained in the limit of
long-late times, that is $p_{0}\rightarrow const.$, should be valid in this case and will allow us to determine
$\ddot{x}_{2}$. An easy power  counting shows that $x_{2}\sim \tilde{x}^{3/2}_{0}$. If we insert the value of
$\rho_{2}$ given at the late-time approximation when the system has reached thermalization \cite{Alex2014}, we find $\ddot{x}_{2}=0$.
 In either case, this means that
the contribution from $x_{2}$ in \Eq(EE-caseII-infty) vanishes. Thus, the contribution from $\delta S_{\Sigma(0)}$ reads
\be
\delta S_{\Sigma(0)}=\frac{5K^{2}}{36}p^{2}_{0}(\tau_{\ast},x_{\ast})\,.
\ee

The contribution form the lower bound of the first term in \Eq(EE-caseII-infty) vanishes as the reader can easily check from
the zeroth-order equations of motion. Going back to \Eq(EE-caseII-S_sigma) and making
 a change of variable from $x_{0}$ to $\rho_{0}$ using \Eq(Alex-EQ) and \Eq(EE-motion) and renaming $y_{0}$ for $x_{0}$,
 we obtain
\bea
S_{\Sigma(2)}\!=\!K^{2}\!\!\int_{0}^{\rho_{m}}
\hspace{-.5cm}\frac{\rho^{3}_{0}d\rho_{0}}{\rho^{3}_{m}\sqrt{\left(1-\rho^{4}_{0}\right)(\rho^{6}_{m}-\rho^{6}_{0})}}
\left[\frac{2}{\rho_{0}}\left(\frac{\rho^{6}_{m}-\rho^{6}_{0}}{1-\rho^{4}_{0}}\right)^{1/2}\!\!\!\hat{\Xi}_{f}
-\frac{\rho^{6}_{m}-\rho^{6}_{0}}{\rho^{4}_{0}\left(1-\rho^{4}_{0}\right)}\hat{A}+\frac{4\rho^{6}_{m}}{\rho^{6}_{0}}\hat{\Sigma}_{b}
+2\hat{\Sigma}_{d}\right]\,.\nonumber\\
\eea
As it is clear from the above expression, it suffers from  infrared divergences. To separate them from the finite part,
we use the asymptotic expansion around the boundary using \Eq(B_A)-\Eq(B_Xi) in the appendix, \footnote{
We have neglected the time derivatives over $p_{0}$.} i.e.
\bea
A&=&-\frac{p^{2}_{0}}{6}+\rho^{2}_{0}a_{2}+\mathcal{O}(\rho^{2}_{0}\ln\rho_{0})\,,\\
\Sigma_{d}&=&-\rho^{2}_{0}\frac{p^{2}_{0}}{12}+\rho^{4}_{0}d_{4}+\mathcal{O}(\rho^{4}_{0}\ln\rho_{0})\,,\\
\Sigma_{b}&=&-\rho^{2}_{0}\frac{p^{2}_{0}}{12}+\rho^{4}_{0}b_{4}+\mathcal{O}(\rho^{4}_{0}\ln\rho_{0})\,,\\
\Xi&=&-\rho_{0}\frac{p_{0}\partial_{x}p_{0}}{9}+\rho^{2}_{0}f_{2}+\mathcal{O}(\rho^{2}_{0}\ln\rho_{0})\,,
\eea 
to find the finite contribution,
\bea\label{Sfin}
S_{\Sigma(2)}^{fin}&=&K^{2}\int_{0}^{\rho_{m}}\frac{d\rho_{0}}{18\rho^{3}_{m}\left(-1+\rho^{4}_{0}\right)^{2}
\left(\rho^{6}_{0}-\rho^{6}_{m}\right)}\times
\nonumber\\&&\hspace{-1cm}
\Bigg[-36f_{2}\rho_{0}^{4}\left(-1+\rho_{0}^{4}\right)\left(\rho_{0}^{6}-\rho^{6}_{m}\right)
-4p^{'}_{0}p_{0}\left(1-\rho_{0}^{4}\right)^{3/2}\left(\rho^{6}_{m}-\rho^{6}_{0}\right)^{3/2}
\nonumber\\&&\hspace{-1cm}
-3\rho_{0}\sqrt{\left(-1+\rho_{0}^{4}\right)\left(\rho^{6}_{0}-\rho^{6}_{m}\right)}
\Big[p^{2}_{0}\rho^{4}_{0}\left(-2+\rho^{4}_{0}\right)+6a_{2}\left(\rho^{6}_{0}-\rho^{6}_{m}\right)
\nonumber\\&&\hspace{1cm}
-12\left(-1+\rho^{4}_{0}\right)\left(d_{4}\rho^{6}_{0}+2b_{4}\rho^{6}_{m}\right)\Big]
\Bigg]\,,
\eea
and in the above, we are using the compact notation for $p^{'}_{0}\equiv \partial_{\rho_{0}}p_{0}$ based on the chain rule.  Since infinitesimal change in $x_{0}$, also varies $\tau_{0}$, the
derivative acts on both arguments of $p_{0}(\tau_{0},x_{0})$.  

Similarly, the divergent part reads
\be
S_{\Sigma(2)}^{div}=-K^{2}\int_{\epsilon}^{\rho_{m}}d\rho_{0}
\frac{p^{2}_{0}\left(-1+2\rho^{4}_{0}\right)\rho^{3}_{m}}{6\rho_{0}\left(-1+\rho^{4}_{0}\right)^{3/2}\left(\rho^{6}_{0}-\rho^{6}_{m}\right)^{1/2}}\,,
\ee
with  $\epsilon$ to regulate the integral. To regularize  the divergent term the following  counter term is added
\be
S_{\Sigma(2)}^{counter}=\frac{K^{2}}{6}\,p^{2}_{0}(\tau_{\ast},x_{\ast})\int_{\epsilon}^{\rho_{m}}
\frac{d\rho_{0}}{\rho_{0}}\,,
\ee
together with a  finite contribution to make  the regularization  scheme independent, 
\be\label{Scor}
S_{cor}=-\frac{1}{6}K^{2}p^{2}_{0}(\tau_{\ast},x_{\ast})\log\rho_{m}\,.
\ee

Preparing \Eq(Sfin)-\Eq(Scor) for numerics with the usual change of variable of $\rho_{0}=\rho_{m}\left(1-q^{2}\right)$, 
the final expression including all terms,
\be
S_{\Sigma(2)}=S^{fin}_{\Sigma(2)}+S^{div}_{\Sigma(2)}+S_{counter}+S_{cor}+\delta S_{\Sigma(0)}\,,
\ee
will take the form
\bea
S_{\Sigma(2)}&=&K^{2}\int_{0}^{1}
\frac{qdq}{9\rho^{8}_{m}\left(1-\rho^{4}_{m}\left(-1+q^{2}\right)^{4}\right)^{2}
\left(-1+\left(1-q^{2}\right)^{6}\right)}\times
\nonumber\\&&
\Bigg[
-36\rho^{10}_{m}f_{2}\left(-1+\rho^{4}_{m}\left(-1+q^{2}\right)^{4}\right)\left(-1+(1-q^{2})^{6}\right)
\left(1-q^{2}\right)^{4}
\nonumber\\&&
-4p^{'}_{0}p_{0}\rho^{9}_{m}\left(1-\rho^{4}_{m}(1-q^{2})^{4}\right)^{3/2}
\left(1-(1-q^{2})^{6}\right)^{3/2}
\nonumber\\&&
-3\rho^{4}_{m}\left(1-q^{2}\right)\sqrt{\left(-1+(1-q^{2})^{6}\right)
\left(-1+(1-q^{2})^{4}\rho^{4}_{m}\right)}
\times
\nonumber\\&&
\Big[p^{2}_{0}\rho^{4}_{m}\left(1-q^{2}\right)^{4}\left(-2+\left(1-q^{2}\right)^{4}\rho^{4}_{m}\right)
+6a_{2}\rho^{6}_{m}(-1+(1-q^{2})^{6})
\nonumber\\&&
-12\rho^{6}_{m}(-1+\rho^{4}_{m}(1-q^{2})^{4})(d_{4}(1-q^{2})^{6}+2b_{4})\Big]
\Bigg]
\nonumber\\&&
-K^{2}\int_{0}^{1}qdq
\frac{p^{2}_{0} \left(-1+2\left(1-q^{2}\right)^{4}\rho^{4}_{m}\right)}
{3\left(1-q^{2}\right)\left(-1+\rho^{4}_{m}\left(1-q^{2}\right)^{4}\right)^{3/2}
\left(-1+(1-q^{2})^{6}\right)^{1/2}}
\nonumber\\&&
+\frac{K^{2}}{3}p^{2}_{0}(t_{\ast},x_{\ast})\left(\int_{0}^{1}\frac{qdq }{1-q^{2}}
-\half\log\rho_{m}
+\frac{5}{12}\right)\,,
\eea
with $p^{'}_{0}\equiv \partial_{\rho_{0}}p_{0}$.

\begin{figure}[!ht]
\begin{subfigure}{.5\textwidth}
{\includegraphics[width=7.0cm]{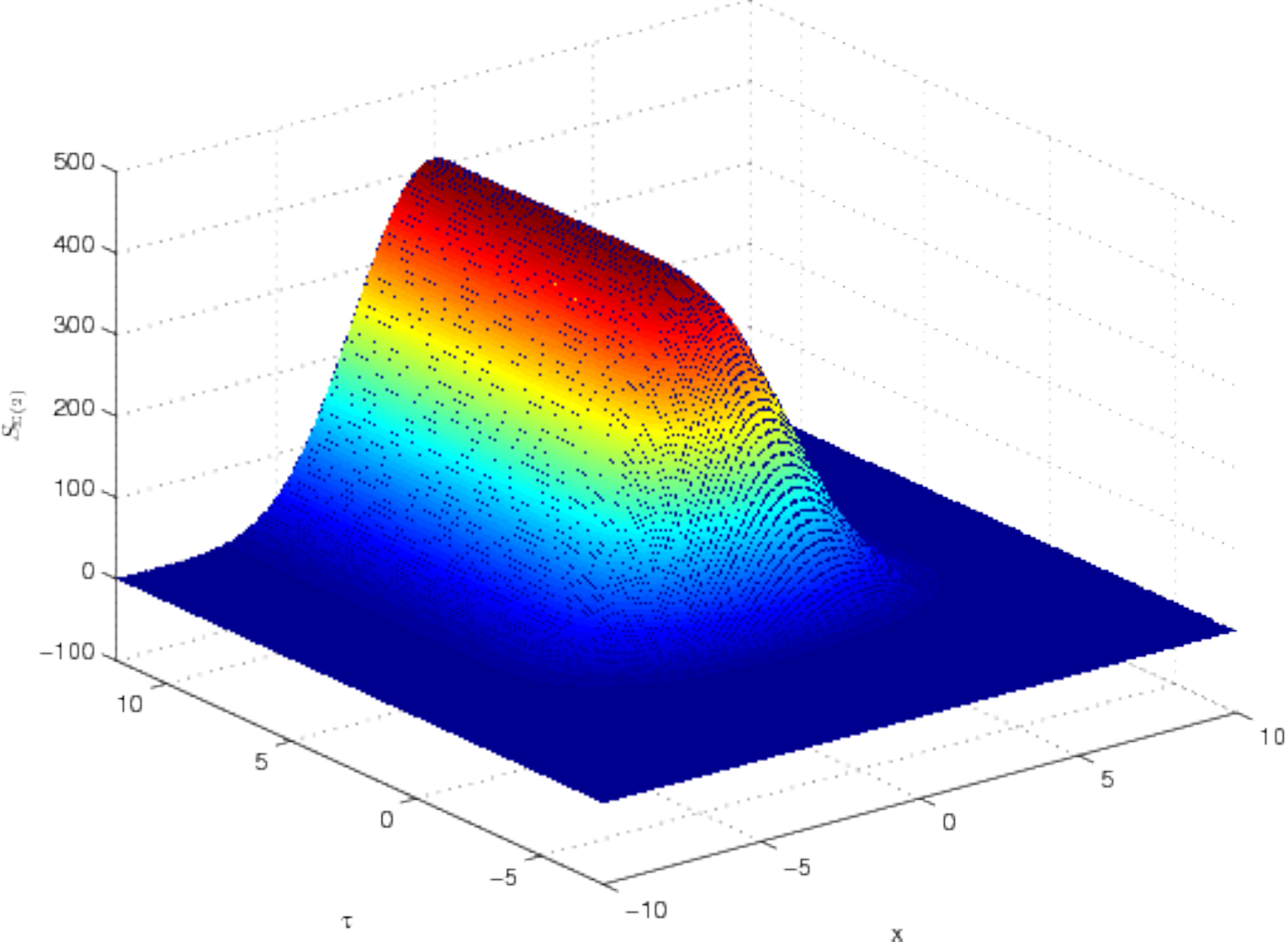}}
  \caption{}
  \label{fig:EEB2cc1}
\end{subfigure}%
\begin{subfigure}{.5\textwidth}
{\includegraphics[width=7.0cm]{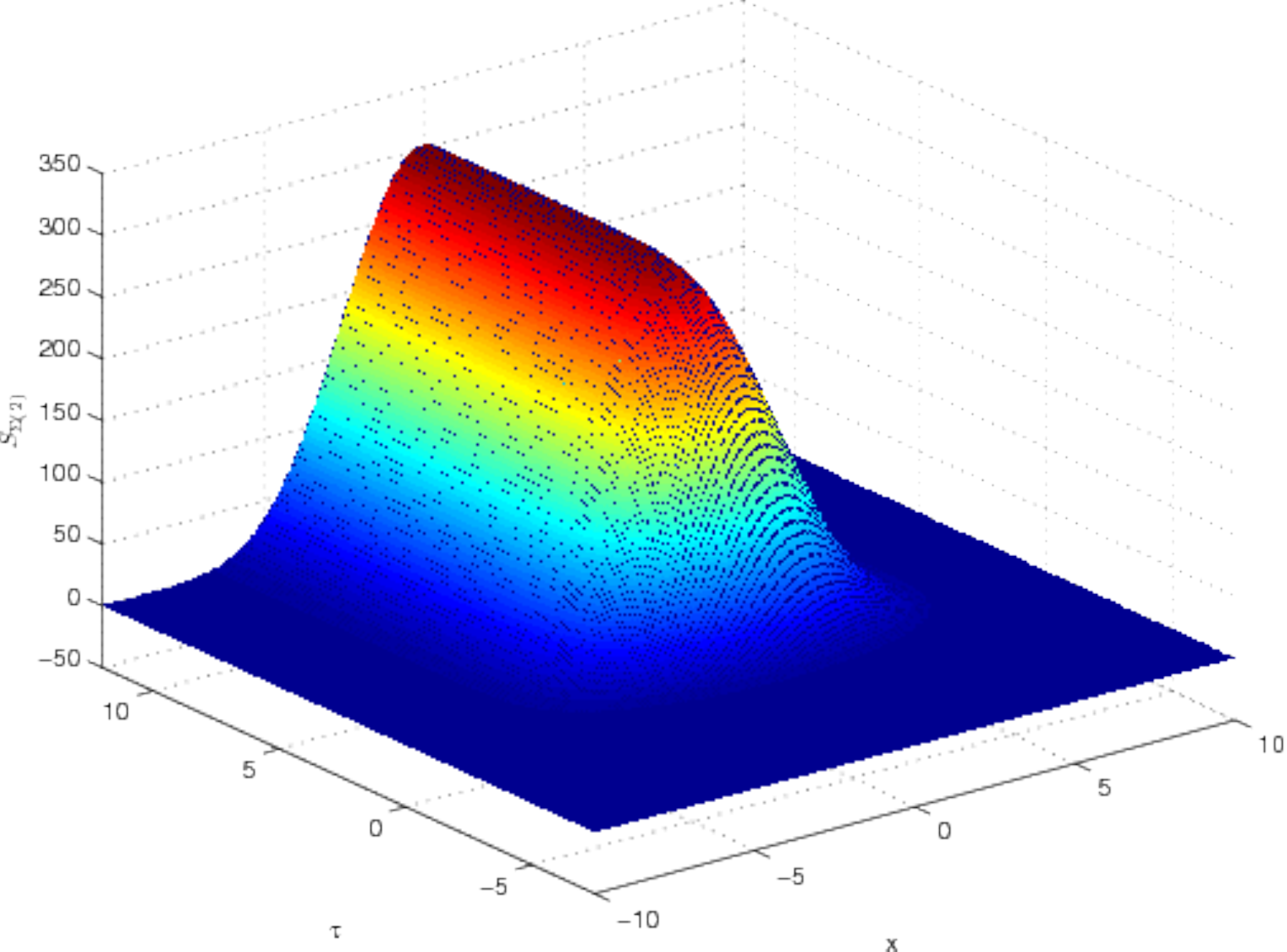}}
  \caption{}
  \label{fig:EEB2cc5}
\end{subfigure}

\begin{subfigure}{.5\textwidth}
{\includegraphics[width=7.0cm]{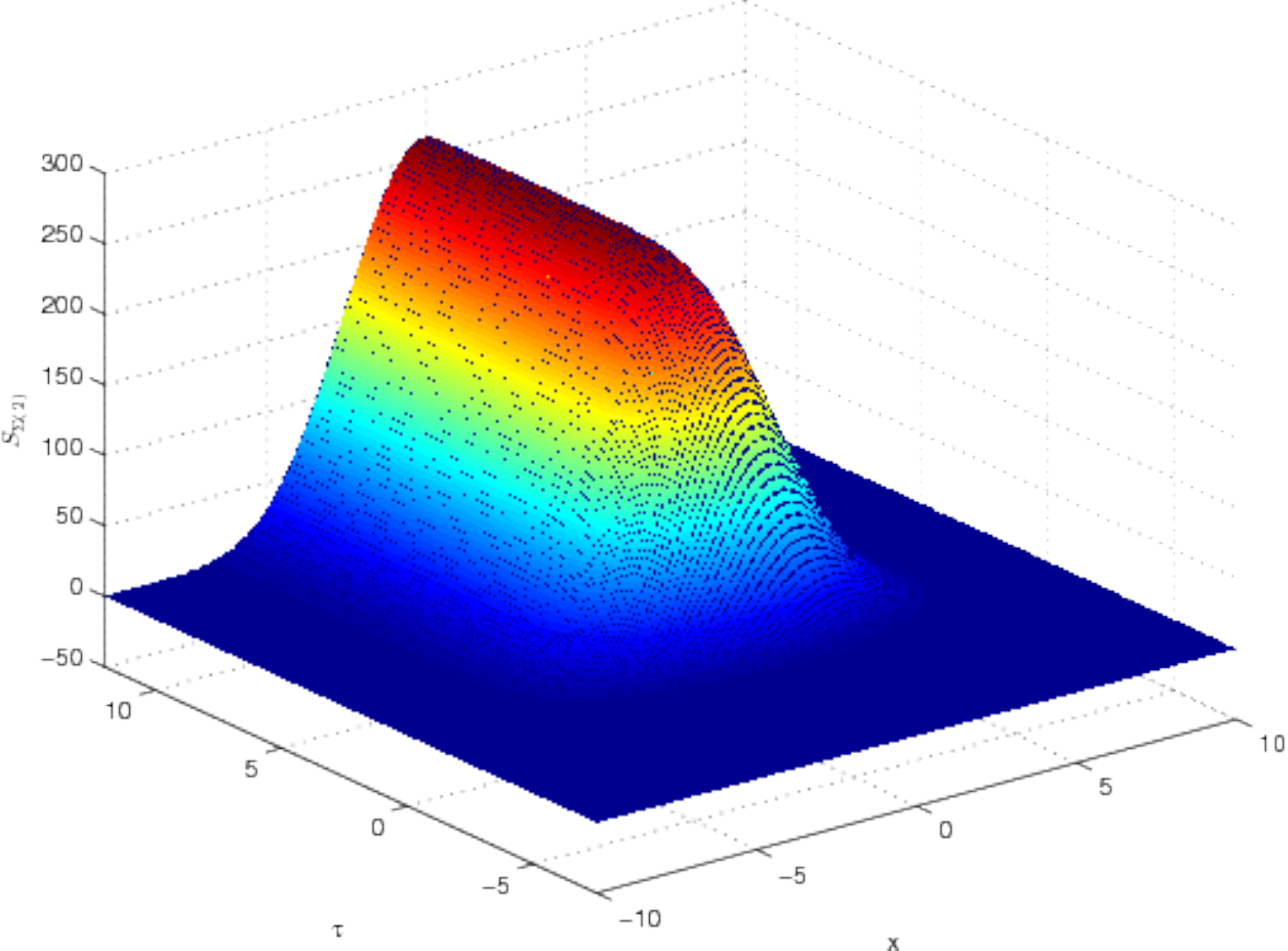}}
  \caption{}
  \label{fig:EEB2cc}
\end{subfigure}%
\begin{subfigure}{.5\textwidth}
{\includegraphics[width=7.0cm]{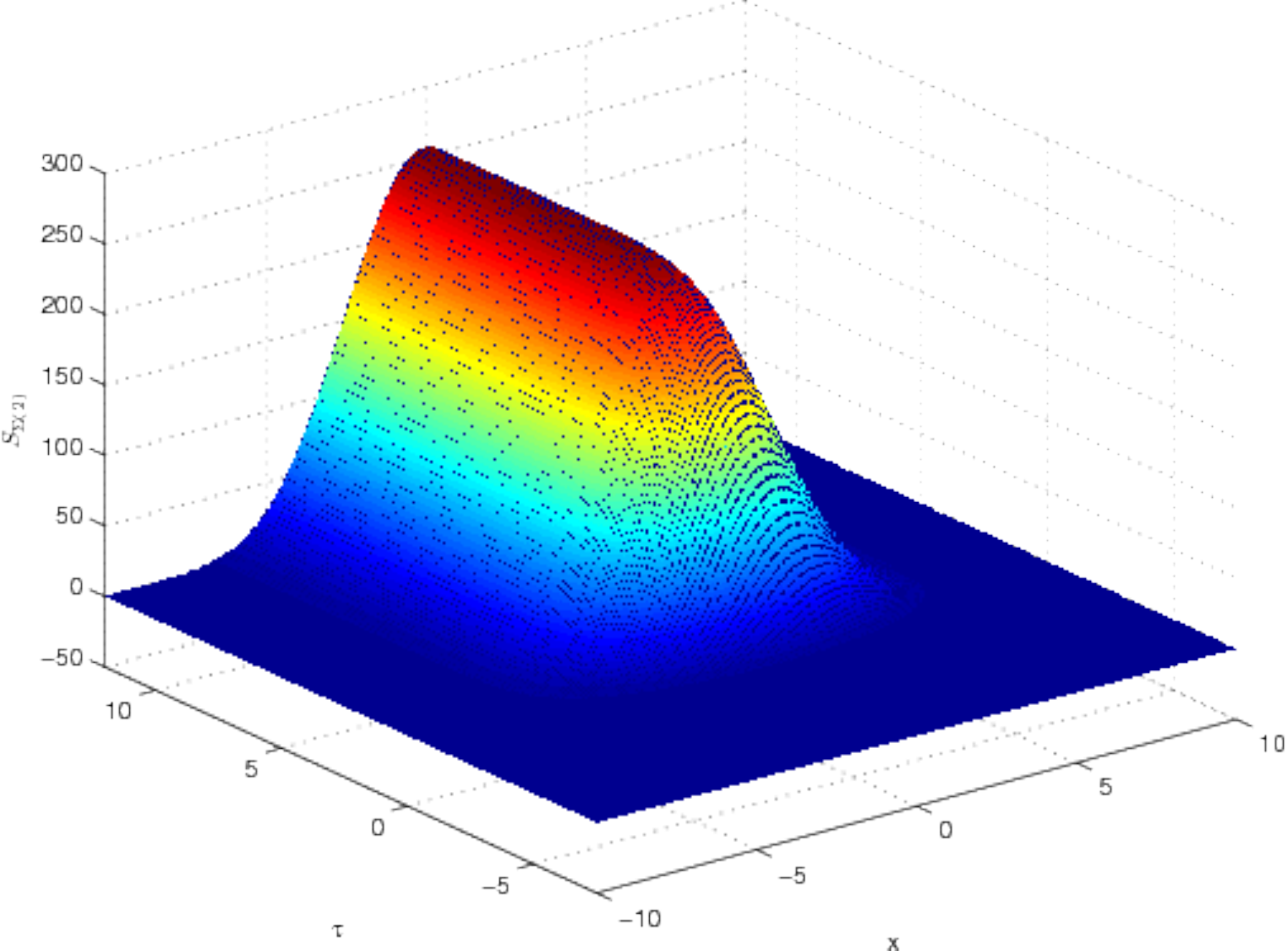}}
  \caption{}
  \label{fig:EEB2cc999}
\end{subfigure}
\caption{In the above figures, time evolutions  of $S_{\Sigma(2)}$ for Case II, are depicted. From (a)-(d), we increase the value of $\rho_{m}$ to reach the maximum thermalization. Fixed tuning parameters such as $\alpha=1$ and $\sigma=\sqrt{L_{x}}$ together with $N_{x}=N_{\rho}=20$ Chebyshev points have been used. The number of time steps for the fourth order Runge-Kutta has been $7810$.}
\label{fig:51}
\end{figure}

\begin{figure}[!ht]
\begin{subfigure}{.5\textwidth}
{\includegraphics[width=7.0cm]{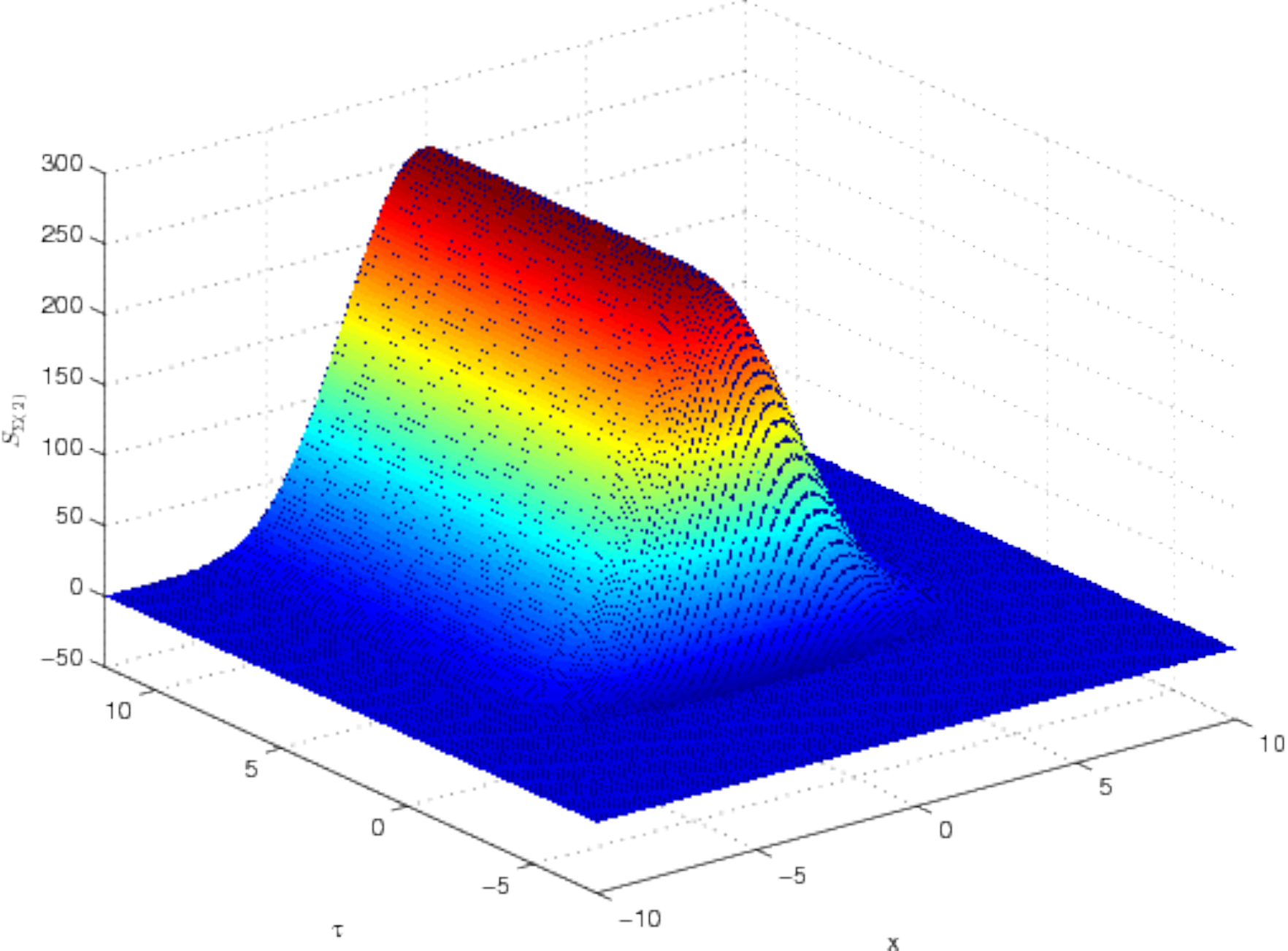}}
  \caption{}
  \label{fig:EEB3a2999}
\end{subfigure}
\begin{subfigure}{.5\textwidth}
{\includegraphics[width=7.0cm]{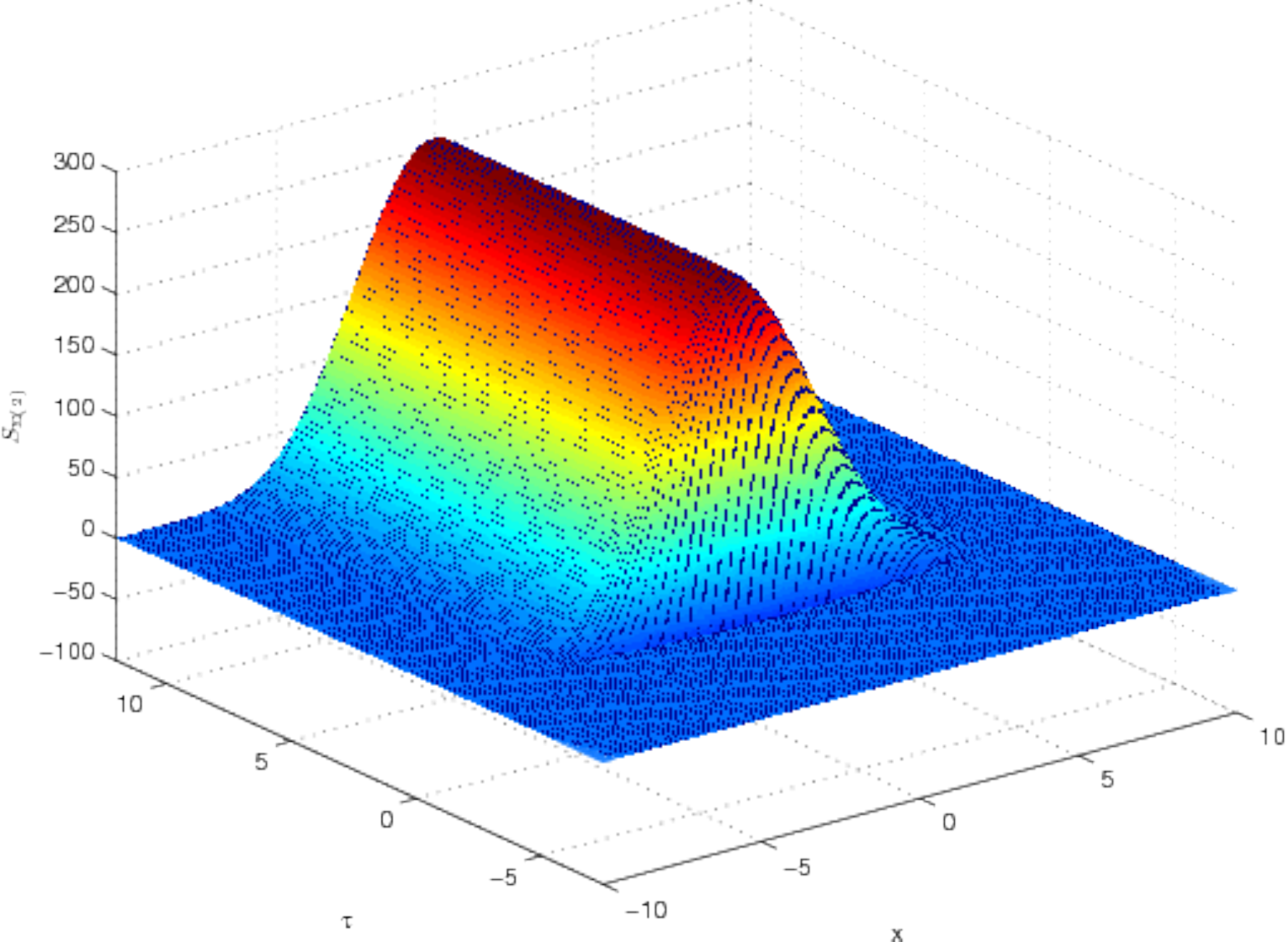}}
  \caption{}
  \label{fig:EEB3b2999}
\end{subfigure}

\begin{subfigure}{.5\textwidth}
{\includegraphics[width=7.0cm]{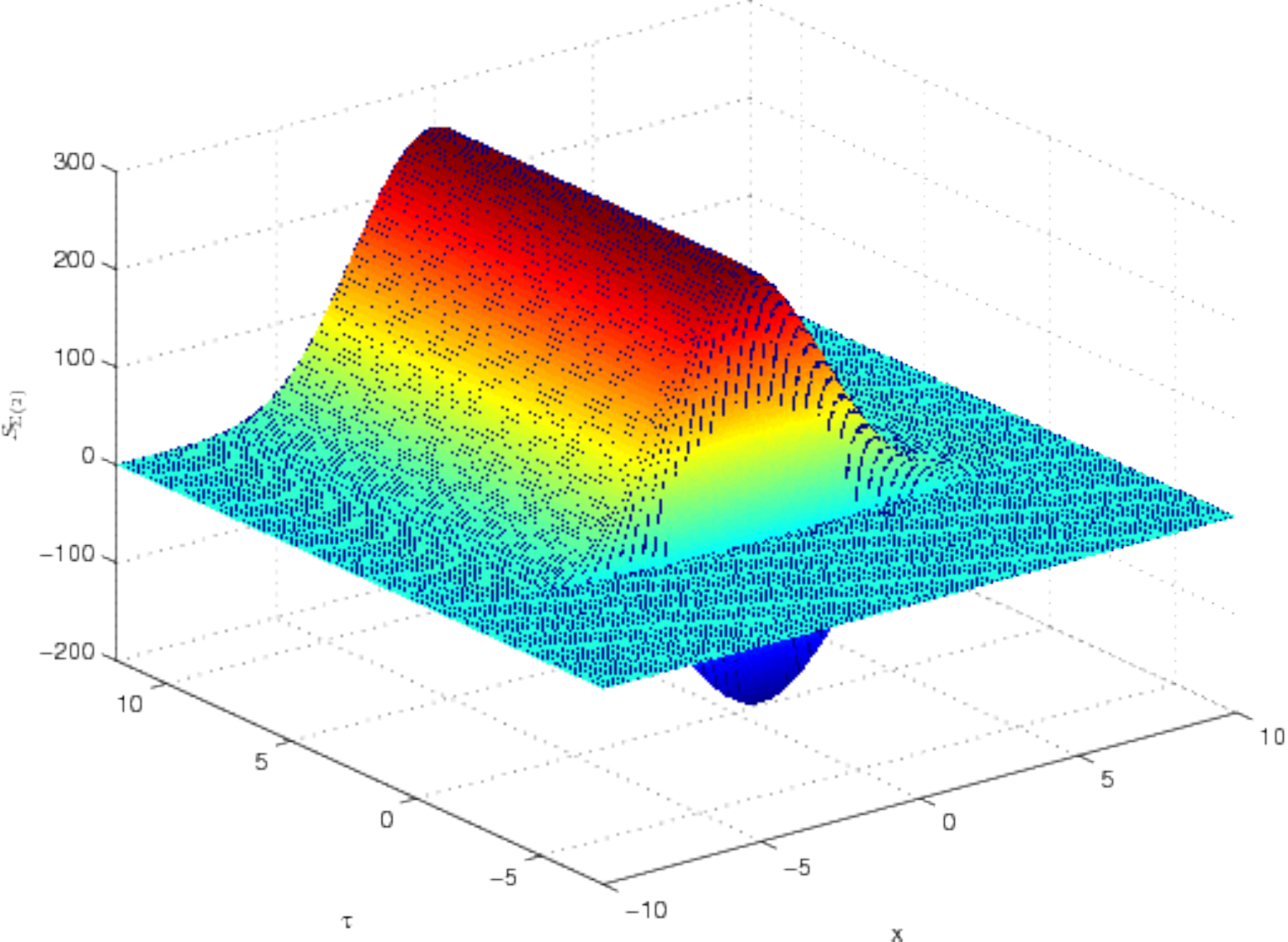}}
  \caption{}
  \label{fig:EEB3c2999}
\end{subfigure}
\begin{subfigure}{.5\textwidth}
{\includegraphics[width=7.0cm]{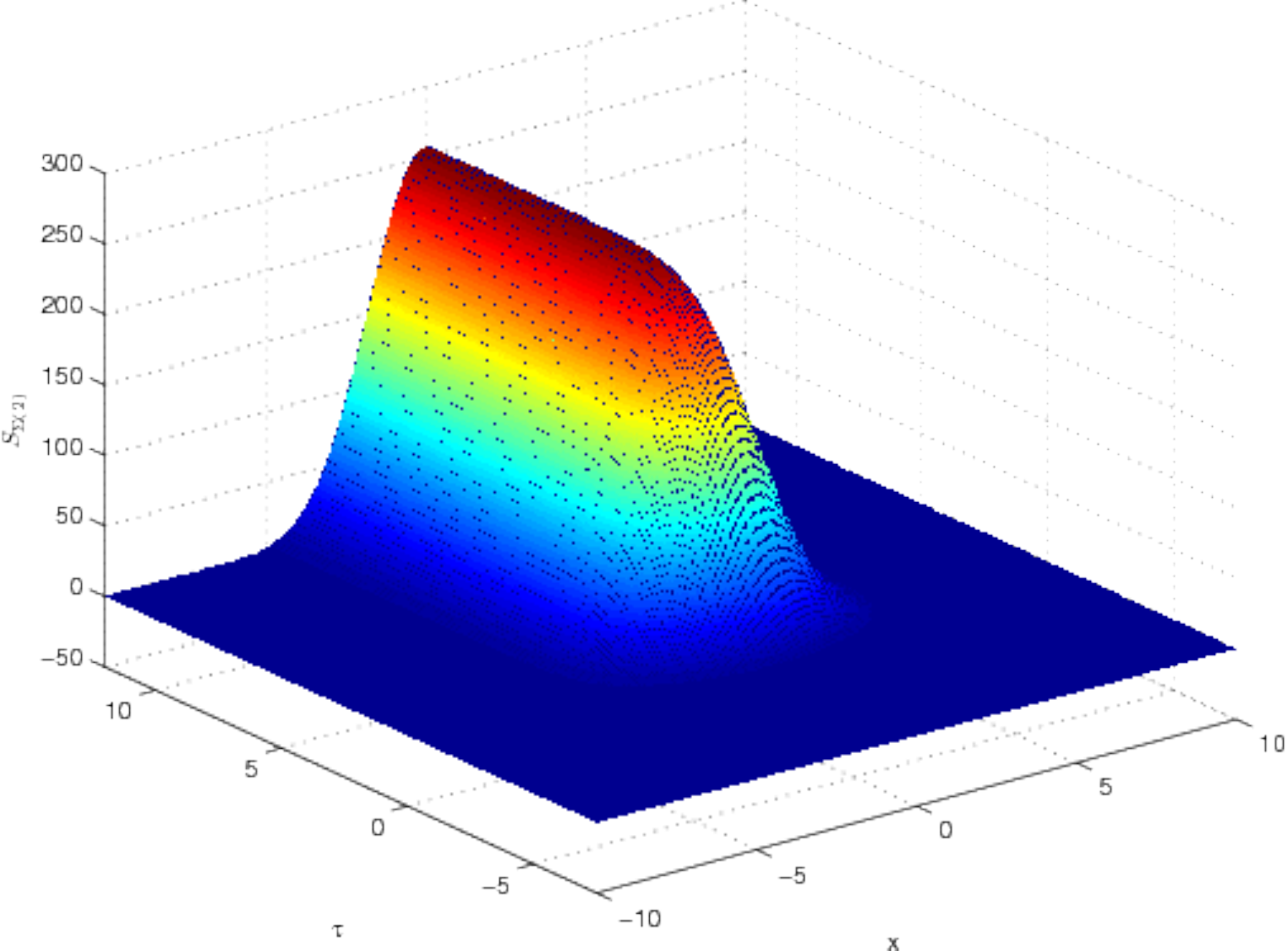}}
  \caption{}
  \label{fig:EEB4a2999}
\end{subfigure}

\begin{subfigure}{.5\textwidth}
{\includegraphics[width=7.0cm]{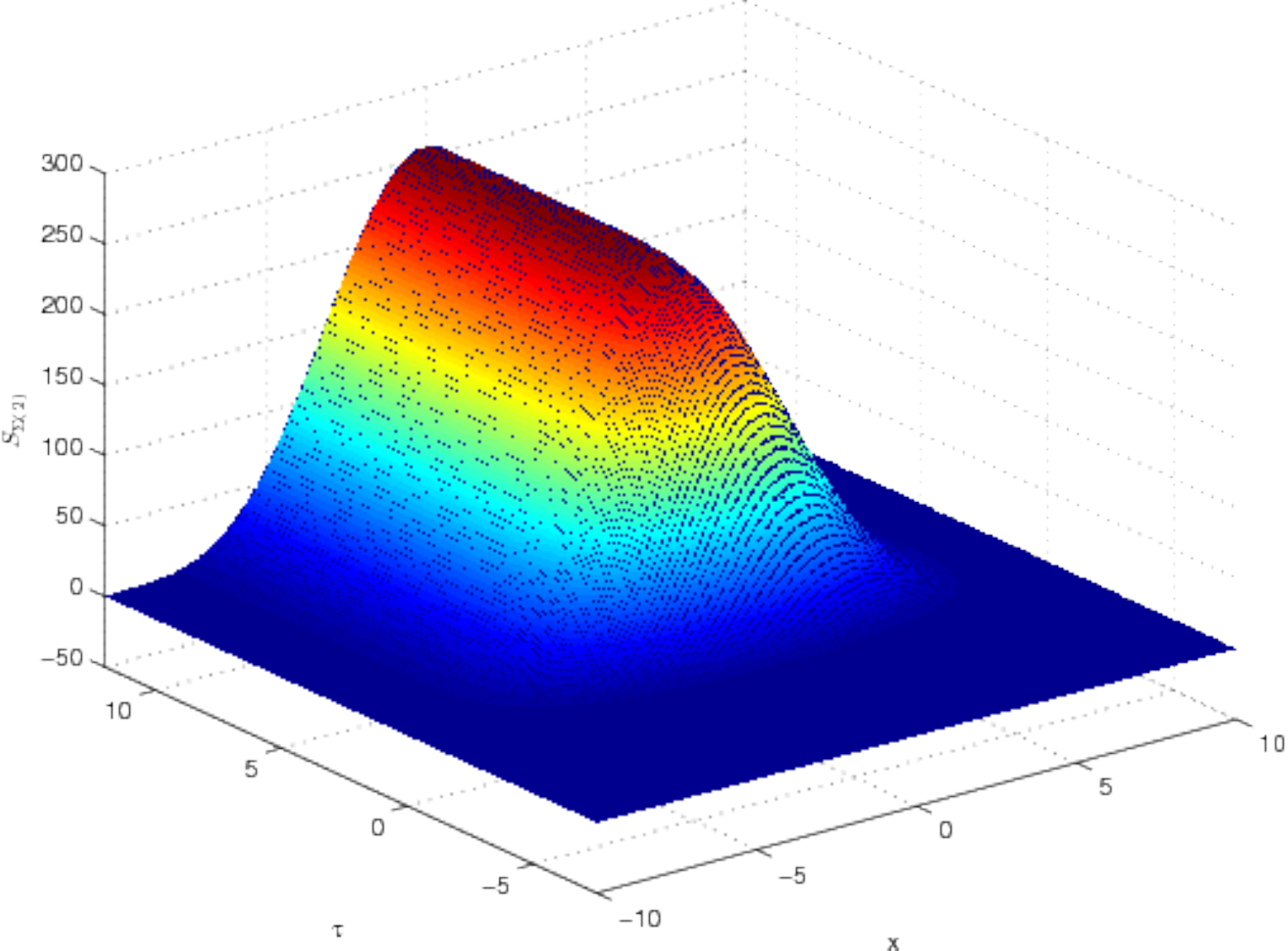}}
  \caption{}
  \label{fig:EEB4b2999}
\end{subfigure}
\begin{subfigure}{.5\textwidth}
{\includegraphics[width=7.0cm]{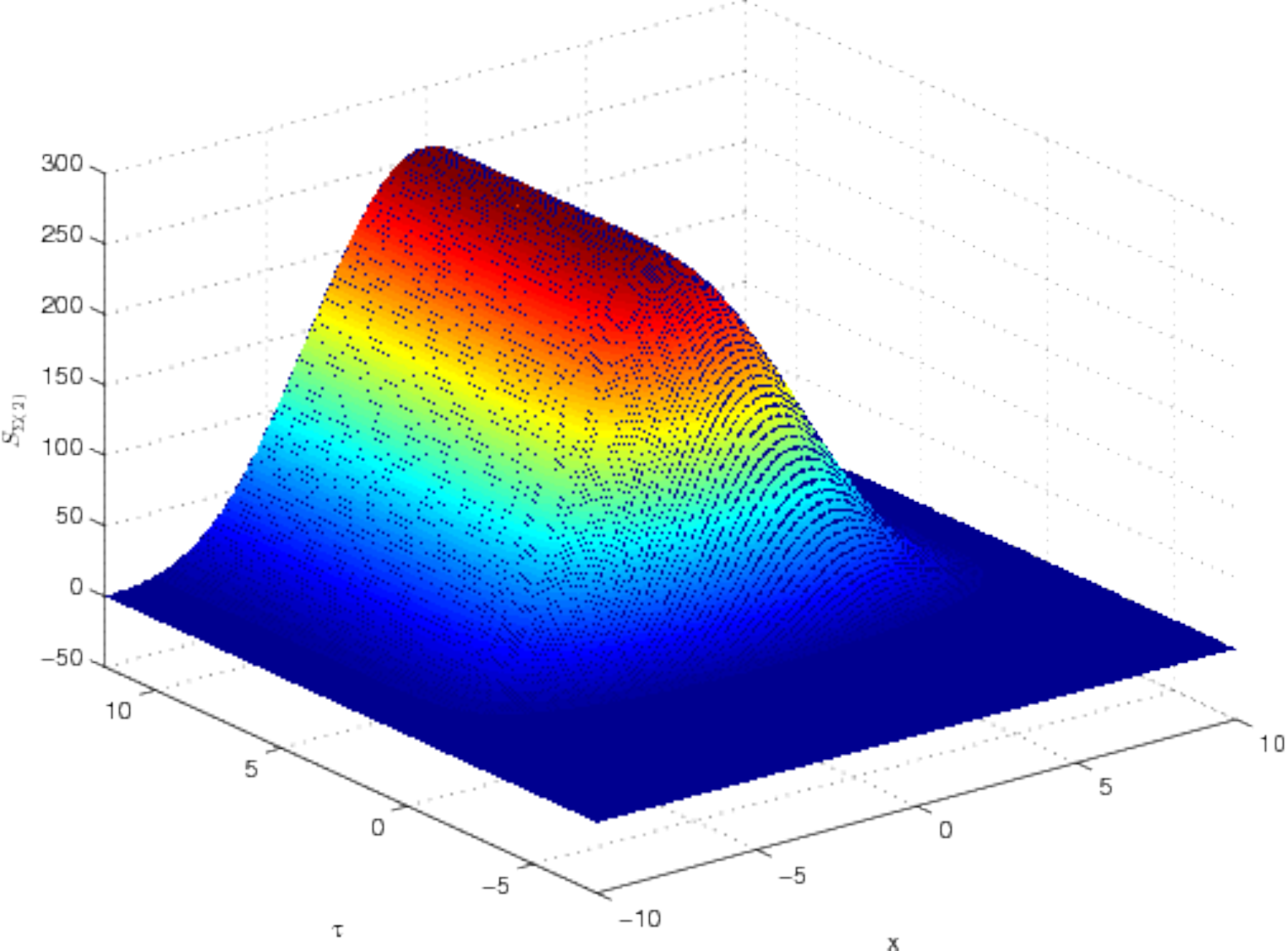}}
  \caption{}
  \label{fig:EEB4c2999}
\end{subfigure}
\caption{Plots of the time evolution  of the entanglement entropy in Case II.
In (a)-(c), the value of $\alpha$ has been reduced while in (d)-(f), we are increasing the tuning parameter $\sigma$. The numerical setup is identical to the last figure.}
\label{fig:63}
\end{figure}

Figures \ref{fig:EEB2cc1}-\ref{fig:EEB4c2999} represent $S_{\Sigma(2)}$, the perturbation to the total EE at $\mathcal{O}(l^{2})$, in the $x-\tau$ plane. They are parts of our main results as they have not been reported in any from to the best of our knowledge and perhaps 
represent the most insightful aspects of EE.

The first thing to notice is the way  profiles for EE change when we vary $\rho_{m}$. This is apparent by comparing Figures \ref{fig:EEA2cc1}-\ref{fig:EEA2cc999} in the last section against  Figures \ref{fig:EEB2cc1}-\ref{fig:EEB2cc999}. A small dip appears at $\tau\sim 0$ in Figures \ref{fig:EEB2cc1}-\ref{fig:EEB2cc999} that its magnitude grows as we reduce the value of $\alpha$. 
As  it is shown in \ref{fig:EEB2cc1}-\ref{fig:EEB2cc999}, by gradually increasing  values of $\rho_{m}$, the maximum of the late-time saturated value for EE reduces. In Figures \ref{fig:EEB3a2999}-\ref{fig:EEB3c2999}, we vary the tuning parameter $\alpha$ from $\frac{1}{2}$ to $\frac{1}{8}$. This causes the dip to get a pinching shape along the $\tau$ direction. Similarly, we can change $\sigma$ which increases the size of the dip side ways along the $x$ axis. These are shown in Figures \ref{fig:EEB4a2999}-\ref{fig:EEB4c2999}.

\clearpage

\clearpage

Comparing these figures with those given in the last section makes it easy to interpret the physics behind EE.  In the last section, we found an approximate length for the correlation length. This will allow us to concentrate on the pair of entangled quasiparticles from an arbitrary Cauchy surface within this length. Our system has a  strip geometry and in Case I the boundary is at $[-y_{m},y_{m}]$ and it is extended to infinity in the $x$ direction whereas in Case II, boundaries are at $[-x_{m},x_{m}]$ and is extended to infinity along the $y$ axis. The direction of inhomogeneity is along the $x$ axis in both cases. The EE originates from entangled quasiparticles that have the chance to reach the boundaries of the system. In Case I, the quench produces the quasiparticles out of the vacuum and Figures 
\ref{fig:EEA3a2999}-\ref{fig:EEA4c2999}, show that pairs that are created at $x=0$ have the highest chance to reach the boundaries at $[-y_{m},y_{m}]$ assuming they dispatch in opposite directions. Equivalently, as much as they are off the symmetry axis their chances are lower and so is their contribution to the EE. Note that what we are plotting are the perturbations of EE at $\mathcal{O}(l^{2})$. This situation can be compared with Case II, where quasiparticles that are produced at $x=0$ and want to reach the boundaries at $[-x_{m},x_{m}]$ have to overcome the Gaussian disturbance. This can be put in simple words using Cardy's suggestion \cite{0808.0116} to define an entanglement entropy current. In Case I, the current induced by the quench is along the axis of the produced quasiparticles. In contrast, the latter current is perpendicular
to the path of the quasiparticle pairs in Case II and explains the presence of the dip in Figures \ref{fig:EEB3a2999}-\ref{fig:EEB4c2999}.

\section{Conclusions}

Throughout this article, we studied various observables such as apparent horizon, two-point Wightman functions and entanglement entropy (EE) to study the physics of thermalization. Our method to derive the system far from equilibrium was the generalization of the setup described by Butchel et al. in \cite{Alex2014} for quenches and we made it inhomogeneous. We then solved the corresponding coupled equations of motion using spectral method outlined by Chesler and Yaffe \cite{Chesler:2013lia}.

Study of the apparent  horizon as a  local observable, showed the presence of excitations out of the vacuum of $\mathcal{N}=4$ SYM,
created by the mass gap that our quench produces. Different behaviors of these excitations or ``quasiparticles'' were observed by varying the
quench tuning parameters such as the width of the Gaussian profile, $\sigma$ or the time scale of the quench $\alpha$. It was shown that  profiles
of the apparent horizon for values of $\alpha\sim 1$ were very similar to profiles of the quench but for $\alpha\sim 0$, a universal behavior was emerging. Increasing $\sigma$ showed that the mass gap excitations would fill up the available space. 

Having an extra nontrivial spatial direction on the boundary allowed us to consider different scenarios that we depicted in Figure \ref{fig:Two-point}. In both Case I and II, the correction to the correlation function at $\mathcal{O}(l^{2})$, where $l$ is the order of the backreaction, was considered. Corrections to the Wightman function in Case I were symmetric along $x=0$ axis unlike Case II. The latter had a contribution from one of the components of the metric that was an odd function in $x$. In both cases, corrections undergo a phase transition that is seen by  change of sign. 
Since the correlator measures the interference of an infinite number of momentum modes \cite{0808.0116}, by speculating about our figures, we could parametrize the path of these modes departing from an arbitrary initial time until their interference by $\tau=x/v_{max}$. Our plots were suggesting that  our quenches belong to the class of $v_{max}\sim 1$. The study of the correlation functions in both Case I and II also revealed that the physics of thermalization is not diffusive (or at least it is very negligible) as far as we could compare the amplitudes in the two sets of figures.

Similar to the Wightman correlation functions, we used the extra nontrivial spatial direction to study EE in various strip boundary setups. These cases were the extensions of the configurations mentioned in Figure \ref{fig:Two-point}. As we increased the depth in which the minimal surface could probe in the bulk, the EE's evolution followed the profile of the source on the boundary more closely.  In Case I, the fingerprint of the quasiparticles reaching their ``horizon'' could be seen as a slight wiggle on the surface of the EE in the $x-\tau$ plane. The setup in Case II gives a completely different profile for the EE. This latter configuration was an interesting part of our paper due to its novelty and a description in terms of the entanglement current of Cardy et al. \cite{0808.0116} and could illuminate the underlying physics.
We think this result requires further investigation in different setups such as entangling hemisphere.

As we mentioned above, our study confirmed that the underlying physics of thermalization is not of a diffusive nature at strong couplings, although defining quantities such as currents seem to be inevitable. In fact, physics of thermalization after a quench in many ways is very similar to the physics of far-from-equilibrium isotropization. Consider the two priory different problems, where the first one explains the equilibration of $\mathcal{N}=4$ SYM in the following holographic setup \cite{Chesler:2013lia} 
\be\label{iso}
ds^{2}=2d\tau dr-A(\tau,r)\,d\tau^{2}+\Sigma^{2}(\tau,r)e^{-2B(\tau,r)}\,dx_{L}+\Sigma^{2}(\tau,r)e^{B(\tau,r)}\,d\mathbf{x}_{T}\,,
\ee
with $r\equiv 1/\rho$ (inverse) radius of the bulk, $A(\tau,r)$, $\Sigma(\tau,r)$ are the warp factors and $B(\tau,r)$ is a function that parametrizes the isotropization with respect to the longitudinal and transverse planes. And the second one is our quench problem with a more simplified background considered in \cite{Alex2014},
\be\label{shortmetric}
ds^{2}=2d\tau dr-A(\tau,r)\,d\tau^{2}+\Sigma(\tau,r)^{2}\,d\mathbf{x}^{2}\,.
\ee
Upon insertion of \Eq(iso) and \Eq(shortmetric) in Einstein equations, the equations of motion will take a specific form \cite{Chesler:2013lia, Alex2014}. We list those of the isotropization problem on the left-hand side and those of the quench on the right-hand side,
\bea
\Sigma \dot{\Sigma}^{'}+2\Sigma^{'}\dot{\Sigma}-2\Sigma^{2}&=&0\,,
\quad\quad
\Sigma \dot{\Sigma}^{'}+2\Sigma^{'}\dot{\Sigma}-2\Sigma^{2}+\frac{1}{12}m^{2}\phi^{2}\Sigma^{2}=0\,,
\\
\Sigma\dot{B}^{'}+\frac{3}{2}\left(\Sigma^{'}\dot{B}+B^{'}\dot{\Sigma}\right)&=&0\,,
\quad \quad
\Sigma\dot{\phi}^{'}+\frac{3}{2}\left(\Sigma^{'}\dot{\phi}+\phi^{'}\dot{\Sigma}\right)-\half\Sigma m^{2}\phi=0\,,
\\
A^{''}+3B^{'}\dot{B}-12\frac{\Sigma^{'}\dot{\Sigma}}{\Sigma^{2}}+4&=&0\,,
\quad\quad 
A^{''}+\phi^{'}\dot{\phi}-12\frac{\Sigma^{'}\dot{\Sigma}}{\Sigma^{2}}+4-\frac{1}{6}m^{2}\phi^{2}=0\,,
\\
\ddot{\Sigma}+\half\left(\dot{B}^{2}\Sigma-A^{'}\dot{\Sigma}\right)&=&0\,,
\quad\quad 
\ddot{\Sigma}+\half\left(\frac{\dot{\phi}^{2}\Sigma}{3}-A^{'}\dot{\Sigma}\right)=0\,,
\\
\Sigma^{''}+\half \left(B^{'}\right)^{2}\Sigma&=&0\,,
\quad\quad 
\Sigma^{''}+\frac{1}{6} \left(\phi^{'}\right)^{2}\Sigma=0\,.
\eea
In the above, we used of $h^{'}\equiv \partial_{r}h$ and $\dot{h}\equiv\partial_{\tau}h+\half\partial_{r}h$.
To make a connection between the two lists of equations on the right and left-hand sides, we realize that by choosing a symmetry factor  $B\equiv\frac{\phi}{\sqrt{3}}$, apart from extra mass terms\footnote{Although the mass terms played a key role in our quenches, we could argue that we start our simulation from a rather nontrivial initial data and then study the evolution without turning on any quenches.}, the two sets  of coupled differential
equations are identical.

\section{Future direction}
Another important aspect of the study of the quantum quenches is their universal scaling behavior \cite{Buchel,AMN}. It has been
shown that for relatively fast quenches, expectation value of the boundary operator scales according to its original source.
Explicitly this means that from the expansion of the scalar field  in the
Eddington-Finkelstein frame
\be\label{scalarexpansion}
\phi(\tau,\rho)=\rho\, p_{0}(\tau)+\rho^{2}\partial_{\tau}p_{0}+\rho^{3}\,p_{2}(\tau)+\mathcal{O}(\rho^{2}\ln\rho)\,,
\ee
if the coupling in \Eq(massdeform) behaves according to $\lambda=\lambda_{0}\left(\frac{\tau}{\delta \tau}\right)^{\kappa}$
the normalizable part of the scalar field in \Eq(scalarexpansion) will turn out to be \cite{Buchel,AMN}:
\be\label{scalemode}
p_{2}(\tau)\sim \delta \tau^{-2}\left(\frac{\tau}{\delta \tau}\right)^{\kappa-2}\,,
\ee
with $\delta \tau$ being the characteristic time that is relevant for the
problem. To find \Eq(scalemode), the limit of $\delta \tau\rightarrow0$ has been taken and information regarding the four dimensional
fermionic operator with $\Delta=3$  has been used. 
Furthermore,  the origin of this  behavior is a direct consequence of  the causality.
Along the same line, we can ask if the above universality is preserved or not analytically in the inhomogeneous case.

An easy way to partially answer the above question is the following; for fast quenches nonlinearities and higher order backreactions can
be neglected since in a short time, perturbations can't propagate through the whole bulk space \cite{Buchel}.
Therefore one expects that an intuitive answer in the neighborhood of the boundary should work. 
Neglecting logarithmic corrections and higher order terms for simplicity,  the boundary terms could be written as
\bea
\label{scalephi}\phi&=&l\left(\rho\, p_{0}
+\rho^{2}\,\partial_{\tau}p_{0}
+\rho^{3}\,p_{2}\right)\,,
\\
\label{scaleA}A&=&\frac{1}{\rho^2}-\rho^2+l^{2}\left(-\frac{1}{6}p_{0}^{2}
+\rho^{2}\,a_{2}\right)\,,
\\
\label{scalesigmad}\Sigma_{d}&=&\frac{1}{\rho}+l^{2}\left(-\rho^{2}\,\frac{p^{2}_{0}}{12}
-\rho^{3}\,\frac{p_{0}\partial_{\tau}p_{0}}{9}
+\rho^{4}\,d_{4}\right)\,,
\\
\label{scalesigmab}\Sigma_{b}&=&\frac{1}{\rho}+l^{2}\left(-\rho^{2}\,\frac{p^{2}_{0}}{12}
-\rho^{3}\,\frac{p_{0}\partial_{\tau}p_{0}}{9}
+\rho^{4}\,b_{4}\right)\,,
\\
\label{scalesigmaxi}\Xi&=&l^{2}\left(-\rho\,\frac{p_{0}\partial_{x}p_{0}}{9}
+\rho^{2}\,f_{2}\right)\,,
\eea
where in the above $p_{0}$, $p_{2}$, $a_{2}$, $b_{4}$, $d_{4}$ and $f_{2}$ depend on $(\tau,x)$. 
An identical argument that was mentioned to reproduce \Eq(scalemode), still implies to \Eq(scalephi). This is due to the absence of any spacial derivative in the right-hand side at that specific order. While the scaling behavior in \Eq(scaleA), \Eq(scalesigmad) and \Eq(scalesigmab) are
 suppressed, a new feature appears in the field $\Xi$. But $\Xi\ll1$, so it's backreaction on the other components imply that  the universality breaks in a very naive way. A more convincing answer to the above question requires an analytic derivation.

\section*{Acknowledgments}
I am indebted to the organizers of the workshop on ``Numerical Holography'' at CERN, December 2014. Specially
Larry Yaffe  and Michal Heller. I have been grateful to have stimulating discussions with  Matthias Blau, Konstantinos Siampos, 
and Dimitrios Giataganas. I also acknowledge discussions, in the early stages on the subject, with Mohamad Aliakbari and Hajar Ebrahim. The author
gratefully acknowledges referee's feedbacks that boosted the quality of the paper.
This work was  supported by the Swiss National Science Foundation
(SNF) under grant 200020-155935 and Natural Sciences and Engineering Council of Canada.

\section{Appendix}
\subsection{Setup}\label{setupindex}
As mentioned before the problem at hand is a scalar field on an AdS-black brane spacetime. Starting with the following ansatz 
for the metric in an infalling observer's picture (Eddington-Finkelstein coordinates), it reads
\bea\label{metric}
ds^{2}_{5}=-A(\tau,r,x)d\tau^{2}+\Sigma_{d} (\tau,r,x)^{2}dx^{2}+\Sigma_{b} (\tau,r,x)^{2}d\vec{y}^{2}+2\Xi(\tau,r,x)d\tau dx+2drd\tau\,.\nonumber\\
\eea
 Our five-dimensional Einstein-Hilbert action with 
a negative cosmological constant is given by
\be
S_{5}=\frac{1}{16\pi G_{5}}\int d^{5}x\sqrt{-g}\left(R+12-\half \left(\partial\phi\right)^{2}-\half m^{2}\phi^{2}
+\mathcal{O}(\phi^3)\right)\,,
\ee
where we have neglected higher order interactions. We may also use inverse of the bulk radius defined by $\rho= 1/r$ and $x$ is the special direction that we apply the inhomogeneity. 
As a wave packet $\phi(\tau,\rho,x)$ is prepared on the boundary, it will evolve according to the equations of
motion and all other fields will be affected by the inhomogeneity. In the following, we will suppress such functionality, $(\tau,r,x)$, to simplify the notation.

Here is how the setup works; the scalar field is zero at the beginning as we turn on the quench at $\tau=-\infty$.
At a region around $\tau=0$, the mass coupling of the fermionic operator with $m^{2}=-3$ is switched on,
this change in the boundary conditions alters the profiles of the fields in the dual bulk space. Classical excitations
of the scalar field collapsing into the black hole will backreact on the metric. 
Eventually, at the asymptotic future, all the bulk fields will have a new
equilibrium, thermalized or partially thermalized configurations. If the final configuration  is static and globally thermalized,
the black hole has a new temperature and  correspondingly a new size consistent with the initial data at the asymptotic past and the boundary conditions.

We focus on $m^2=-3$, the scalar field is then mapped to
to a dual fermionic mass operator  $\Delta=3$ in a mass-deformed and thermal $\mathcal{N}=2^{*}$ gauge theory in $d=4$ flat spacetime.
As argued in \cite{Alex2014}, high temperature quenches $m/T\ll1$ are dual to the perturbative scalar
field in the background geometry. At the leading order, the static non-equilibrium equation for $\phi$  is given by
\be\label{hyperbolic}
\frac{m^{2}}{\rho^{2}}\phi_{equil.}
-\partial^{2}_{x}\phi_{equil.}
+\frac{1}{\rho}\left(3+\rho^{4}\right)\partial_{\rho}\phi_{equil.}
-\left(1-\rho^{4}\right)\partial^{2}_{\rho}\phi_{equil.}
=0\,.
\ee 
The solution to the above equation is  the profile for the scalar field that corresponds the the equilibrium configuration
at the asymptotic future. Unless $\partial_{x}\phi_{equil.}=0$, there is no
 analytic solution in terms
of the hyperbolic functions \cite{Alex2014} for \Eq(hyperbolic),
\be
\phi_{equil.}(\rho)=l\pi^{-1/2}\Gamma\left(\frac{3}{4}\right)^{2}
 \,_{2}F_{1}\left(\frac{3}{4},\frac{3}{4},1,1-\rho^{4}\right)\,\rho^{3}\,,
\ee
and information about the final general profile will be available through numerics or through
approximations in extreme regimes \cite{Balasubramanian:2013rva}. For further applications of \Eq(metric) refer to  \cite{Dimitrios} where they study the physics of anisotropy.

\subsection{Backreaction}\label{Backreaction}
A simple study of the EOMs shows that if the fluctuations of the scalar field are at the scale of $l$, then the effect from
backreaction appears at $l^2$. Therefore   for simplicity, we consider an expansion of the form
\bea
\phi(\tau,\rho,x)&\!=\!&l\, \hat{\phi}(\tau,\rho,x)+\mathcal{O}(l^3)\,,\\
A(\tau,\rho,x)&\!=\!&\frac{1}{\rho^2}-\rho^2+l^2\hat{A}(\tau,\rho,x)+\mathcal{O}(l^4)\,,\\
\Sigma(\tau,\rho,x)&\!=\!&\frac{1}{\rho}e^{l^2\hat{\Sigma}(\tau,\rho,x)}+\mathcal{O}(l^4)\,,\\
\Xi(\tau,\rho,x)&\!=\!&l^{2}\hat{\Xi}+\mathcal{O}(l^3)\,.
\eea
in the above, we mean $\Sigma\in\{\Sigma_{d},\Sigma_{b}\}$.

We can classify the equations into two categories; evolution equations
and constraints. 
Given some initial state or profile for the field, constraints
allow us to extract the value of the dependent fields on the former initial profiles through out the domain
of the computation. On the other hand,  evolution equations
permit the evolution of the initial state into later times. According to this distinction,
the following constraints and evolution equations are obtained. The Klein-Gordon equation of motion for the scalar field,
\be\label{ugly-phi}
\frac{m^{2}}{\rho^{2}}\phi
-\partial^{2}_{x}\phi
+3\frac{\partial_{\rho}\phi}{\rho}
+\rho^{3}\partial_{\rho}\phi
-\partial^{2}_{\rho}\phi
+\rho^{4}\partial^{2}_{\rho}\phi
-3\frac{\partial_{\tau}\phi}{\rho}
+2\partial_{\tau}\partial_{\rho}\phi=0\,,
\ee
that gives the evolution of the the scalar field. Then constraint for the combination of $\Sigma_{d}+2\Sigma_{b}$ will  be
\be
\partial^{2}_{\rho}(\Sigma_{d}+2\Sigma_{b})+\half(\partial_{\rho}\phi)^{2}
=0\,.
\ee
knowing  the profiles for $\Sigma_{b}$ and $\phi$ allows us to find $\Xi$ by the constraint,
\be
\partial^{2}_{\rho}\Xi
-4\frac{\Xi}{\rho^{2}}
+\frac{\partial_{\rho}\Xi}{\rho}
+\frac{\partial_{x}\phi\partial_{\rho}\phi}{\rho^2}
+4\frac{\partial_{\rho}\partial_{x}\Sigma_{b}}{\rho^2}
=0\,.
\ee
Similar description also hold for determining the value of the warp factor $A$ in the whole domain of the computation,
\bea
&&
\partial^{2}_{\rho}A
+\frac{m^2\phi^{2}}{3\rho^{4}}
-\frac{\partial_{\rho}A}{\rho}
-\frac{2}{\rho^{3}}\partial_{\rho}\left[\Sigma_{d}+2\Sigma_{b}\right]
-2\rho\partial_{\rho}\left[\Sigma_{d}+2\Sigma_{b}\right]
+\partial_{\rho}\partial_{x}\Xi
    \nonumber\\&&
+\frac{2}{\rho^{3}}\partial_{\tau}\left[\Sigma_{d}+2\Sigma_{b}\right]
-\frac{\partial_{\rho}\phi\partial_{t}\phi}{\rho^{2}}
-\frac{2}{\rho^{2}}\partial_{\tau}\partial_{\rho}\left[\Sigma_{d}+2\Sigma_{b}\right]
=0\,.
\eea
After determining the initial profiles for all the fields according to the above constraints, the set of 
 coupled evolution equations for $\Sigma_{d}$ and $\Sigma_{b}$,
\bea
&&
-2A
-\frac{m^{2}}{6}\frac{\phi^{2}}{\rho^{2}}
+\rho\Xi
+\rho^{5}\partial_{x}\Xi
-\partial^{2}_{x}\Sigma_{b}
+\rho\partial_{\rho}A+\frac{\partial_{\rho}\Sigma_{d}}{\rho}
-\rho^{3}\partial_{\rho}\Sigma_{d}
+5\frac{\partial_{\rho}\Sigma_{b}}{\rho}
\nonumber\\&&
-\rho^{3}\partial_{\rho}\Sigma_{b}
-\partial^{2}_{\rho}\Sigma_{b}
+\rho^{4}\partial^{2}_{\rho}\Sigma_{b}
-\frac{\partial_{\tau}\Sigma_{d}}{\rho}
-5\frac{\partial_{\tau}\Sigma_{b}}{\rho}
+2\partial_{\tau}\partial_{\rho}\Sigma_{b}=0\,,
\eea
together with
\bea
&&
\frac{m^{2}}{6}\frac{\phi^{2}}{\rho^{2}}
-\frac{m^{2}}{6}\rho^{2}\phi^{2}
-\rho\left(1+\rho^{4}\right)\partial_{x}\Xi
+\frac{\rho^{2}}{2}\partial^{2}_{x}A
-\frac{\rho}{2}\left(1-\rho^{4}\right)\partial_{\rho}A
\nonumber\\&&
-\frac{1}{\rho}\partial_{\rho}\left[\Sigma_{d}+2\Sigma_{b}\right]
+\rho^{7}\partial_{\rho}\left[\Sigma_{d}+2\Sigma_{b}\right]
+\frac{\rho^{2}}{2}(1-\rho^{4})\partial^{2}_{\rho}A
-\frac{3}{2}\rho\partial_{t}A
+\left(\frac{1}{\rho}+\rho^{3}\right)\partial_{\tau}\Sigma_{d}
\nonumber\\&&
+2\left(\frac{1}{\rho}+\rho^{3}\right)\partial_{\tau}\Sigma_{b}
-\half (\partial_{\tau}\phi)^{2}
+\rho^{2}\partial_{\tau}\partial_{x}\Xi
-\partial^{2}_{\tau}\left[\Sigma_{d}+2\Sigma_{b}\right]=0\,,
\eea
permits to solve for future profiles of the fields. Finally, the constraint and evolution equation
 for $\Xi$, are given by
\bea
&&\label{ugly-A}
2A
+\frac{m^{2}}{6}\frac{\phi^{2}}{\rho^{2}}
-2\rho\partial_{x}\Xi
+2\rho(1+\rho^{4})\partial_{x}\Xi
+\frac{1}{2}(\partial_{x}\phi)^{2}
+2\partial^{2}_{x}\Sigma_{b}
-\rho\partial_{\rho}A
-\frac{2}{\rho}\partial_{\rho}\left[2\Sigma_{d}+\Sigma_{b}\right]
\nonumber\\&&
+2\rho^{3}\partial_{\rho}\Sigma_{b}
+\rho^2\partial_{\rho}\partial_{x}\Xi
+\partial^{2}_{\rho}\Sigma_{d}
-\rho^{4}\partial^{2}_{\rho}\Sigma_{d}
+\frac{2}{\rho}\partial_{\tau}\left[2\Sigma_{d}+\Sigma_{b}\right]
-\partial_{\tau}\partial_{\rho}\Sigma_{d}=0\,,
\eea
to be solved with
\bea\label{ugly-xie}
&&
-4\frac{\Xi}{\rho^{2}}
+4\rho^{2}\Xi
+\frac{\partial_{x}A}{\rho}
+\frac{\partial_{\rho}\Xi}{\rho}
-\rho^{3}\partial_{\rho}\Xi
-\partial_{\rho}\partial_{x}A
+(1-\rho^{4})\partial^{2}_{\rho}\Xi
-2\frac{\partial_{\tau}\Xi}{\rho}
\nonumber\\&&
+\frac{\partial_{x}\phi\partial_{t}\phi}{\rho^{2}}
+4\frac{\partial_{\tau}\partial_{x}\Sigma_{b}}{\rho^{2}}
-\partial_{\tau}\partial_{\rho}\Xi=0\,.
\eea

Focusing on the fermionic operator as discussed in \cite{Buchel:2012gw}, throughout our
 computation we will assume $m^{2}=\Delta(\Delta-d)=-3$, where $\Delta$ is the conformal dimension of the 
scalar field $\phi(\tau,\rho,x)$. Now that we have both the constraints and the evolution equations, it's important to find the
boundary expansion on the AdS$_{5}$ that  follows from the Einstein equations by successive iteration of the solutions.
The few interesting terms of the expansion of each field  are listed and will be used extensively through out the paper\footnote{Similar to \cite{Alex2014}, we make an implicit gauge choice 
in writing the following boundary expansions since metric components are invariant under residual diffeomorphism.}

\bea\label{B_phi}
\hat{\phi}&=&\rho p_{0}
+\rho^{2}\partial_{\tau}p_{0}
+\rho^{3}\left[p_{2}
-\half\ln \rho\left(\partial^{2}_{x}p_{0}
-\partial^{2}_{\tau}p_{0}\right)\right]
+\rho^{4}\left(\partial_{\tau}p_{2}
-\frac{1}{3}\partial^{3}_{\tau}p_{0}\right)
\nonumber\\&&\hspace{1cm}
-\frac{\rho^{4}\ln \rho}{2}\left(\partial_{\tau}\partial^{2}_{x}p_{0}
-\partial^{3}_{\tau}p_{0}\right)
+\mathcal{O}(\rho^{5})\,,
\\\label{B_A}
\hat{A}&=&-\frac{1}{6}p_{0}^{2}
+\rho^{2}\left(a_{2}+\frac{\ln \rho}{18}\left[(\partial_{x}p_{0})^{2}
+3(\partial_{\tau}p_{0})^{2}+p_{0}\Big(\partial^{2}_{x}p_{0}
-3\partial^{2}_{\tau}p_{0}\Big)\right]\right)
+\mathcal{O}(\rho^{3})\,,
\\\label{B_Sigmaa}
\hat{\Sigma_{d}}&=&-\frac{1}{12}\rho^{2}p^{2}_{0}
-\frac{1}{9}\rho^{3}p_{0}\partial_{\tau}p_{0}
+\rho^{4}\left(d_{4}+\frac{\ln \rho}{72}\left[-4(\partial_{x}p_{0})^{2}
+p_{0}\left(5\partial^2_{x}p_{0}-3\partial^{2}_{\tau}p_{0}\right)\right]\right)
+\mathcal{O}(\rho^{5})\,,\nonumber\\\\
\label{B_Sigmab}
\hat{\Sigma_{b}}&=&-\frac{1}{12}\rho^{2}p^{2}_{0}
-\frac{1}{9}\rho^{3}p_{0}\partial_{\tau}p_{0}
+\rho^{4}\left(b_{4}+\frac{\ln \rho}{72}\left[2(\partial_{x}p_{0})^{2}
+p_{0}\left(2\partial^2_{x}p_{0}
-3\partial^{2}_{\tau}p_{0}\right)\right]\right)
+\mathcal{O}(\rho^{5})\,,\nonumber\\\\
\label{B_Xi}
\hat{\Xi}&=&-\frac{1}{9}\rho p_{0}\partial_{x}p_{0}
+\rho^{2}\left(f_{2}+\frac{\ln \rho}{12}\Big[p_{0}\partial_{\tau}\partial_{x}p_{0}
-2\partial_{x}p_{0}\partial_{\tau}p_{0}\Big]\right)
+\mathcal{O}(\rho^{3})\,.
\eea
Note that in practice, we have worked out the above expansion to $\mathcal{O}(\rho^{8})$.  Further, we should draw the attention of the reader to 
the normalizable terms such as $\left\{p_{2}, a_{2},f_{2},d_{4},b_{4}\right\}$. These coefficients are the response 
of the fields to the alterations in the system.

\subsection{2D Chebyshev lattice}
\addsubsubsection{General overview}
In what follows, we  do the computations as symbolic as possible. Our goal here has been to achieve relatively very small rounding errors through successive operations that have been carried out. The fact
that smooth functions can be approximated in a creative way by polynomial interpolation in Chebyshev points and the use of
Fast Fourier Transform, allow us to use new sort of polynomials called Chebyshev polynomials. To do the numerics in a stable and effective way, accuracy to within roughly machine precision can be achieved using spectral methods.

In the interval of $0<\rho<1$, a convenient basis of expansion in terms of the Chebyshev polynomials 
$T_{n}(z)\equiv\cos\left(n\cos^{-1}z\right)$, will have the form
\be
g(\rho)=\sum_{n=0}^{M}\alpha_{n}T_{n}(2\rho-1)\,,
\ee
which is nothing other than  rewriting  the Fourier expansion with a change of variable $\theta\equiv\cos^{-1}(2\rho-1)$. 
In a general approach, pseudospectral or collocation method, one finds the expansion coefficients $\alpha_{n}$ by inserting
the above truncated series into the differential equation of interest and turn  the problem into an eigenvalue problem.
We should point out that although in the conventional Fourier transformation one is interested in equally spaced
lattices, in the spectral method, we avoid this primitive setup and instead use basis function that are 
matched by the position of the maximums/minimums and endpoints of the M'th Chebyshev basis. In our case for the interval $[0,1]$, these are given by
\be
\rho_{m}=\frac{1}{2}\left(1-\cos\frac{m\pi}{M}\right)\,.
\ee
with the knowledge of $\alpha_{n}$, we reconstruct the whole function \{$g_{m}\equiv g(\rho_{m})$\} from the collocation grid
points.

The range $x\in[0, 1]$ is the most convenient one to use but sometimes the other option,
$z\in[-1, 1]$, is required. The map between the two sets is given by
$x=\frac{1}{2}\left(1+z\right)$ and this leads to a shifted\footnote{The map for the 
general case of $x\in[a,b]$ can be constructed similarly using  $s=\frac{2x-(a+b)}{b-a}$ for $x\in[-1,1]$.} Chebyshev polynomial \cite{Mason}
\be
T^{\ast}_{n}(x)=T_{n}(z)=T_{n}(2x-1)\,.
\ee
We will use this latter set for the spectral grid in the $x$ direction where we need the boundary in the range $[-L_{x},L_{x}]$ .

The  concept of Chebyshev points can be extended to  differential operators thus, we will be working with Chebyshev differential matrices later on. Meanwhile, there are various
interesting identities \cite{Guo} for the Chebyshev polynomials that will be useful throughout this appendix. They satisfy
\be
\sqrt{1-x^2}\frac{d}{dx}\left(\sqrt{1-x^{2}}\right)+n^{2}T_{n}(x)=0\,,
\ee 
and their explicit integral evaluates to
\be
\int_{-1}^{1}dxT_{n}(x)=-\frac{2}{n^{2}-1}\quad \text{for even $n$}\,,
\ee
while zero for any odd $n$. At the boundaries they satisfy 
\bea
T_{n}(x=\pm1)&=&(\pm1)^{n}\,,\quad \left.\frac{dT_{n}}{dx}\right|_{x=\pm1}=(\pm1)^{n+1}n^{2}\,,
\nonumber\\
\left.\frac{d^{2}T_{n}}{dx^{2}}\right|_{x=\pm1}&=&\frac{1}{3}(\pm1)^{n}(n-1)n^{2}(n+1)\,.
\eea

\addsubsubsection{2D aspects}
The above one-dimensional boundary value problem can be extended to higher dimensions. To be specific, 
here we use a 2D setup.  For such a problem, we naturally set up a grid based on Chebyshev points  in
each direction independently. This is usually called a \textit{tensor product grid}. It's interesting to note that in comparison with  an equally spaced
grid, Chebyshev grid is $2/\pi$ times as dense in the middle and in our current 
2D setup this ratio becomes $\left(2/\pi\right)^{2}$. Thus the majority of the grid points lie near the boundaries.
As the enforcing boundary condition is applied at $\rho=0$, this will enhance the resolution. Therefore, the tensor product construction of a spectral grid is the natural way to go. This can easily be done by tensor product in
linear algebra, for instance for two matrices $A$ and $B$ the \textit{Kronecker product} is given by $A\otimes B$.
That is, if $A$ and $B$ are matrices of dimensions $p\times q$ and $r\times s$ respectively, then $A \otimes B$ is a matrix of
dimension $pr\times qs$ with $p\times q$ block forms, where each $i$ and $j$ block has the value of $a_{ij}B$.

With a data set represented symbolically as $(v_{1},v_{2},\cdots,v_{10})^{T}$, we can use the 1D
representation of the differential operators to find a representation of of its counterpart in 2D in the following way
\be
L_{N_{\rho}\times N_{x}}=I_{N_{\rho}}\otimes D_{N_{x}}+D_{N_{\rho}}\otimes I_{N_{x}}\,.
\ee
Using the above representation, it's also possible to derive $D^{2}_{N}$ of the Laplace operator on the above lattices. In 
principle, we could have used the polar coordinates but we stick to the choice of the Cartesian one since we are imposing
the boundary condition exactly at $\rho=0$ and we can't avoid any creative trick to avoid this point.  One extra complication with
respect to the 1D setup is the issue of corner compatibility which states that
\be
\alpha_{\pm}(x=L_{max})=\beta_{+}(\rho=0\,\, \text{and}\,\, 1)\,,\quad \alpha_{\pm}(x=-L_{max})=\beta_{-}(\rho=0\,\,\text{and}\,\, 1) \,.
\ee
In the above, we assume that the boundary values for $\rho=0$ and $\rho=1$ are given by $\alpha_{+}(x)$ and $\alpha_{-}(x)$ respectively and
 the corresponding boundary values on vertical walls at $x=\pm L_{max}$ are equal to $\beta_{\pm}(\rho)$. 
The effect of these corner conditions becomes prominent when we calculate derivatives of the fields.

\begin{figure}[ht!]
\centering
{\includegraphics[width=8cm]{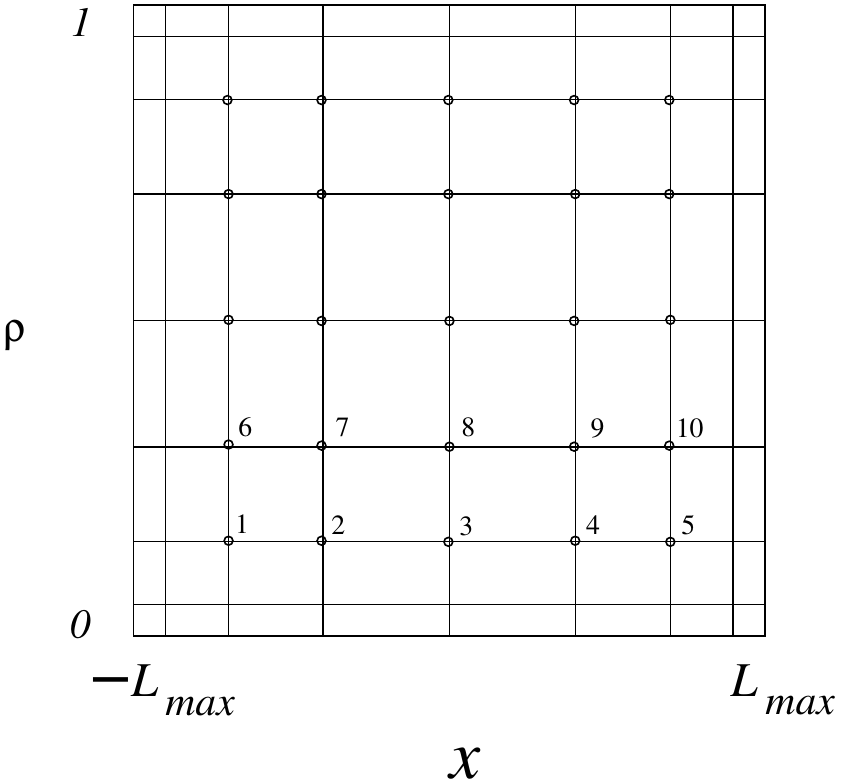}}
{\caption{A tensor product grid; there are two spacial directions. $x$ is the direction of the 
inhomogeneity and $\rho$ is the bulk radius. The numbers at each site represent the  lexicographic representation of the grid points while doing the operation as a tensor grid.}\label{fig:test}}
\end{figure}

After discretizing the problem in  rectangular Cartesian coordinates, we use the generalization of the
pseudo-spectral method in 2D. For instance a function, $f(\rho,x)$, has an expansion as  linear combinations of Chebyshev polynomials,
\be\label{cheby-expansion}
f^{N,L}(\rho,x)=\sum_{l=0}^{L}\sum_{n=0}^{N}\hat{\rho}_{ln}T_{n}(\rho)T_{l}(x)\,,
\ee
here the $\hat{\rho}_{ln}$s are the 2D spectra of $f(\rho,x)$. In addition $N$ and $L$ are 
the number of collocation points in $\rho$ and $x$ coordinates. In vectorial notation, we rewrite the Chebyshev
 polynomials in $x$ and $\rho$ directions
\be
\left(\mathbf{T}_{x}\right)_{l\lambda}=(-1)^{\lambda}\cos\left(l\lambda\frac{\pi}{L}\right)\,,\quad
\left(\mathbf{T}_{\rho}\right)_{n\nu}=(-1)^{\nu}\cos\left(n\nu\frac{\pi}{N}\right)\,.
\ee
Based on Figure \ref{fig:test}, the representation for the general solution will then  can be selected as
\be\label{lattice-var}
F=\left(\underbrace{f_{00},f_{10},\cdots,f_{L0}},\underbrace{f_{01},f_{11},\cdots,f_{L1}},
\underbrace{\cdots},\underbrace{\cdots},\underbrace{f_{0N},f_{1N},\cdots,f_{LN}}\right)^{T}\,.
\ee
These are $(N+1)$ blocks of $(L+1)$ quantities and each block corresponds to a position in the $\rho$
 coordinates. In this representation \Eq(cheby-expansion) will
take the compact form of 
\be
F=\left(\mathbf{T}_{\rho}\otimes \mathbf{T}_{x}\right)\hat{F}\,,
\ee
which is suitable for our notation throughout the rest of this appendix.

\subsection{Coupled equations}
Our first step in the numerical code, is to make the following definitions 
\bea
\label{defpi}
\pi(\tau,\rho,x)&=&\partial_{\tau}\hat{\phi}(\tau,\rho,x)+\frac{\rho^4-1}{2}\,\partial_{\rho}\hat{\phi}(\tau,\rho,x)\,,
\\
\label{defbeta}
\beta(\tau,\rho,x)&=&\partial_{\tau}\hat{\Sigma}_{d}(\tau,\rho,x)+\frac{\rho^4-1}{2}\,\partial_{\rho}\hat{\Sigma}_{d}(\tau,\rho,x)\,,
\\
\label{defgamma}
\gamma(\tau,\rho,x)&=&\partial_{\tau}\hat{\Sigma}_{b}(\tau,\rho,x)+\frac{\rho^4-1}{2}\,\partial_{\rho}\hat{\Sigma}_{b}(\tau,\rho,x)\,,
\\
\label{defchi}
\chi(\tau,\rho,x)&=&\partial_{\tau}\hat{\Xi}(\tau,\rho,x)+\frac{\rho^4-1}{2}\,\partial_{\rho}\hat{\Xi}(\tau,\rho,x)\,,
\eea
that transform \Eq(ugly-phi)-\Eq(ugly-A) into more compact forms
\bea
&&\label{source_pi}
\partial_{\rho}\pi-\frac{3}{2}\frac{\pi}{\rho}=-J_{\phi}\,,
\\
\label{source_phi}
&&\partial^{2}_{\rho}(\Sigma_{d}+2\Sigma_{b})+\half(\partial_{\rho}\phi)^{2}=0\,,
\\
&&\label{source_Xi}
\partial^{2}_{\rho}\Xi
-4\frac{\Xi}{\rho^{2}}
+\frac{\partial_{\rho}\Xi}{\rho}
+\frac{\partial_{x}\phi\partial_{\rho}\phi}{\rho^2}
+4\frac{\partial_{\rho}\partial_{x}\Sigma_{b}}{\rho^2}
=0\,,
\\
&&\label{source_beta}
\partial_{\rho}\beta-\frac{1}{\rho}\left[2\beta+\gamma\right]=-J_{\Sigma_{d}}\,,
\\
&&\label{source_gamma}
\partial_{\rho}\gamma-\frac{1}{2\rho}\left[\beta+5\gamma\right]=-J_{\Sigma_{b}}\,,
\\
\label{source_A}
&&\partial^2_{\rho}A
-\frac{\partial_{\rho}A}{\rho}
+\frac{-2}{\rho^2}\partial_{\rho}\left[\beta+2\gamma\right]
+\frac{2}{\rho^3}\left[\beta+2\gamma\right]
-\pi\frac{\partial_{\rho}\phi}{\rho^2}=-J_{a}\,,
\\
&&\label{source_chi}
\partial_{\rho}\chi+
2\frac{\chi}{\rho}
-4\frac{\partial_{x}\gamma}{\rho^2}
-\pi\frac{\partial_{x}\phi}{\rho^2}
+2\rho^2\left(1-\frac{1}{\rho^4}\right)\partial_{\rho}\partial_{x}\Sigma_{b}
=-J_{\Xi}\,,
\eea
with the sources on the right-hand sides of the above equations defined according to
\bea
J_{\phi}
&=&
\frac{m^{2}}{2}\frac{\phi}{\rho^2}
-\half \partial^{2}_{x}\phi
-\frac{3}{4}\rho^{3}\left(1-\frac{1}{\rho^{4}}\right)\partial_{\rho}\phi\,,
\\
\label{current-Jsigmad}
J_{\Sigma_{d}}
&=&
-A
-\frac{m^{2}}{12}\frac{\phi^2}{\rho^2}
+\rho\partial_{x}\Xi
-\frac{1}{4}(\partial_{x}\phi)^2
-\partial^{2}_{x}\Sigma_{b}
+\half \rho\partial_{\rho}A
-\frac{\rho^{3}}{2}\left(1-\frac{1}{\rho^{4}}\right)\partial_{\rho}\left[2\Sigma_{d}+\Sigma_{b}\right]
\nonumber\\&&
-\half \rho^{2}\partial_{\rho}\partial_{x}\Xi\,,
\\
\label{current-Jsigmab}
J_{\Sigma_{b}}
&=&
-A
-\frac{m^{2}}{12}\frac{\phi^2}{\rho^2}
+\half \rho\partial_{x}\Xi
-\half\partial^{2}_{x}\Sigma_{b}
+\half \rho\partial_{\rho}A
-\frac{\rho^{3}}{4}\left(1-\frac{1}{\rho^{4}}\right)\partial_{\rho}\left[\Sigma_{d}+5\Sigma_{b}\right]\,,
\\
J_{a}
&=&
\frac{m^2}{3}\frac{\phi^{2}}{\rho^4}
+\rho\left(1-\frac{1}{\rho^4}\right)\partial_{\rho}\left[\Sigma_{d}+2\Sigma_{b}\right]
+\frac{\rho^2}{2}\left(1-\frac{1}{\rho^4}\right)(\partial_{\rho}\phi)^{2}
+\partial_{\rho}\partial_{x}\Xi
\nonumber\\&&
+\rho^2\left(1-\frac{1}{\rho^4}\right)\partial^2_{\rho}\left[\Sigma_{d}+2\Sigma_{b}\right]\,,
\\
J_{\Xi}
&=&
-4\rho^{2}\left(1-\frac{1}{\rho^{4}}\right)\Xi
-\frac{\partial_{x}A}{\rho}
-2\rho^{3}\partial_{\rho}\Xi
+\frac{\rho^{2}}{2}\left(1-\frac{1}{\rho^{4}}\right)\partial_{x}\phi\partial_{\rho}\phi
+\partial_{\rho}\partial_{x}A
\nonumber\\&&
-\half\left(1-\rho^4\right)\partial^{2}_{\rho}\Xi\,.
\eea

We point out a few comments about the above equations.
They are listed chronologically, that's we start solving the coupled differential equations starting
from \Eq(source_pi) and  end in \Eq(source_chi).  The equation of motion for the scalar field is not
coupled to the other metric components. This is due to the choice of cutoff that we have imposed on the backreaction.  
From the boundary expansion, it's clear that the $x$ dependence of $\Sigma$s do not factorize. Therefore, 
$x$ dependence of $\phi$ must not factorize according to Eq.\,(\ref{source_phi}). 
As it's clear from Eq.\,(\ref{source_phi}), knowing the value of the scalar field $\phi_{0}$, everywhere
in the bulk, only gives the information about the combination of $\Sigma_{d}+2\Sigma_{b}$. Moreover, 
the $x$ dependency of $\Sigma_{d}+2\Sigma_{b}$ will be trivial since the derivatives act on the $\rho$ direction.

\addsubsubsection{Extra identities}
In addition to the above differential equations, in this subsection, we derive identities that are useful when we are applying the boundary conditions on the fields. 

Summation of Eq.\,(\ref{source_beta}) and Eq.\,(\ref{source_gamma}) gives $\beta+2\gamma$ as a function of 
$\Sigma_{d}+2\Sigma_{b}$, that is
\be
\partial_{\rho}\left[\beta+2\gamma\right]-\frac{3}{\rho}\left[\beta+2\gamma\right]=-J_{\Sigma_{d}+2\Sigma_{b}}\,,
\ee
with $J_{\Sigma_{d}+2\Sigma_{b}}$ that reads
\bea
J_{\Sigma_{d}+2\Sigma_{b}}&=&
-3A
-\frac{m^{2}}{4}\frac{\phi^2}{\rho^2}
+2\rho\partial_{x}\Xi
-\frac{1}{4}(\partial_{x}\phi)^2
-2\partial^{2}_{x}\Sigma_{b}
+\frac{3}{2} \rho\partial_{\rho}A
\nonumber\\&&
-\frac{3\rho^{3}}{2}\left(1-\frac{1}{\rho^{4}}\right)\partial_{\rho}\left[\Sigma_{d}+2\Sigma_{b}\right]
-\half \rho^{2}\partial_{\rho}\partial_{x}\Xi\,,
\eea
but the presence of $\partial^{2}_{x}\Sigma_{b}$ requires some extra knowledge of $\Sigma_{b}$.
Furthermore,  from Eq.\,(\ref{source_Xi}) we can solve for $\partial_{\rho}\partial_{x}\Sigma_{b}$ and insert
it in Eq.\,(\ref{source_chi}) to obtain
\bea
\partial_{\rho}\chi+
2\frac{\chi}{\rho}
-4\frac{\partial_{x}\gamma}{\rho^2}
-\pi\frac{\partial_{x}\phi}{\rho^2}
=-J_{\chi_{b}}\,,
\eea
with
\be
J_{\chi_{b}}=-2\rho^2\left(1-\frac{1}{\rho^4}\right)\Xi-\frac{5}{2}\rho^{3}\partial_{\rho}\Xi+\frac{1}{2\rho}\partial_{\rho}\Xi
-\frac{\partial_{x}A}{\rho}+\partial_{\rho}\partial_{x}A\,,
\ee
and again in the above, extra knowledge of $\partial_{x}\gamma$ will be necessary to solve for $\chi$.

In addition to the above constraints, we also have
\be
d_{4}+2b_{4}
+\frac{1}{4}p_{0}p_{2}
+\frac{1}{32}p_{0}\partial^{2}_{x}p_{0}
+\frac{1}{6}(\partial_{\tau}p_{0})^{2}
-\frac{1}{32}p_{0}\partial^{2}_{\tau}p_{0}=0\,,
\ee
and
\bea\label{evolution_a}
&&2\partial_{x}f_{2}
-\half p_{2}\partial_{\tau}p_{0}
+\frac{5}{18}\partial^{2}_{x}p_{0}\partial_{\tau}p_{0}
+\frac{3}{2}\partial_{\tau}a_{2}
+\half p_{0}\partial_{\tau}p_{2}
+\frac{13}{72}\partial_{x}p_{0}\partial_{\tau}\partial_{x}p_{0}
+\frac{11}{72}p_{0}\partial_{\tau}\partial^{2}_{x}p_{0}
\nonumber\\&&
+\frac{1}{12}\partial_{\tau}p_{0}\partial^{2}_{\tau}p_{0}
-\frac{1}{3}p_{0}\partial^{3}_{\tau}p_{0}=0\,,
\eea
which means that in order to extract the evolution of  $a_{2}(\tau,x)$, the coefficient in the warp factor,
we have to provide  $\partial_{x}f_{2}$ in addition to the initial condition of $a_{2}(\tau_{0},x)$. 
In the rest of this appendix, we will solve \Eq(source_pi)-\Eq(source_chi) and the above identities numerically.

\subsection{Numerical implementation}

As we mentioned before, in practice, we have a finite number of points available in the inhomogeneous direction.
The cutoff should be chosen with respect to the value of the other parameters such as the size of the 
system or the profile of the source under consideration. We consider rather a  general profile for the source
\cite{Alex2014}, \cite{Balasubramanian:2013rva},
\be\label{p0}
p_{0}(\tau,x)=\frac{1}{2}\left[1+\tanh\left(\frac{\tau}{\alpha}\right)\right]\,e^{-\frac{x^{2}}{\sigma^{2}}}\,,
\ee
and choose the cutoff for the coordinate $x\in \left[-L_{x},L_{x}\right]$ with $L_{x}=10$ and multiple values for 
$\sigma \in \left[\sqrt{L_{x}},\sqrt{1.5L_{x}} \right]$ and $\alpha \in \left[\frac{1}{8},1\right]$. Each of these parameters  simulates
a different physical scenario. Parameter $\alpha$ is the scale of the time variation of the quench, unlike $\sigma$ which is the spacial
scale of the inhomogeneity applied to the system. The shape of $p_{0}$
has been chosen so that at the asymptotic past, the source is zero. In principle for doing the numerical analysis, 
we considered  the time interval $\tau\in \left[\tau_{i},\tau_{f}\right]$ with $\tau_{i}=-7.5$ and 
$\tau_{f}=12$, that works out for our goal similar to \cite{Alex2014}. 

As it was pointed out in Section \ref{Backreaction}, near the boundary we encounter logarithmic divergences
that cause numerical instabilities, to tackle them on the lattice, the standard
method is to isolate the finite contributions. Therefore  it's advisable to make the following  change of variables
\bea\label{phi_connected}
\hat{\phi}(\tau,\rho,x)&=&\hat{\phi}_{log}(\tau,\rho,x)+\phi^{c}(\tau,\rho,x)\,,
\\\label{sigma_connected}
\hat{\Sigma}(\tau,\rho,x)&=&\hat{\Sigma}_{log}(\tau,\rho,x)+\Sigma^{c}(\tau,\rho,x)\,,
\\\label{A_connected}
\hat{A}(\tau,\rho,x)&=&\hat{A}_{log}(\tau,\rho,x)+A^{c}(\tau,\rho,x)\,,
\\\label{xi_connected}
\hat{\Xi}(\tau,\rho,x)&=&\hat{\Xi}_{log}(\tau,\rho,x)+\Xi^{c}(\tau,\rho,x)\,,
\eea
and follow these numerical algorithms that we label them by \textbullet\, below:

\textbullet\,  At $\tau=\tau_{i}$, we have to start with an initial profile for the fields, our choice is
\be\label{initial}
\phi^{0}_{l,n}\equiv \phi^{c}\left(\tau_{i},\rho_{l},x_{n}\right)\,,\quad
\Sigma^{0}_{b\,\, l,n}\equiv \Sigma_{b}^{c}\left(\tau_{i},\rho_{l},x_{n}\right)\,,
\ee
with $\phi^{0}_{l,n}=\Sigma^{0}_{b\,\, l,n}=0$. These two initial profiles at $\tau_{i}$
are sufficient to solve \Eq(source_phi) and \Eq(source_pi) for all points on the lattice at time $\tau_{i}$.
For $\Sigma_{d}$, with  definitions from \Eq(phi_connected), \Eq(sigma_connected) and 
inserting them into \Eq(source_phi), we can see that
\be\label{sigmad0}
\partial^{2}_{\rho}\Sigma^{c}_{d}=\tilde{J}_{\Sigma_{d}}\,,
\ee
with
\be\label{sigmad-log-sources}
\tilde{J}_{\Sigma_{d}}=\partial^{2}_{\rho}\Sigma^{d}_{log}
+2\partial^{2}_{\rho}\Sigma^{c}_{b}+2\partial^{2}_{\rho}\Sigma_{log}^{b}
+\half\left(\partial_{\rho}\phi_{c}+\partial_{\rho}\phi_{log}\right)^{2}\,.
\ee
Then in the above, we'll use the initial profiles of $\phi^{0}_{l,n}$ and $\Sigma^{0}_{b\,\, l,n}$ to replace the 
$\phi_{c}$ and $\Sigma^{b}_{log}$ and solve the above equation for the solution of
$\Sigma_{d}^{c}(t_{0},\rho,x)$, with the boundary conditions
\be
\Sigma^{con.}(\tau,0,x)=0\,,\quad \partial_{\rho}\Sigma^{con.}(\tau,0,x)=0\,,
\ee
that have been derived from \Eq(B_Sigmaa). The matrix form of the differential equation is
\be
\left(I_{x}\otimes D^{2}_{\rho}\right)\Sigma_{d\,\,l,n}=\left(\tilde{J}_{\Sigma_{d}}\right)_{n,l}\,,
\ee
where we impose the boundary conditions in a matrix form since $\Sigma_{d\,\,l,n}$ has a 
form similar to \Eq(lattice-var). As it's clear in \Eq(sigmad-log-sources), in addition to the
 finite contributions of the fields $\Sigma^{c}_{b}$ and $\phi_{c}$ on the right-hand side, we also need 
their logarithmic corrections. In order to subtract the logarithms, we make an expansion over the bulk radius. 
From  \Eq(B_phi)-\Eq(B_Sigmab), we have
\bea
\phi_{log}&=&\log \rho \sum_{i=3}^{8}\frac{\rho^{i}}{(1+\rho)^{1+i}}\mathcal{F}_{i}\left[p_{0}(\tau,x)\right]\,,
\\\label{sigmab-log}
\Sigma^{log}_{b}&\!\!=\!\!&\rho^{2}\log \rho \sum_{i=2}^{5}\frac{\rho^{i}}{(1+\rho)^{1+i}}\mathcal{B}_{1,i}\left[p_{0}(\tau,x),p_{2}(\tau,x)\right]
\!+\!\rho^{2}\left(\log \rho\right)^{2}\sum_{i=4}^{5}\frac{\rho^{i}}{(1+\rho)^{1+i}}\mathcal{B}_{2,i}\left[p_{0}(\tau,x)\right]\,,
\nonumber\\
\\\label{sigmad-log}
\Sigma^{log}_{d}&\!\!=\!\!&\rho^{2}\log \rho \sum_{i=2}^{7}\frac{\rho^{i}}{(1+\rho)^{1+i}}\mathcal{D}_{1,i}\left[p_{0}(\tau,x),p_{2}(\tau,x)\right]
\!+\!\rho^{2}\left(\log \rho\right)^{2}\sum_{i=4}^{7}\frac{\rho^{i}}{(1+\rho)^{1+i}}\mathcal{D}_{2,i}\left[p_{0}(\tau,x)\right]\,,\nonumber\\
\eea
with the coefficients of $\mathcal{F}_{i}$, $\mathcal{B}_{1,i}$, $\mathcal{B}_{2,i}$
, $\mathcal{D}_{1,i}$ and $\mathcal{D}_{2,i}$  rather having a  complicated form to mention here. As it has been mentioned in \cite{Alex2014}, the  upper bound 
for the series can go to infinity but as it's apparent from the first terms of \Eq(sigmab-log) and
\Eq(sigmad-log), they are functions of $p_{2}$, an expansion parameter in  the scalar field 
$\phi$ from \Eq(B_phi) (the normalizable mode). Since we have no information about this coefficient priory to solving the evolution equation
for the scalar field, instead we  use 
\be \label{p2}
p_{2}(\tau,x)=\left.\frac{1}{6}\,\partial^{3}_{\rho}\phi(\tau,\rho,x)\right|_{\rho=1}\,.
\ee
But the error in subtracting the coefficient in $p_{2}(\tau,x)$, stops us from increasing the upper bounds in  \Eq(sigmab-log) and \Eq(sigmad-log). 

\textbullet\, Since we need  time derivatives of $p_{2}(\tau,x)$ for evaluating the coefficients in 
\Eq(sigmab-log)-\Eq(sigmad-log), a time evolution of $\phi(\tau_{i}+\Delta\tau,\rho,x)$ is necessary. To do this, 
first we solve \Eq(source_pi),
\be
\left(I_{x}\otimes D_{\rho}-\frac{3}{2\rho}\right)\pi^{\tau_{i}}_{n,l}=-\left(J_{\phi}\right)_{n,l}\,,
\ee
at $\tau_{i}$ with the boundary condition that reads
\be
 \pi^{c}(\tau_{i},0,x)=- \frac{p_{0}(\tau_{i},x)}{2}\,.
\ee
Then in order to translate it to $\phi_{c}$, we use
\be\label{evo_phi}
\partial_{\tau}\phi^{c}(\tau,\rho,x)=\pi^{c}(\tau_{i},\rho,x)+\frac{1-\rho^{4}}{2}\partial_{\rho}\phi^{c}(\tau_{i},\rho,x)
+k_{log}(\tau_{i},\rho,x)\,,
\ee
with
\be
k_{log}(\tau_{i},\rho,x)=\pi_{log}(\tau_{i},\rho,x)+\frac{1-\rho^{4}}{2}\partial_{\rho}\phi_{log}(\tau_{i},\rho,x)
-\partial_{\tau_{i}}\phi_{log}(\tau_{i},\rho,x)\,.
\ee
The initial condition to solve \Eq(evo_phi) is $\phi_{c}(-\infty,\rho,x)\!=\!0$. Note that the form of $\phi_{log}$ and $\pi_{log}$ are related according to \Eq(phi_connected) and \Eq(defpi). The latter explicitly is given by
\be
\pi_{log}=\log \rho \sum_{i=2}^{7}\frac{\rho^{i}}{(1+\rho)^{i}}\mathcal{P}_{i}\left[p_{0}(\tau,x)\right]\,.
\ee
The evaluation
is done by completing the first Runge-Kutta (RK) step,
\be
k_{1,\phi}=\Delta\tau\left(
\pi^{\tau_{i}}_{n,l}+\half\left(1-\rho^{4}\right)\partial_{\rho}\phi^{\tau_{i}}_{n,l}+k_{log}\right)\,,
\ee
that is accompanied by the following shifts
\be
\tau_{i}\rightarrow\tau_{i}+\half\Delta\tau\,,\quad
\phi^{\tau_{i}}_{n,l}\rightarrow\phi^{\tau_{i}}_{n,l}+\frac{k_{1,\phi}}{2}\,,
\ee
and with these new values for $\tau_{i}$ and $\phi^{\tau_{i}}_{n,l}$, we repeat RK step 1 to find $k_{2,\phi}$. This
 completes RK step 2. In RK step 3, we have
\be
\tau_{i}\rightarrow\tau_{i}+\half\Delta\tau\,,\quad
\phi^{\tau_{i}}_{n,l}\rightarrow\phi^{\tau_{i}}_{n,l}+\frac{k_{2,\phi}}{2}\,,
\ee
and we repeat steps in RK step 1 to find $k_{3,\phi}$. At RK step 4, finally we make the last set of shifts
\be
\tau_{i}\rightarrow\tau_{i}+\Delta\tau\,,\quad
\phi^{\tau_{i}}_{n,l}\rightarrow\phi^{\tau_{i}}_{n,l}+k_{3,\phi}\,,
\ee
to obtain the  value of the scalar field at $\tau=\tau_{i}+\Delta\tau$,
\be
\phi^{\tau+\Delta\tau}_{n,l}=\phi^{\tau}_{n,l}+\frac{1}{6}k_{1,\phi}+\frac{1}{3}k_{2,\phi}+\frac{1}{6}k_{4,\phi}\,.
\ee
This finishes the procedure of evaluating time derivatives of $p_{2}$ based on \Eq(p2). Knowing all the variables
in \Eq(sigmad0) allows us to evaluate $\Sigma_{d\,\,n,l}^{\tau_{i}}\equiv\Sigma_{d}(\tau_{i},\rho,x)$.

\textbullet\,  In order to find  $A^{\tau_{i}}\equiv A(\tau_{i},\rho,x)$, we still need to evaluate 
the value of $\Xi^{\tau_{i}}_{n,l}\equiv \Xi(\tau_{i},\rho,x)$. The values of $\phi^{\tau_{i}}_{n,l}$ and 
$\Sigma^{\tau_{i}}_{b\,\, n,l}$ are enough to do this as we describe in this section. \Eq(source_Xi) on the
 lattice will  be given by
\be\label{forgot}
\left(I_{x}\otimes D^{2}_{\rho}-\frac{4}{\rho^{2}}+\frac{I_{x}\otimes D_{\rho}}{\rho}\right)\Xi^{\tau_{i}}_{n,l}=
-\left(J_{\phi^{\tau_{i}},\Sigma^{\tau_{i}}_{b}}\right)_{n,l}\,,
\ee
where the current $J_{\phi^{\tau_{i}},\Sigma^{\tau_{i}}_{b}}$ are all the terms that include 
 $\phi^{\tau_{i}}$  and $\Sigma^{\tau_{i}}_{b}$ and have been taken to the right-hand side in \Eq(source_Xi). 
We also need the logarithmic part $\Xi^{log}$ subtracted by
\bea
\Xi^{log}=\log \rho \sum_{i=2}^{5}\frac{\rho^{i}}{(1+\rho)^{1+i}}\mathcal{K}_{1,i}\left[p_{0}(\tau,x),p_{2}(\tau,x)\right]
+\left(\log \rho\right)^{2}\frac{\rho^{5}}{(1+\rho)^{6}}\mathcal{K}_{2,5}\left[p_{0}(\tau,x)\right]\,.\nonumber\\
\eea
Once again the boundary condition at $\rho=0$ for solving \Eq(forgot) is given by
\be
\Xi^{con.}_{f}(\tau,0,x)=0\,,\quad \partial_{\rho}\Xi^{con.}_{f}(\tau,0,x)=-\frac{1}{9}p_{0}\partial_{x}p_{0}\,.
\ee

\textbullet\, As we mentioned before, knowing all the values of the fields  $\phi^{\tau_{i}}_{n,l}$,    
 $\Sigma^{\tau_{i}}_{b\,\, n,l}$ and $\Xi^{\tau_{i}}_{n,l}$, we can evaluate  $A^{\tau_{i}}$ in principle from
\Eq(source_A) that has been deduced. Since it's a second-order differential equation
with the two initial conditions that each will increase the size of the arrays (cost of the computation) by a 
factor of $N_{x}\times N_{\rho}$, we will rather replace for $\beta$ and $\gamma$
from \Eq(source_beta) and \Eq(source_gamma) similar to the approach of \cite{Alex2014} in favor
of a more complicated but linear equation for
\be\label{source_Atilde}
\partial_{\rho}\tilde{A}
=
-J_{\tilde{A}}\,,
\ee
with 
\bea\label{Atildedef}
\tilde{A}&\equiv& \partial^{2}_{\rho}A+\frac{2}{\rho}\partial_{\rho}A\,,
\\\label{JAtilde}
J_{\tilde{A}}&\equiv& \partial_{\rho}J_{a}
+\frac{1}{\rho^{2}}\partial_{\rho}\left[\acute{J}_{\Sigma_{d}}+2\acute{J}_{\Sigma_{b}}\right]\,,
\eea
and in the above,  $\acute{J}$ refers to terms that are proportional to $A$ in $J$s and
have been taken to the left-hand side of \Eq(source_Atilde). Our boundary condition that is consistent with \Eq(B_A) reads 
\be\label{boundaryAtilde}
\tilde{A}(t,0,x)=6a_{2}+\frac{5}{18}\left[(\partial_{x}p_{0})^{2}
+3(\partial_{\tau}p_{0})^{2}+p_{0}\Big(\partial^{2}_{x}p_{0}
-3\partial^{2}_{\tau}p_{0}\Big)\right]\,,
\ee
where all the coefficients,  $p_{0}$ and $a_{2}$, are functions of  $(\tau,x)$.  It is possible to rewrite \Eq(source_Atilde)
 in a more illuminating form
\be\label{illum}
\partial_{\rho}\tilde{A}_{c}=-\partial_{\rho}\tilde{A}_{log}-J_{\tilde{A}}\,, 
\ee
with 
\be
\tilde{A}_{c}=\partial^{2}_{\rho}A_{c}+\frac{2}{\rho}\partial_{\rho}A_{c}\,,
\quad
\tilde{A}_{log}=\partial^{2}_{\rho}A_{log}+\frac{2}{\rho}\partial_{\rho}A_{log}\,,
\ee
and $J_{\tilde{A}}$ given in \Eq(JAtilde) with $A_{log}$, having the form
\bea
A^{log}=\log \rho \sum_{i=2}^{5}\frac{\rho^{i}}{(1+\rho)^{1+i}}\mathcal{A}_{1,i}\left[p_{0}(\tau,x),p_{2}(\tau,x)\right]
+\left(\log \rho\right)^{2}\sum_{i=4}^{5}\frac{\rho^{i}}{(1+\rho)^{1+i}}\mathcal{A}_{2,i}\left[p_{0}(\tau,x)\right]\,.\nonumber\\
\eea
The differential equation in \Eq(illum) will accordingly take the simple  matrix form
\be
\left(I_{x}\otimes D_{\rho}\right)\tilde{A}^{\tau_{i}}_{n,l}=-\left(J_{\tilde{A}^{\tau_{i}}}\right)_{n,l}-
\left(\partial_{\rho}\tilde{A}^{\tau_{i}}\right)_{n,l}\,,
\ee
and it is an easy exercise to implement the boundary condition \Eq(boundaryAtilde). Note that
the boundary condition of $\tilde{A}^{\tau_{i}}_{n,l}$ in \Eq(boundaryAtilde), depends on the coefficient $a_{2}$ 
defined in  \Eq(B_A). This means, in order to solve the set of the above equations, we need to provide
an initial profile  
\be
\left(a_{2}^{\tau_{i}}\right)_{n,l}\equiv a_{2}(\tau_{i},\rho,x)\,.
\ee
Our choice is $\left(a_{2}^{\tau_{i}}\right)_{n,l}=0$. Finally we will transform the value obtained from 
$\tilde{A}^{\tau_{i}}_{n,l}$ to $A^{\tau_{i}}_{n,l}$ according to \Eq(Atildedef) by integration.

\textbullet\, At this point, we have access to the value of the scalar field and all the components of the metric
in the whole plane of the lattice but only at the initial time $\tau_{i}$. The goal is to extend our computation
to later times. This being said, on the other hand,   we started the computation at the beginning of our numerical algorithm by introducing
 the initial profile for $\left(\Sigma^{\tau_{i}}_{b}\right)_{n,l}$ by hand. Clearly this initial profile at
different times must  evolve too. This brings us to the coupled equations of \Eq(source_beta)-\Eq(source_gamma)
\bea
\label{source_beta2}
&&\partial_{\rho}\beta-\frac{1}{\rho}\left[2\beta+\gamma\right]=-J_{\Sigma_{d}}\,,
\\
&&\label{source_gamma2}
\partial_{\rho}\gamma-\frac{1}{2\rho}\left[\beta+5\gamma\right]=-J_{\Sigma_{b}}\,,
\eea
with the corresponding assignments in \Eq(defbeta) and \Eq(defgamma),
\bea
\label{defbeta2}
\beta(\tau,\rho,x)&=&\partial_{\tau}\hat{\Sigma}_{d}(\tau,\rho,x)+\frac{\rho^4-1}{2}\,\partial_{\rho}\hat{\Sigma}_{d}(\tau,\rho,x)\,,
\\
\label{defgamma2}
\gamma(\tau,\rho,x)&=&\partial_{\tau}\hat{\Sigma}_{b}(\tau,\rho,x)+\frac{\rho^4-1}{2}\,\partial_{\rho}\hat{\Sigma}_{b}(\tau,\rho,x)\,,
\eea
and the sources $J_{\Sigma_{d}}$ and $J_{\Sigma_{b}}$ that are defined in \Eq(current-Jsigmad) and
 \Eq(current-Jsigmab). Since they are functions of the known fields at $\tau_{i}$, we can solve 
the coupled differential equations with the following boundary conditions
\be
\beta^{c}(\tau,0,x)=\gamma^{c}(\tau,0,x)=0\,.
\ee
Since on the lattice, we deal with  finite variables occasionally, we will be sloppy about mentioning the subindex $c$ for
the scalar field and various metric components.

Splitting the finite and logarithmic corrections in \Eq(source_beta2) and \Eq(source_gamma2), they take the form
\bea\label{betalog}
\partial_{\rho}\beta_{c}-\frac{1}{\rho}\left(\gamma_{c}+2\beta_{c}\right)
&=&
-J_{\Sigma_{d}}-\partial_{\rho}\beta_{log}+
\frac{1}{\rho}\left(2\beta_{log}+\gamma_{log}\right)\,,
\\
\label{gammalog}
\partial_{\rho}\gamma_{c}-\frac{1}{2\rho}\left(\beta_{c}+5\gamma_{c}\right)
&=&
-J_{\Sigma_{b}}-\partial_{\rho}\gamma_{log}
+\frac{1}{2\rho}\left(\beta_{log}+5\gamma_{log}\right)\,,
\eea
where the logarithmic correction to $\beta$ and $\gamma$ are calculated from \Eq(defbeta),
\bea
\beta^{log}&=&\log \rho \sum_{i=3}^{6}\frac{\rho^{i}}{(1+\rho)^{i}}\tilde{\mathcal{D}}_{1,i}\left[p_{0}(\tau,x),p_{2}(\tau,x)\right]
+\left(\log \rho\right)^{2}\sum_{i=5}^{6}\frac{\rho^{i}}{(1+\rho)^{i}}\tilde{\mathcal{D}}_{2,i}\left[p_{0}(\tau,x)\right]\,,\nonumber\\
\\
\gamma^{log}&=&\log \rho \sum_{i=3}^{6}\frac{\rho^{i}}{(1+\rho)^{i}}\tilde{\mathcal{B}}_{1,i}\left[p_{0}(\tau,x),p_{2}(\tau,x)\right]
+\left(\log \rho\right)^{2}\sum_{i=5}^{6}\frac{\rho^{i}}{(1+\rho)^{i}}\tilde{\mathcal{B}}_{2,i}\left[p_{0}(\tau,x)\right]\,.
\nonumber\\\eea
In the matrix form, we can rewrite \Eq(source_beta2)-\Eq(source_gamma2), in the following way
\bea
\begin{pmatrix}
I_{x}\otimes D_{\rho} & \!-\!\frac{1}{\rho} 
\\
\!-\!\frac{1}{2\rho} & I_{x}\otimes D_{\rho}\!-\!\frac{5}{2\rho} 
\end{pmatrix}
\begin{pmatrix}
\beta^{\tau_{i}}
\\
\gamma^{\tau_{i}}  
\end{pmatrix}_{n,l}  =-
\begin{pmatrix}
\tilde{J}_{\Sigma^{\tau_{i}}_{d}}
\\
\tilde{J}_{\Sigma^{\tau_{i}}_{b}}
\end{pmatrix}_{n,l}\,,  
\eea
with $\tilde{J}_{\Sigma^{\tau_{i}}_{d}}$ and $\tilde{J}_{\Sigma^{\tau_{i}}_{b}}$ that include terms such as 
$\beta_{log}$, $\gamma_{log}$ and their derivative as they appear on the right-hand side of \Eq(betalog)
and \Eq(gammalog). This yields $\beta^{\tau_{i}}_{n,l}$ and $\gamma^{\tau_{i}}_{n,l}$ at
 the initial time $\tau=\tau_{i}$. Now, similar to the procedure mentioned in detail for the scalar field 
$\phi^{\tau_{i}}_{n,l}$, we can perform  4 steps of RK method to evaluate  \Eq(defbeta2)-\Eq(defgamma2) for
 $\tau=\tau_{i}+\Delta\tau$. This is the last stage of our simulation and all the steps that we have done so far will be
repetitively performed until the desired final time $\tau=\tau_{f}$, is reached.

\addsubsubsection{Discretization}\label{Discretization}
In this section, we look at the effect of the discretization and possible sources of numerical artifacts. There are two main sources of numerical artifacts, the chosen number of points on the lattice and the chosen value for the time steps $\Delta\tau$. 
One advantage of having an observable as a function of two coordinates, is that numerical instabilities or artifacts if any are hard to miss. Therefore, the best way is just to change the number of lattice points and compare them.

For simplicity, all of the lattices that we have considered are square lattices with $N_{x}=N_{\rho}$. As an example, we compared $\phi(\tau,\rho,x)$ for lattices with $20\times20$, $30\times30$, $40\times40$ and $50\times50$ at two specific times, early before ($\tau=-3.75$) and long time after turning on the quench ($\tau=12$). 

\begin{figure}[!ht]
\centering
\begin{subfigure}{.5\textwidth}
{\includegraphics[width=7.0cm]{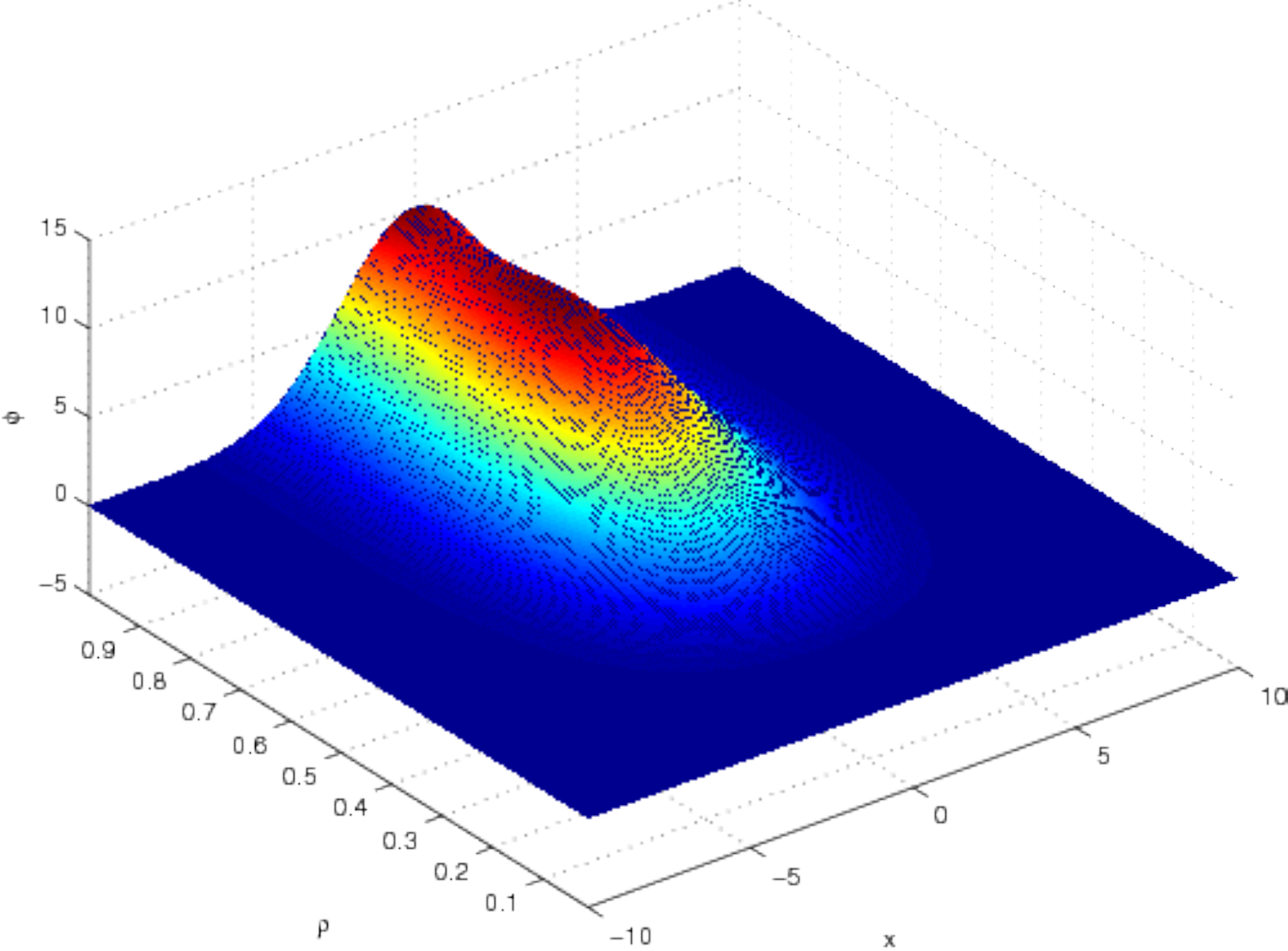}}
\end{subfigure}%
\begin{subfigure}{.5\textwidth}
  \centering
{\includegraphics[width=7.0cm]{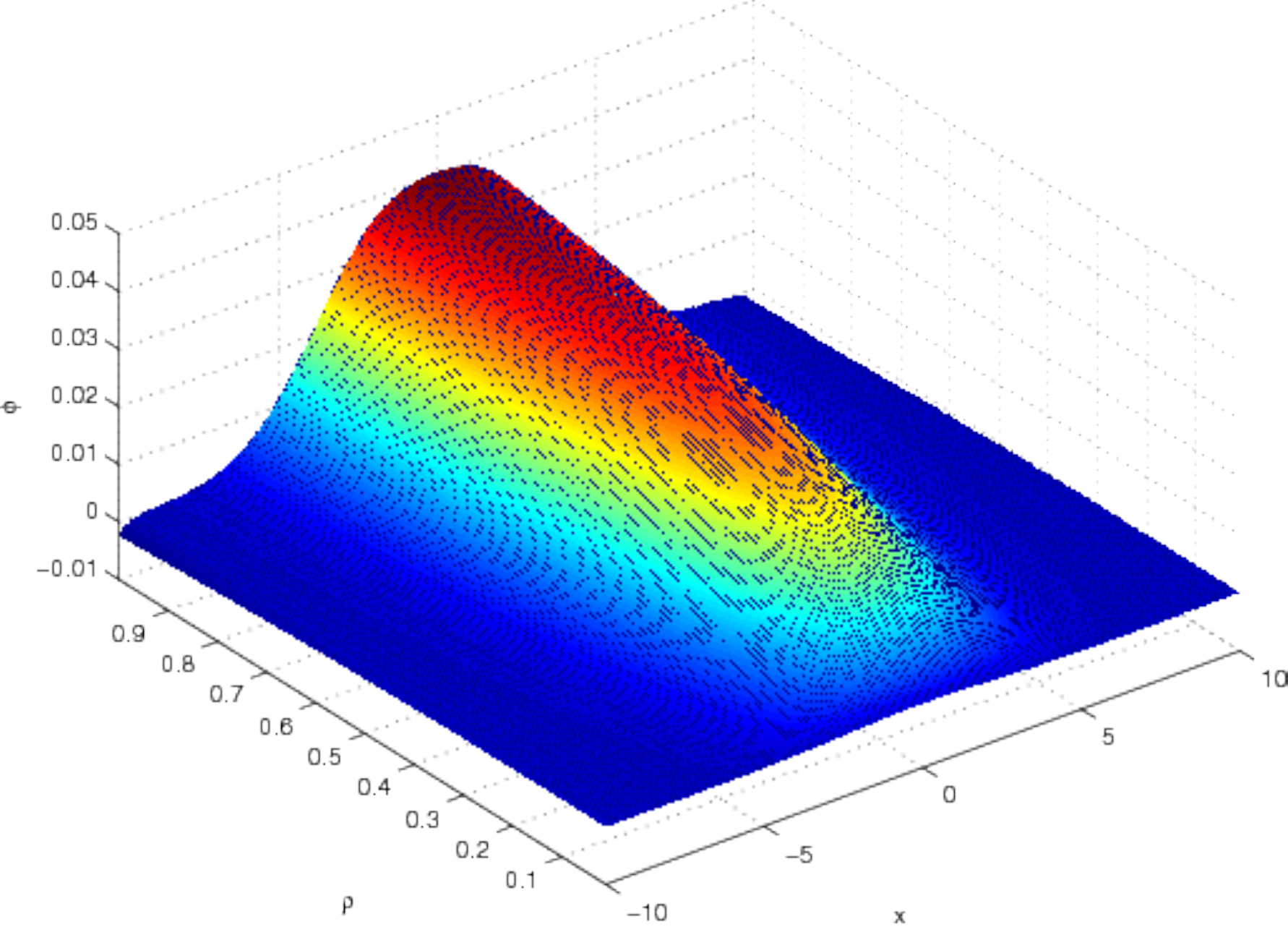}}
\end{subfigure}%

\centering
\begin{subfigure}{.5\textwidth}
{\includegraphics[width=7.0cm]{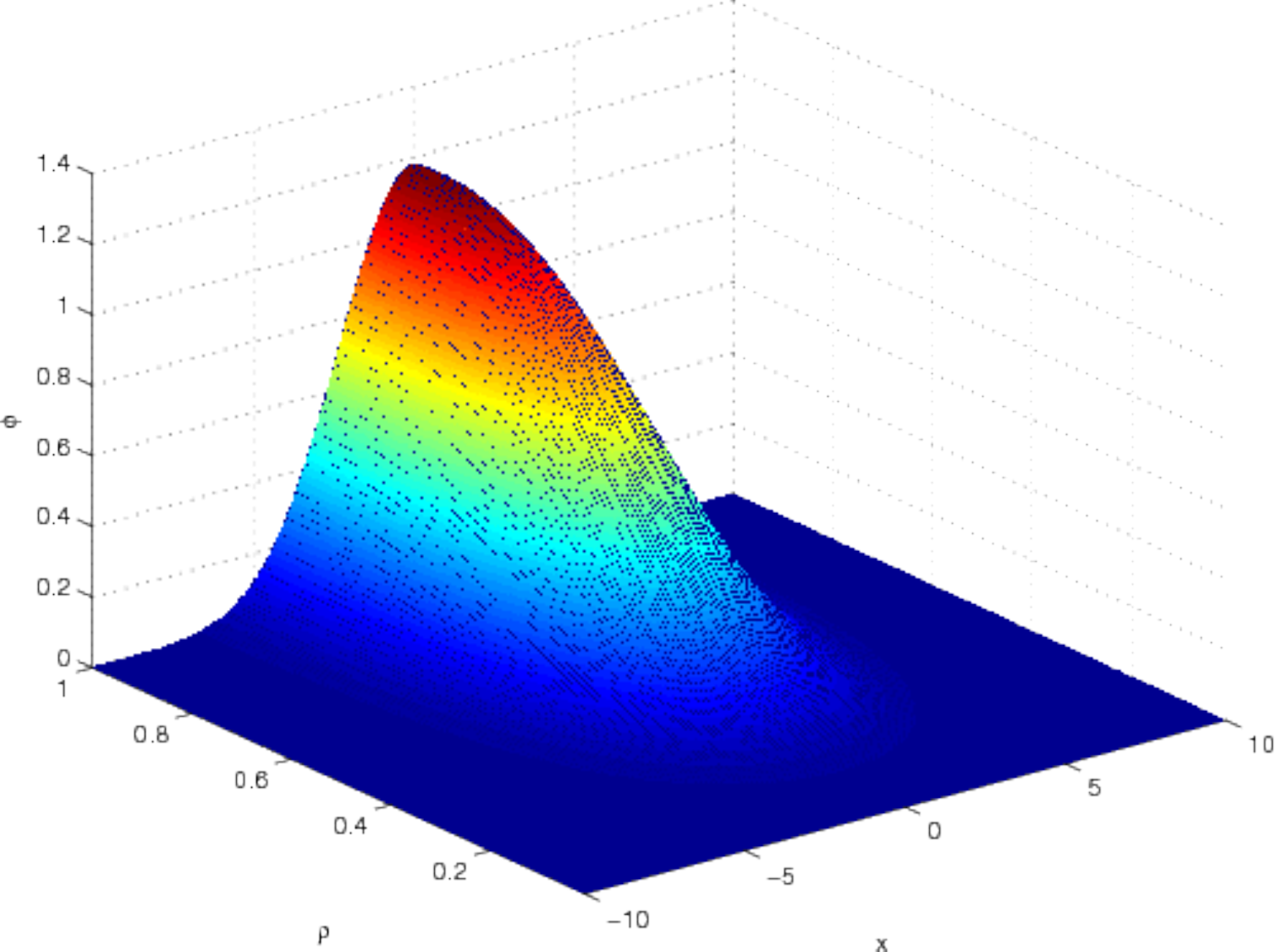}}
\end{subfigure}%
\begin{subfigure}{.5\textwidth}
  \centering
{\includegraphics[width=7.0cm]{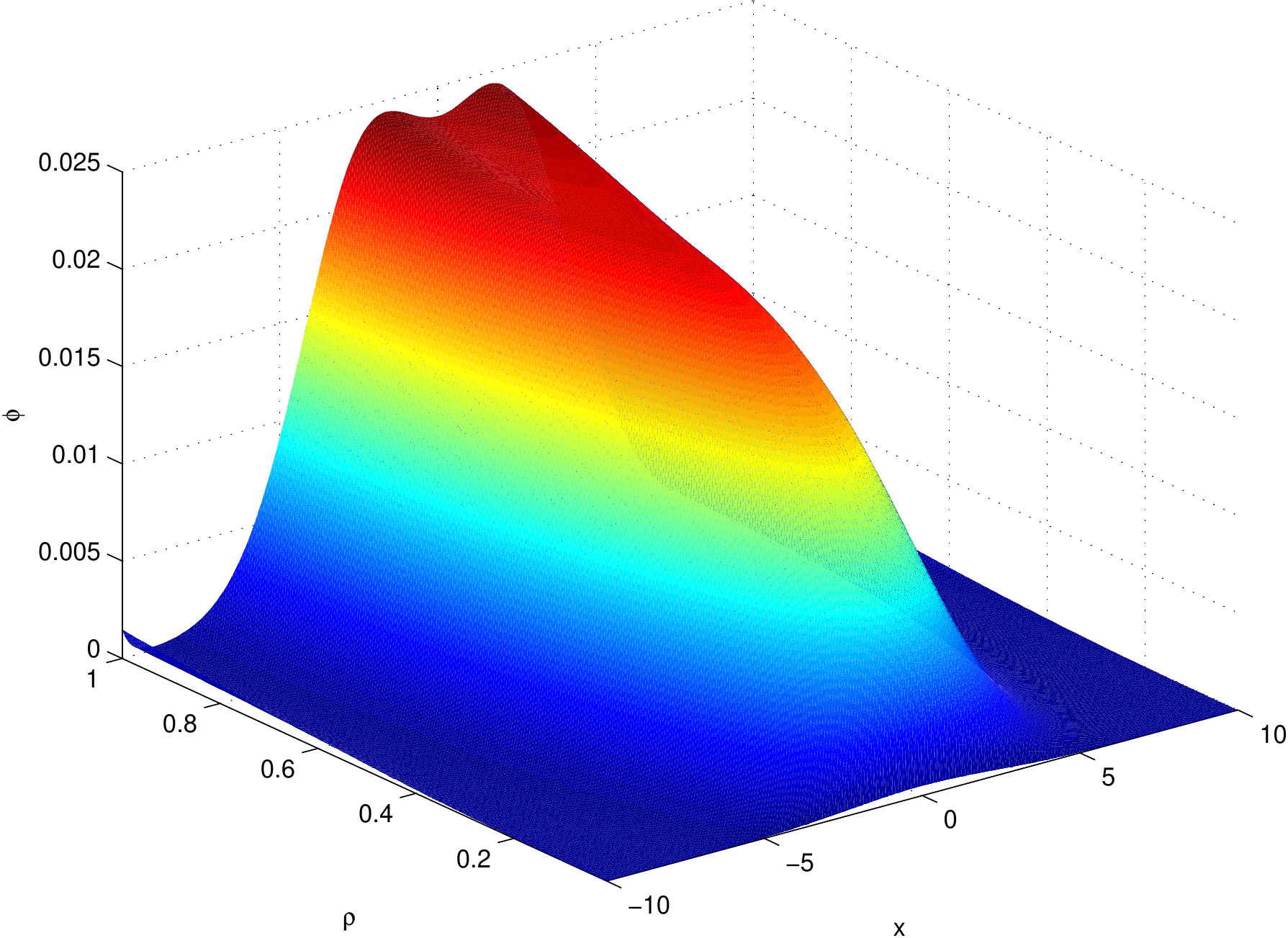}}
\end{subfigure}%

\centering
\begin{subfigure}{.5\textwidth}
{\includegraphics[width=7.0cm]{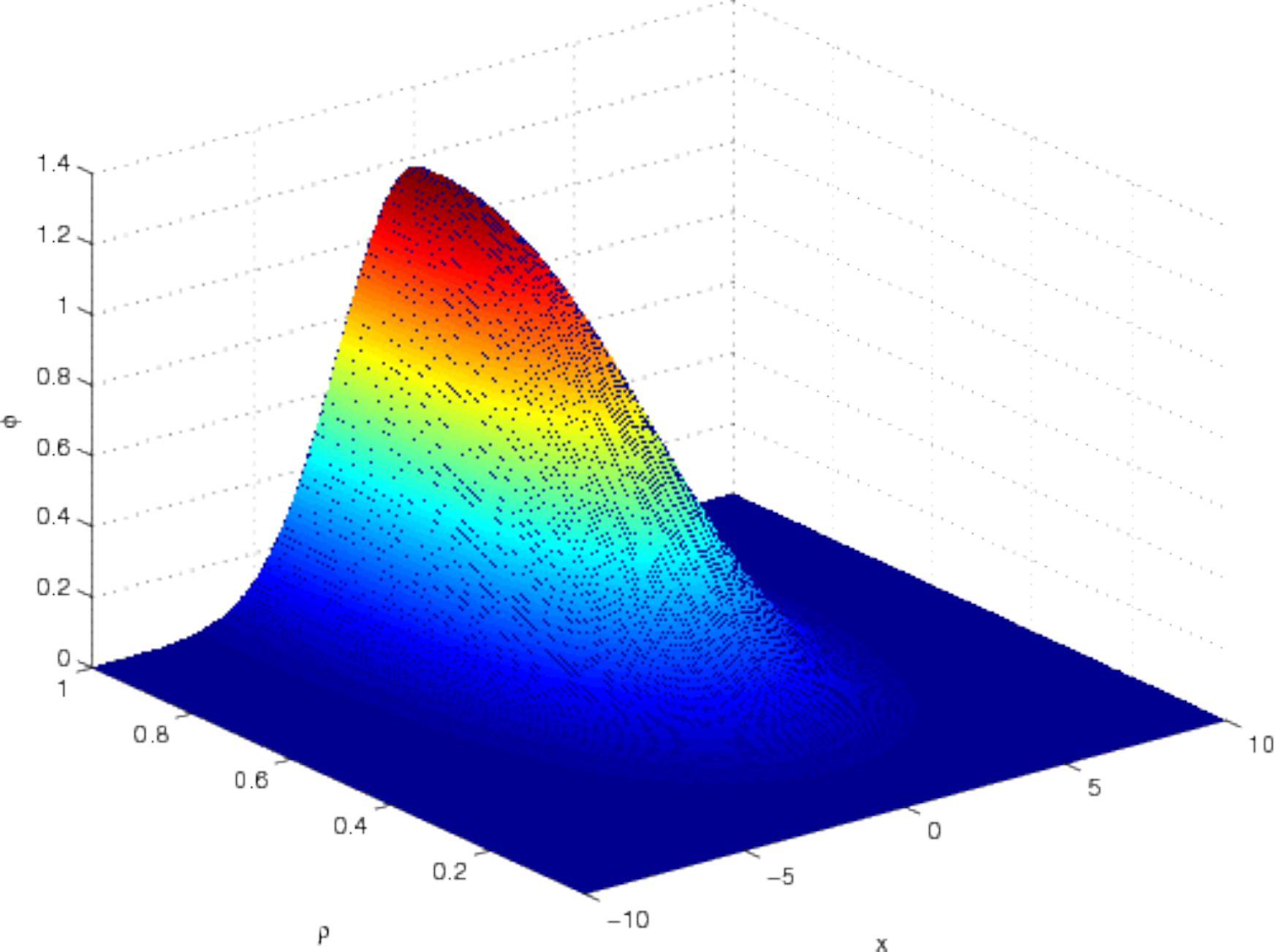}}
\end{subfigure}%
\begin{subfigure}{.5\textwidth}
  \centering
{\includegraphics[width=7.0cm]{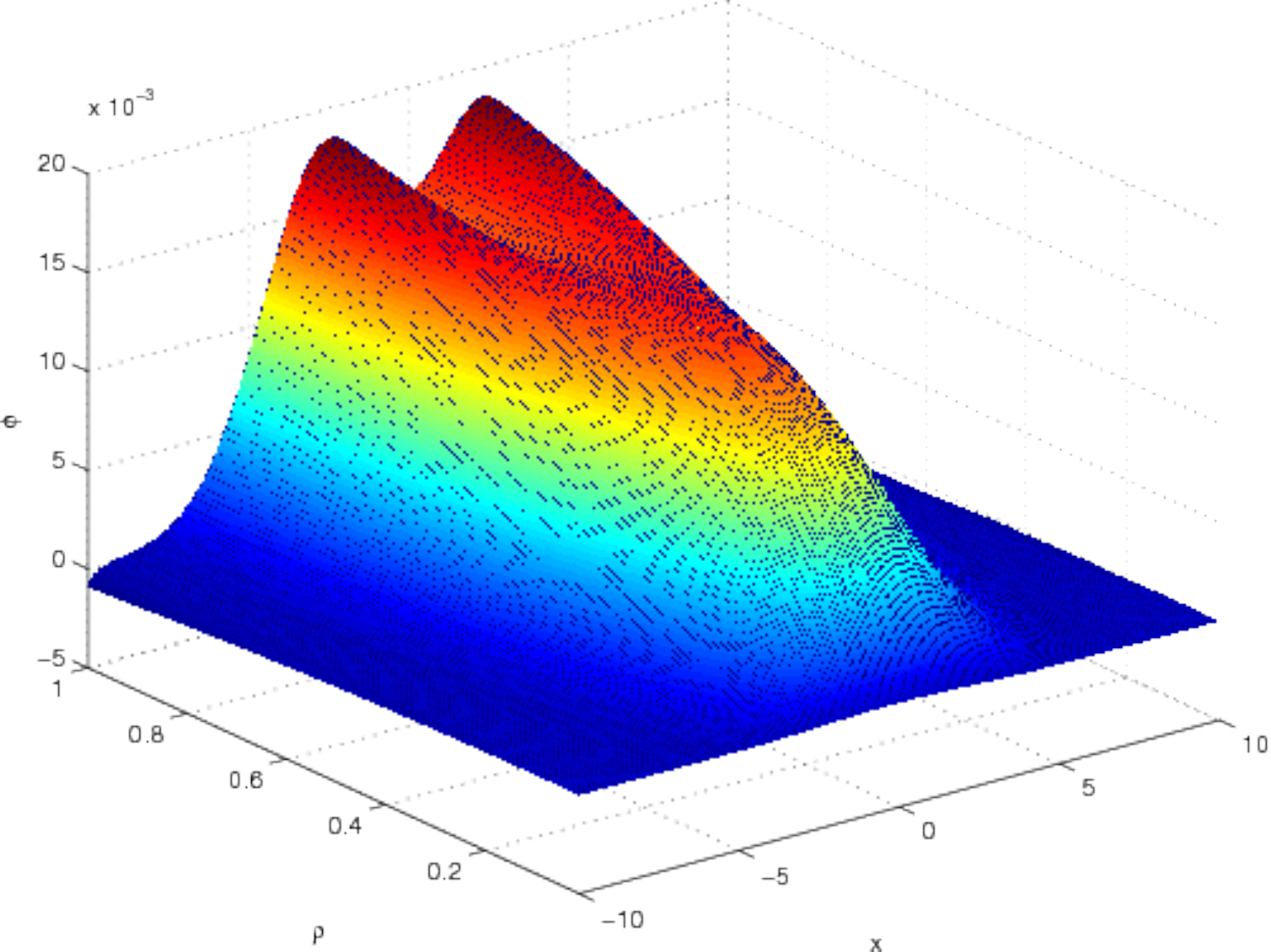}}
\end{subfigure}%

\centering
\begin{subfigure}{.5\textwidth}
  \centering
{\includegraphics[width=7.0cm]{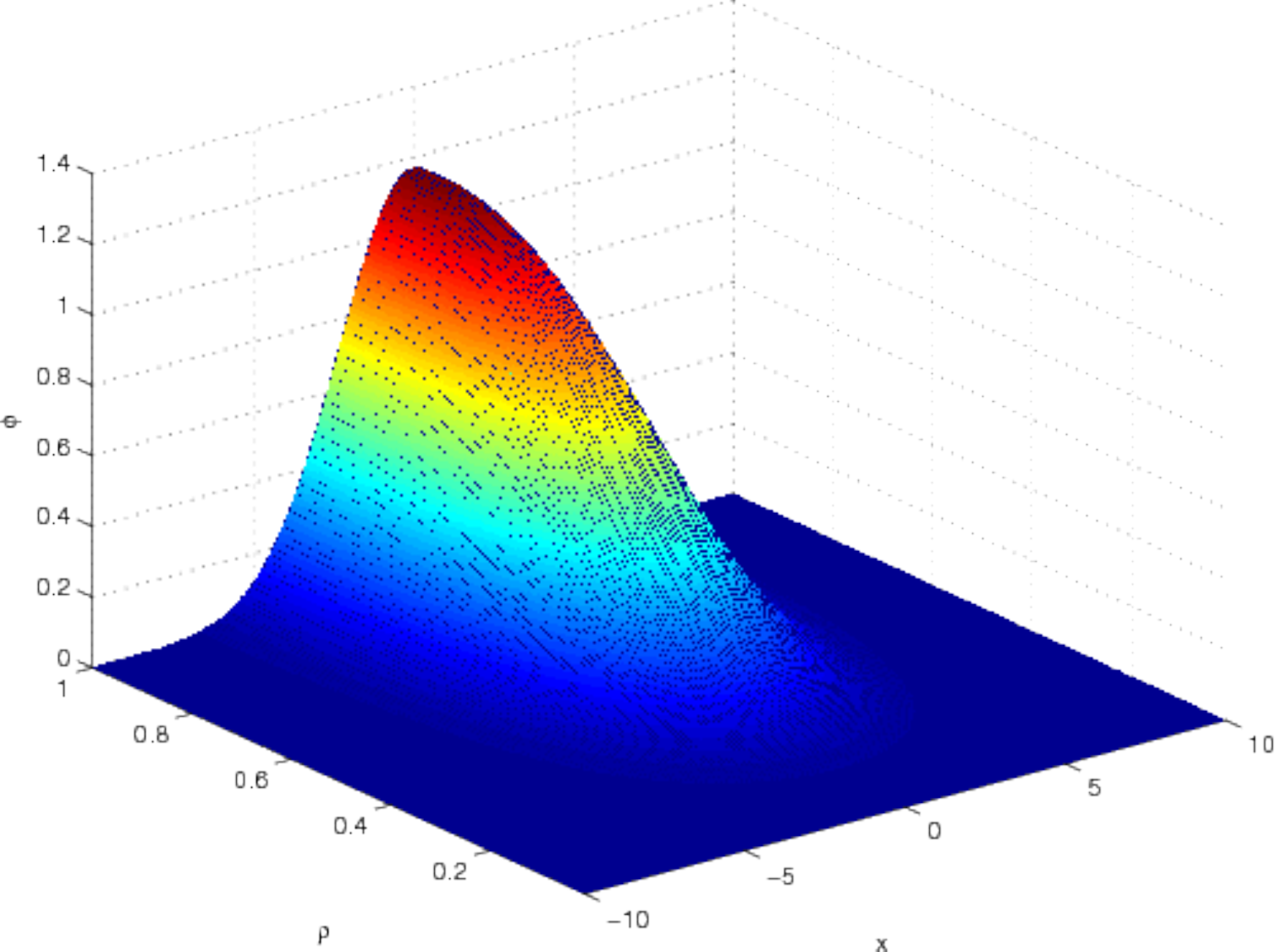}}
\end{subfigure}%
\begin{subfigure}{.5\textwidth}
  \centering
{\includegraphics[width=7.0cm]{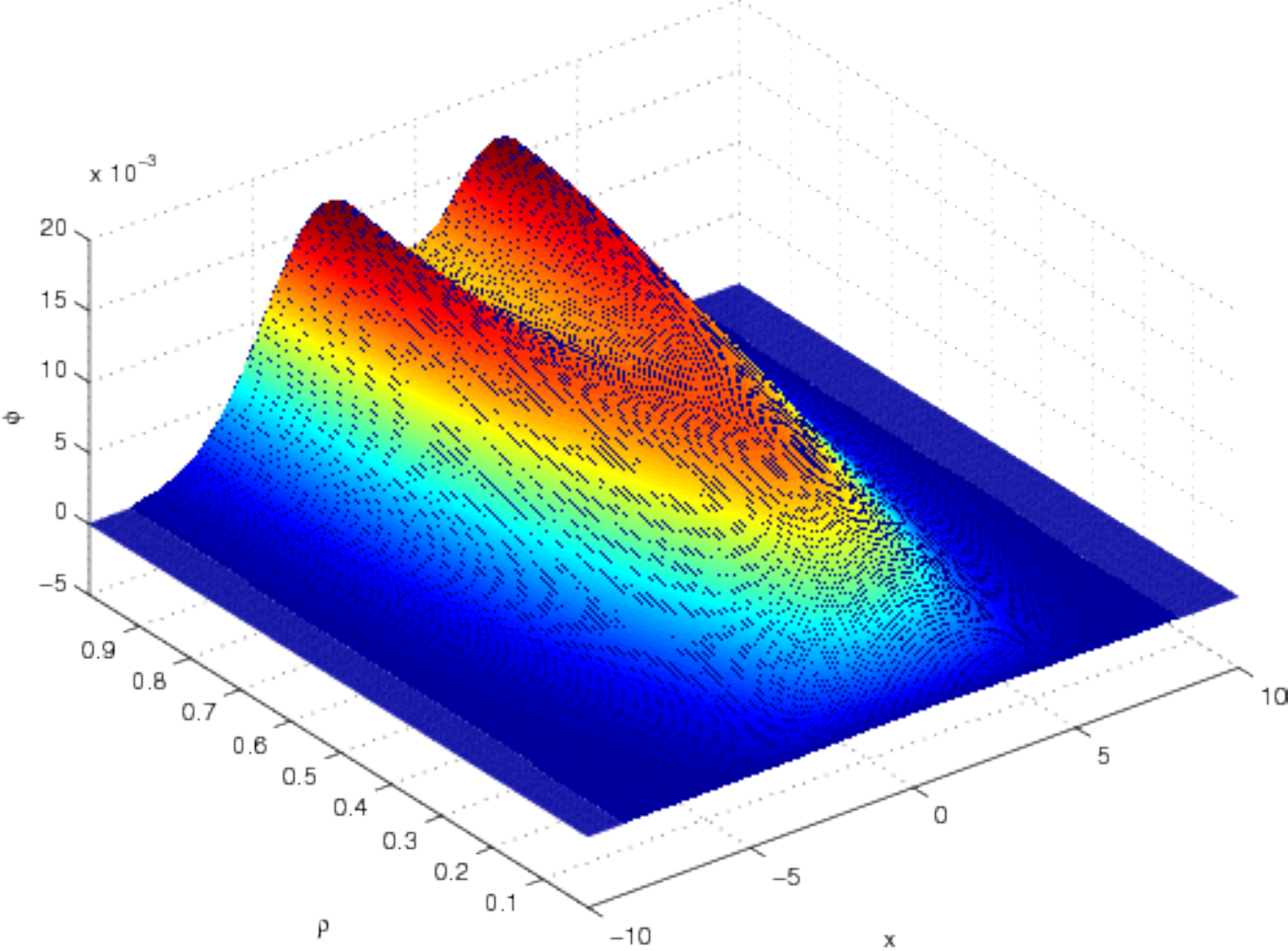}}
\end{subfigure}%
\caption{Plots of $\phi(\tau,\rho,x)$ at two specific times. On the left-hand sides, the quench hasn't been switched on. Specifically these plots show the configuration at $\tau=-3.75$. At some time long after the quench for instance, $\tau=12$, the profiles for $\phi(\tau,\rho,x)$ are shown on the right-hand sides. Dimensions of  lattices from  the first to the last row are respectively given by $20\times20$, $30\times30$, $40\times40$ and $50\times50$. Fixed parameters are $\alpha=1$ and $\sigma=\sqrt{L_{x}}$.}
\end{figure}
\clearpage
The other source of numerical instability is the value chosen for the marching  steps in the Runge-Kutta method. Below, we compare one of the observables computed in the paper, $\mathcal{L}_{2}$, the two-point function for case I, for two different time steps, $\Delta\tau$ and $\frac{\Delta\tau}{2}$.   

\begin{figure}[!ht]
\centering
\begin{subfigure}{.5\textwidth}
{\includegraphics[width=7.0cm]{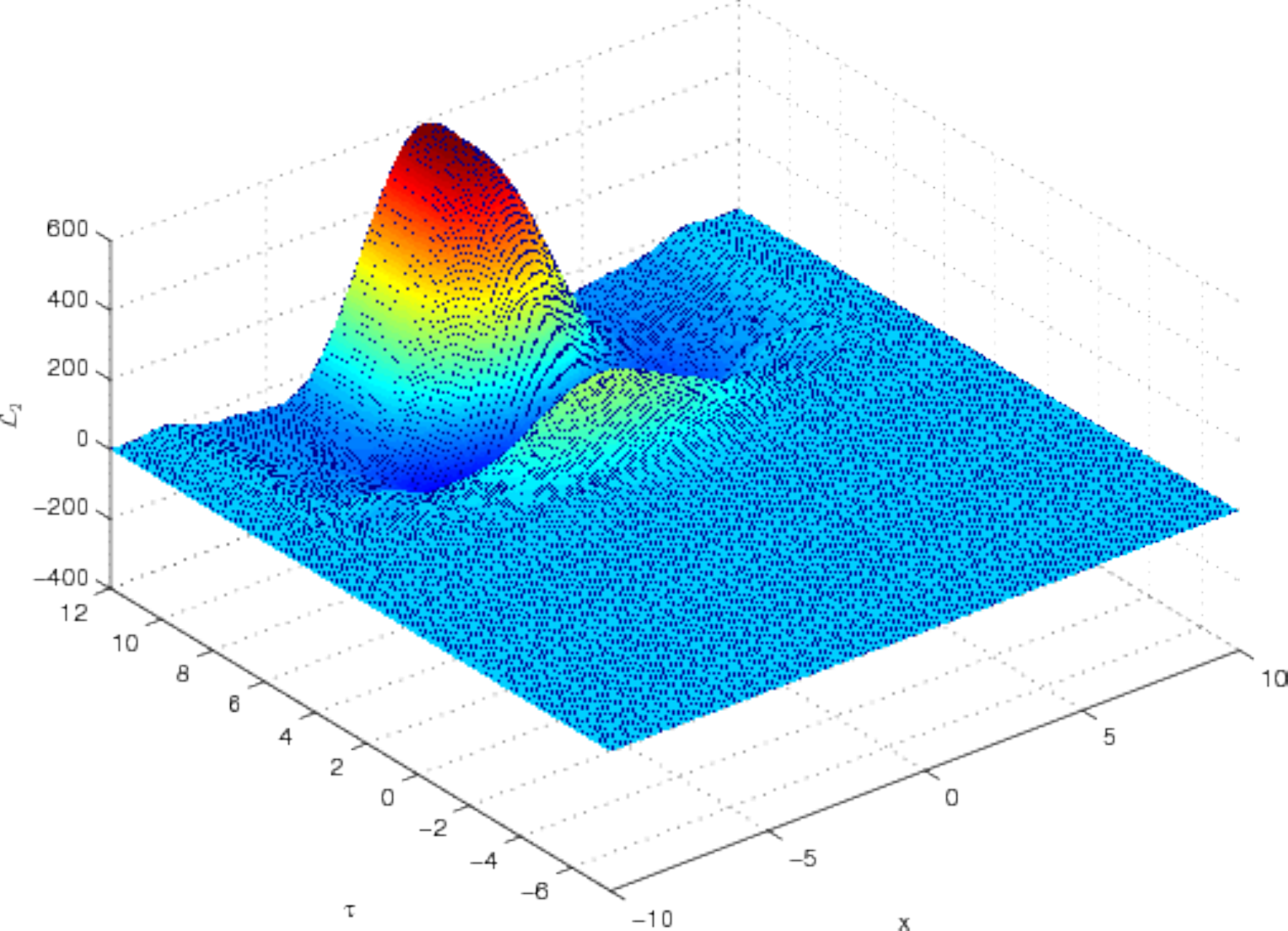}}
  \caption{}
  \label{fig:append1}
\end{subfigure}%
\begin{subfigure}{.5\textwidth}
  \centering
{\includegraphics[width=7.0cm]{L2A20points2cc-crop}}
  \caption{}
  \label{fig:append2}
\end{subfigure}%
\caption{The effect of the numerical instabilities coming from the RK method shown for the two-point function. Here in both diagrams $N_{x}=N_{\rho}=20$. The time step for the left-hand diagram is $\Delta\tau$ and  half of this value for the plot on the right-hand side. }
\end{figure}

Numerical instabilities that are produced this way are specifically dominant for a region near $\rho\sim 1$. Observables in our computation, such as entanglement entropies that depend on  Taylor expansions of  the metric near $\rho\sim 0$ are the least affected quantities by these instabilities. This is mainly due to the fact that we have calculated their expansions upto $\mathcal{O}(\rho^{8})$ analytically using the Einstein equations. This in part allows us using fewer number of lattice points for simulating them.

This check is the most time consuming part. For a lattice of $N_{x}=N_{\rho}=20$ points, this takes roughly a month. For a lattice of $N_{x}=N_{\rho}=30$ this process takes  two months. Different observables for various parameters have been executed on different nodes. It takes three days to produce a single plot for a parallelized code on a sixteen-core node.

\begin{figure}[!ht]
\centering
{\includegraphics[width=15.0cm]{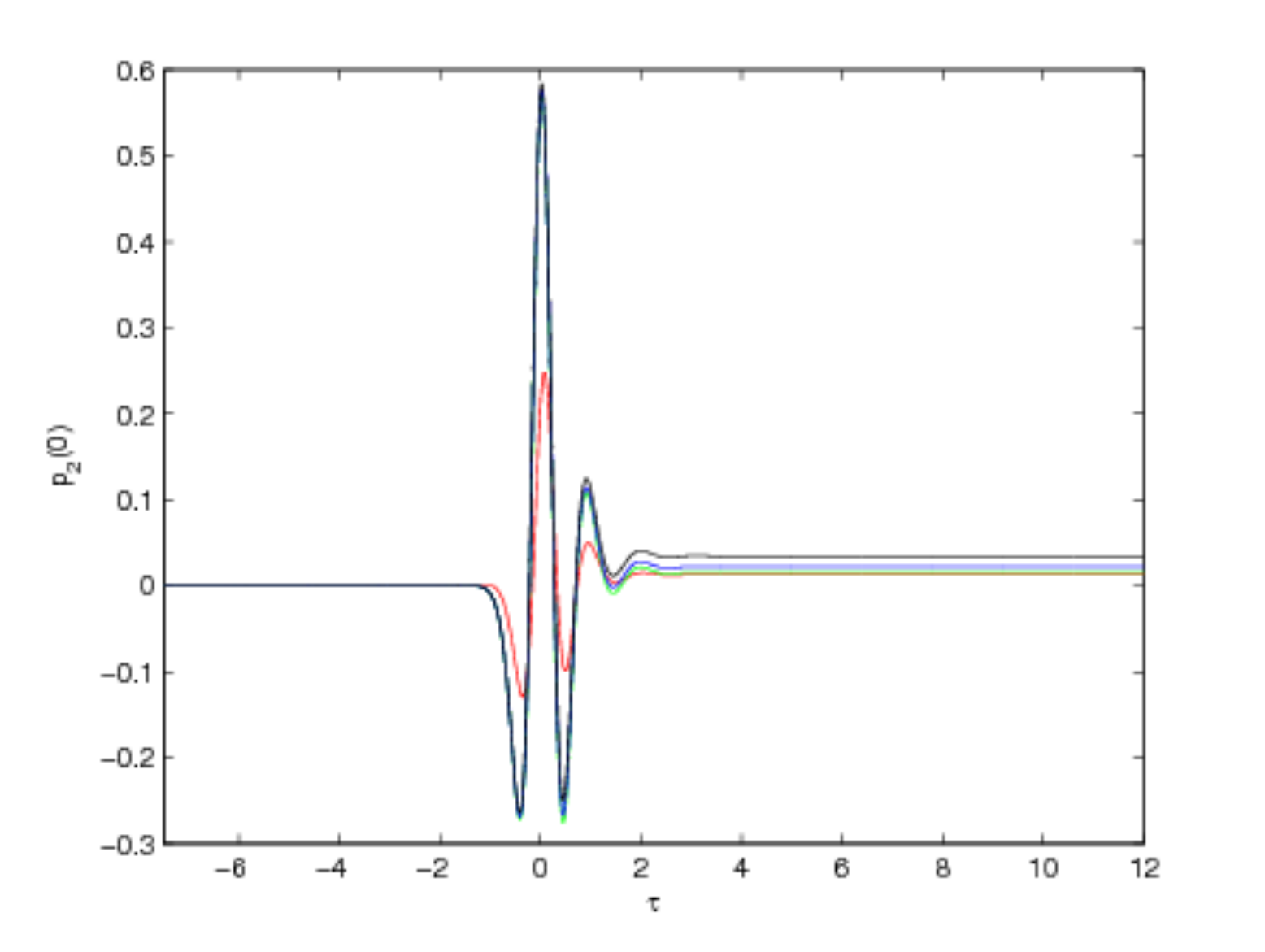}}
\caption{ Evolution of the normalizable mode of the scalar field on various lattice sizes. Black color for $N_{x}=N_{\rho}=20$, blue for $N_{x}=N_{\rho}=30$, green for $N_{x}=N_{\rho}=40$ and
red for $N_{x}=N_{\rho}=50$.}
 \label{fig:phi_plot}
\end{figure}

\addsubsubsection{Thermalization}\label{Thermalization}
In the same category, we look at the effects of the lattice artifacts on the thermalization. The normalizable mode in the expansion of
the bulk scalar allows us to observe this since this is the response to the mass gap. Practically, the numerical algorithm was designed to stop when the
standard deviation from the mean value goes below $~10^{-11}$ while in the trend towards thermalization. We plot the dynamical evolution of this component of the scalar 
field for various lattice sizes in Figure \ref{fig:phi_plot}. The standard deviations from the mean values at late times are given in table below.

\begin{center}
\begin{tabular}{ |p{2cm}|p{2cm}||p{2cm}| }
 \hline
 \multicolumn{3}{|c|}{Measure of thermalization } \\
 \hline
 $N_{x}$ or $N_{\rho}$ & $N_{\tau}$ &standard \hspace{0.9cm} deviation\\
 \hline
 $20$   & $7810$  & $8.9 \times 10^{-12}$\\
 $30$   & $17560$ & $3.5 \times 10^{-13}$\\
 $40$   & $31210$ & $1.7 \times 10^{-13}$\\
 $50$   & $41272$ & $2.1 \times 10^{-13}$\\
 \hline
\end{tabular}

\end{center}

\end{document}